\documentclass[fleqn,10pt]{wlscirep}
\usepackage[utf8]{inputenc}
\usepackage[T1]{fontenc}

\usepackage[version=3]{mhchem} 
\usepackage{subcaption}
\usepackage[colorinlistoftodos]{todonotes}
\usepackage{enumitem}
\usepackage{mathtools}
\usepackage{afterpage}

\title{H$_2$-roaming dynamics in the formation of H$_{3}^{+}$ following two-photon double ionization of ethanol and aminoethanol}

\author[1]{Aaron Ngai}
\author[1, *]{Sebastian Hartweg}
\author[2]{Jakob D. Asmussen}
\author[3]{Björn Bastian}
\author[4, 5]{Matteo Bonanomi}
\author[6]{Carlo Callegari}
\author[6]{Miltcho Danailov}
\author[6]{Michele di Fraia}
\author[7]{Raimund Feifel}
\author[1]{Sarang Dev Ganeshamandiram}
\author[8]{Sivarama Krishnan}
\author[9]{Aaron LaForge}
\author[1]{Friedemann Landmesser}
\author[2]{Ltaief Ben Ltaief}
\author[1]{Moritz Michelbach}
\author[6]{Nitish Pal}
\author[6]{Oksana Plekan}
\author[1]{Nicolas Rendler}
\author[6]{Lorenzo Raimondi}
\author[1]{Fabian Richter}
\author[1]{Audrey Scognamiglio}
\author[1]{Tobias Sixt}
\author[7]{Richard J. Squibb}
\author[10]{Katrin Dulitz}
\author[1]{Frank Stienkemeier}
\author[2]{Marcel Mudrich}

\affil[1]{Institute of Physics, Albert-Ludwigs-Universität Freiburg, Freiburg, Germany}
\affil[2]{Department of Physics and Astronomy, Aarhus University, Denmark}
\affil[3]{Wilhelm-Ostwald-Institut für Physikalische und Theoretische Chemie, Universität Leipzig, Germany}
\affil[4]{Dipartimento di Fisica Politecnico, Milano, Italy}
\affil[5]{Istituto di Fotonica e Nanotecnologie (CNR-IFN) Milano, Italy}
\affil[6]{Elettra — Sincrotrone Trieste S.C.p.A., Basovizza, Trieste, Italy}
\affil[7]{Department of Physics, University of Gothenburg, Gothenburg, Sweden}
\affil[8]{Department of Physics, Indian Institute of Technology Madras, Chennai, India}
\affil[9]{Department of Physics, University of Connecticut, Storrs, Connecticut, US}
\affil[10]{Institut für Ionenphysik und Angewandte Physik, Universität Innsbruck, 6020 Innsbruck, Austria}
\affil[*]{Corresponding author: sebastian.hartweg@physik.uni-freiburg.de}


\begin{abstract}
Roaming reactions involving a neutral fragment of a molecule that transiently wanders around another fragment before forming a new bond are intriguing and peculiar pathways for molecular rearrangement.
Such reactions can occur for example upon double ionization of small organic molecules, and have recently sparked much scientific interest. 
We have studied the dynamics of the H$_2$-roaming reaction leading to the formation of H$_3^+$ after two-photon double ionization of ethanol and 2-aminoethanol, using an XUV-UV pump-probe scheme.
For ethanol, we find dynamics similar to previous studies employing different pump-probe schemes, indicating the independence of the observed dynamics from the method of ionization and the photon energy of the disruptive probe pulse.
Surprisingly, we do not observe a kinetic isotope effect in ethanol-D$_6$, in contrast to previous experiments on methanol where such an effect was observed.
This distinction indicates fundamental differences in the energetics of the reaction pathways as compared to the methanol molecule.
The larger number of possible roaming pathways compared to methanol complicates the analysis considerably.
In contrast to previous studies, we additionally analyze a broad range of dissociative ionization products, which feature distinct dynamics from that of H$_{3}^{+}$ and allow initial insight into the action of the disruptive UV-probe pulse.
\end{abstract}

\begin{document}

\setcounter{page}{0}
\afterpage{%
	\thispagestyle{empty}
    This article may be downloaded for personal use only. Any other use requires prior permission of the author and Springer Nature.
    
    This article appeared in [Ngai \textit{et al.}, "H$_2$-roaming dynamics in the formation of H$_{3}^{+}$ following two-photon double ionization of ethanol and aminoethanol", \textit{Sci. Rep.} \textbf{15}, 3201 (2025)] and may be found at (\url{https://doi.org/10.1038/s41598-024-84531-9}).
}
\clearpage
\clearpage

\flushbottom
\maketitle

\thispagestyle{empty}

\section*{Introduction}
The tri-atomic hydrogen cation H$_3^+$ belongs to the most abundantly produced cations in interstellar space and is considered a key driver of interstellar chemistry\cite{Oka_2006_PNatlAcadSciUSA}. 
It is believed that most H$_3^+$ is formed by ion-neutral reactions between molecular hydrogen and its cation $\textrm{H}_2+\textrm{H}_2^+\rightarrow\textrm{H}_3^++\textrm{H}$, a reaction first observed in hydrogen plasmas\cite{Hogness_1925}.
Some small organic molecules also produce H$_3^+$ ions in their fragmentation induced by single and double ionization\cite{Eland_1992,Eland_1996_RapidCommunMassSp,Furukawa_2005,Okino_2006_ChemPhysLett,Hoshina_2008_JChemPhys}. Such fragmentation reactions induced by impact of charged particles\cite{De_2006_PhysRevLett} or energetic photons \cite{Piling_2007_MonNotRAstronSoc} have been suggested as additional source of H$_3^+$ ions in outer space.
While the production of H$_3^+$ from ionized organic molecules was originally attributed mostly to the abstraction from methyl (CH$_{3}$) groups, these tri-atomic cations have also been observed after the ionization of molecules that do not contain
methyl groups\cite{Hoshina_2008_JChemPhys,Ideböhn_2022,Mebel_2008}.
In organic molecules without CH$_3$ groups, the production of H$_3^+$ requires significant rearrangement of chemical bonds and the transfer of hydrogen atoms over large distances.
Such reactions have recently been explained by a roaming mechanism that also significantly contributes to the formation of H$_3^+$ from molecules that do contain a CH$_{3}$ group\cite{Ekanayake_2017_SciRep}.
The proposed roaming mechanism proceeds via the dissociation of a neutral molecular H$_2$ fragment from a CH$_3$ or CH$_2$ group of the parent dication as RH(H$_{2}$)$^{++}\rightarrow\text{RH}^{++}+\text{H}_{2}$\cite{Ekanayake_2018_NatComm}.
The neutral molecular fragment stays in the vicinity of the remaining RH$^{++}$ until it abstracts a proton from the dicationic moiety $\textrm{RH}^{++}+\textrm{H}_2\rightarrow\textrm{R}^++\textrm{H}_3^+$. 
The resulting H$_3^+$ cation and now only singly-charged cationic fragment R$^+$ subsequently undergo a Coulomb explosion leading to a high kinetic energy release.
The roaming reaction in the formation of H$_3^+$ from ionized organic molecules has received significant scientific attention since its discovery, in part due to the similarity of the final proton transfer step to the astrochemically highly relevant bimolecular reaction between H$_2$ and H$_2^+$\cite{Hogness_1925}.
The reaction has, for example, been studied in methanol\cite{Ekanayake_2017_SciRep,Luzon_2017_PhysChemChemPhys,Luzon_2019_JPhysChemLett,Livshits_2020_CommunChem, Ando_2018_CommunChem, Gope_2022_SciAdv, Nakai_2013_JChemPhys}, larger alcohols\cite{Ekanayake_2018_NatComm,Bittner_2022_JChemPhys,Wang_2023_JPhysChemLett,Gope_2021_NatSci,Gope_2023_PhysChemChemPhys, Wang_2020_JPhysChemA}, thiols\cite{Ekanayake_2018_JChemPhys} and other small molecules\cite{Ideböhn_2022,Hoshina_2008_JChemPhys,Zhang_2019_PhysRevA,Zhang_2019_JChemPhys,Majima_2014_PhysRevA,Townsend_2021_FrontPhys}, and even clusters\cite{Wang_2021}, involving various ionization mechanisms.
Similar roaming reactions, following non-traditional reaction pathways far from the typical minimum energy trajectories, avoiding tight transition states, have previously been studied in neutral molecules, with the unimolecular photodissociation of formaldehyde being the prototypical example\cite{Townsend_2004_Science,Endo_2020}.

Studies of the dynamics of H$_3^{+}$ formation from alcohols have so far employed near-infrared (NIR) single-color\cite{Ekanayake_2017_SciRep,Ekanayake_2018_NatComm}, extreme ultraviolet (XUV) single-color\cite{Wang_2023_JPhysChemLett} as well as XUV-IR two-color\cite{Livshits_2020_CommunChem} pump-probe schemes. 
While the photon energies used in these schemes differ, they all describe the observed dynamics in the framework of a disruptive probe.
This means that the probe pulse depletes the H$_3^+$ yield by acting on some reaction intermediate. 
If this reaction intermediate is susceptible to the action of the probe pulse during the entirety of the roaming process, i.e. from double ionization to the final formation of H$_3^+$, the observed dynamics corresponds directly to the dynamics of the roaming reaction.
But without knowledge of the reaction intermediate and the action of the probe pulse, this is however not clear a priori.
The possibility that the disruptive probe captures only part of the dynamics occurring in the dicationic fragment, or also acts on the final products, cannot be easily excluded.

The complexity of dynamical studies of the roaming process is further enhanced by the fact that reactions involving a roaming neutral H$_2$ do not exclusively produce H$_3^+$. 
For example, in direct competition with the proton transfer leading to formation of H$_3^{+}$, the roaming H$_2$ can transfer an electron to the RH$^{++}$ dication, leading to the formation and Coulomb explosion of H$_2^+$ and RH$^+$\cite{Gope_2022_SciAdv,Wang_2023_JPhysChemLett}.
The roaming H$_2$ can, in principle, also contribute to the formation of H$_3$O$^+$ ions \cite{Kling_2019_NatComm} from doubly ionized alcohols.
Finally, the neutral H$_2$ can also just wander off, leaving the dicationic RH$^{++}$ to follow its own unimolecular dissociation processes.
As long as the neutral roamer cannot be observed in real time, as it was possible for the roaming hydrogen atom in the dissociation of formaldehyde\cite{Endo_2020}, it may be advantageous to not focus solely on the dynamical yield of H$_3^+$ ions, but to also discuss the dynamics in comparison to other fragment ions as will be subject to this work.

We present here a study of the fragmentation of ethanol and 2-aminoethanol after two-photon double ionization with XUV pulses from the Free-Electron Laser (FEL) FERMI. 
We probe the dynamics disruptively with a UV laser pulse of 392~nm (3.16~eV); let us note that this photon energy is intermediate between that of the NIR and XUV pulses used in previous studies\cite{Ekanayake_2018_NatComm, Ekanayake_2018_JChemPhys, Wang_2023_JPhysChemLett}.
H$_3^+$ formation in doubly-ionized ethanol has been studied less extensively than in methanol, but there exist previous studies for comparison that potentially allow to isolate effects of the different probe energies.
We detect pump-probe delay-dependent mass spectra and observe transiently depleted or enhanced ion yields for various fragment ions, providing a general picture of the fragmentation dynamics of ethanol after single and double ionization.
We compare data for FEL pulses of photon energies above and below the single-photon double ionization threshold, verifying that the dominating formation pathway of molecular dications is a two-photon process, also accessible below the single-photon double-ionization threshold.
We also compare ethanol to 2-aminoethanol to observe a chemical substitution effect, as the latter differs from the former only through an additional NH$_{2}$ group, that can, for example, accept or donate protons.

\section*{Results and Discussion}

\begin{figure}[ht]
    \includegraphics[width=\linewidth]{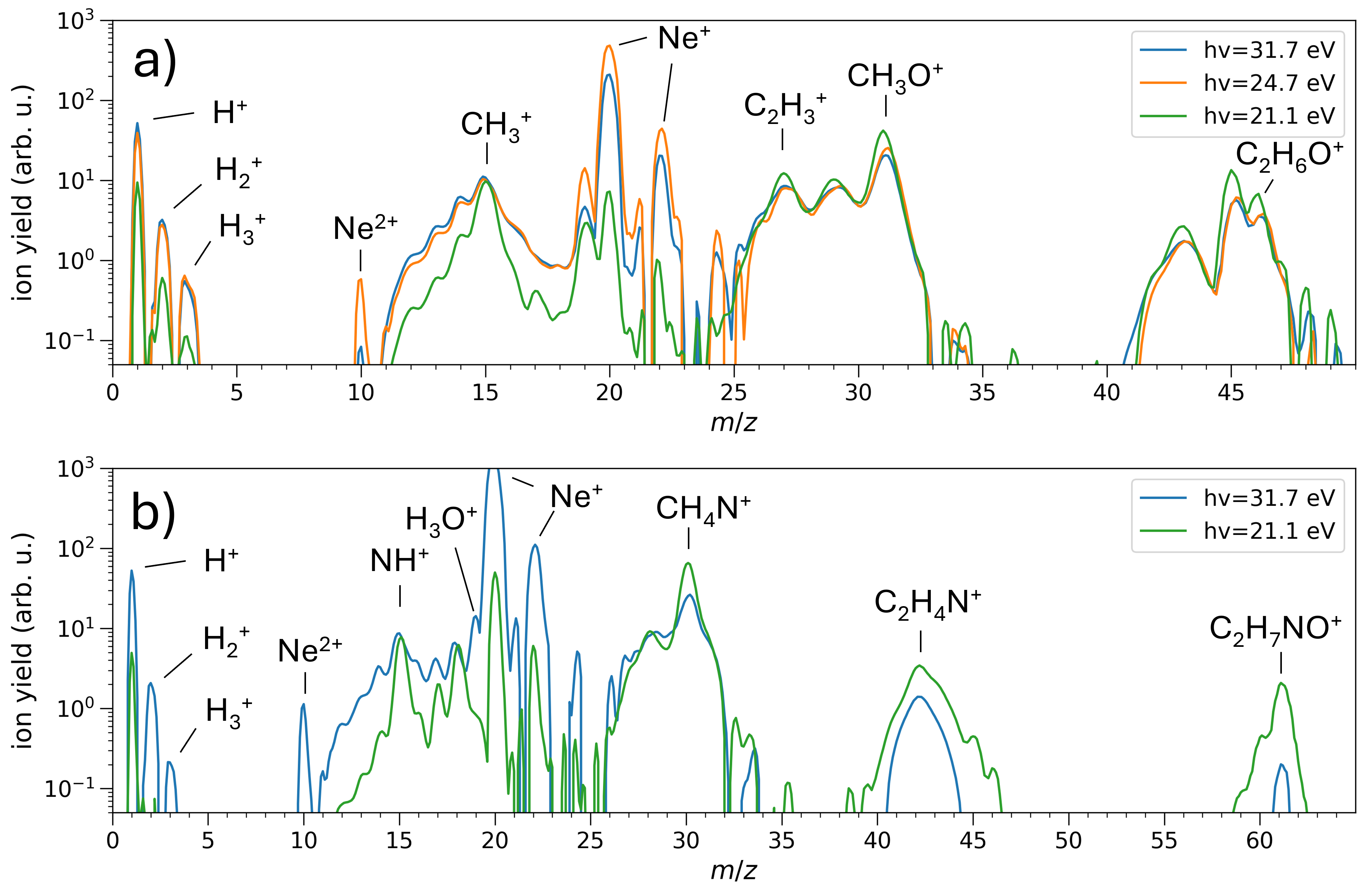}
    \caption[]{\label{fig:ethanol_mass_spectrum} Typical mass spectra recorded after the photoionization of a) ethanol and b) 2-aminoethanol using XUV pulses at different photon energies.}
\end{figure}

Typical mass spectra obtained after the photoionization of ethanol at above (31.7\,eV) and below (24.7\,eV, 21.1\,eV) the lowest vertical double ionization energy (29.6\,eV)~\cite{Linusson_2009_PhysRevA} are shown in the top panel of Fig.~\ref{fig:ethanol_mass_spectrum}.
The mass spectrum of ethanol after photoionization is well known and all mass peaks have been observed and described before~\cite{Niwa_1982_IntJMassSpectrom}. 
In addition to the known peaks, the mass spectra recorded in this study show pronounced peaks at $m/z=20$ and $m/z=22$ arising from the neon carrier gas used in the molecular beam expansion.
Ringing effects caused by these intense signals may prevent the observation of possible weak signals within the range 20$\leq m/z\leq$26. 
We additionally give yields for all observed ion masses relative to the ethanol parent ion in Table~\ref{tab:fragment_abundances}.
To obtain ion yields for partially overlapping peaks in the mass spectra, we separate the individual contributions using masking filters derived from Gaussian fits (see supplementary material).

\begin{table}
    \centering
    \caption[]{\label{tab:fragment_abundances} Relative ion yields for ethanol (C$_{2}$H$_{6}$O, C$_{2}$D$_{6}$O) upon XUV ionization.}
    \begin{tabular}{c | c | c | c | c || c | c | c}
        \hline\hline
        $m/z$ from & possible & relative yield at & relative yield at  & relative yield at & $m/z$ from & possible & relative yield at \\
        C$_{2}$H$_{6}$O & identity & $hv=21.1$\,eV & $hv=24.7$\,eV & $hv=31.7$\,eV & C$_{2}$D$_{6}$O & identity & $hv=31.7$\,eV\\
        \hline\hline
1  & H$^{{+}}$                       & 50(3)    & 128.6(6) & 185.4(6)  & 2  & D$^{{+}}$                       & 108.6(3) \\
2  & H$_{{2}}^{{+}}$                 & 5.0(4)   & 20.3(2)  & 25.3(1)   & 4  & D$_{{2}}^{{+}}$                 & 14.22(8) \\
3  & H$_{{3}}^{{+}}$                 & 1.0(2)   & 6.9(2)   & 5.97(8)   & 6  & D$_{{3}}^{{+}}$                 & 2.61(6)  \\
12 & C$^{{+}}$                       & 3.9(4)   & 9.0(2)   & 10.32(8)  & 12 & C$^{{+}}$                       & 6.77(5)  \\
13 & CH$_{{1}}^{{+}}$                & 13.1(9)  & 20.0(1)  & 25.38(8)  & 14 & CD$_{{1}}^{{+}}$                & 16.95(6) \\
14 & CH$_{{2}}^{{+}}$                & 41(2)    & 63.1(4)  & 83.4(2)   & 16 & CD$_{{2}}^{{+}}$                & 53.5(1)  \\
   &                                 &          &          &           &    & O$^{{+}}$                       & ``''     \\
15 & CH$_{{3}}^{{+}}$                & 204(1)   & 104.1(2) & 116.5(2)  & 18 & CD$_{{3}}^{{+}}$                & 94.1(1)  \\
16 & O$^{{+}}$                       & 16.6(7)  & 46.3(3)  & 47.1(1)   &    &                                 &          \\
19 & H$_{{3}}$O$^{{+}}$              & 46(1)    & 137(4)   & 53.7(2)   & 22 & D$_{{3}}$O$^{{+}}$              & —        \\
27 & C$_{{2}}$H$_{{3}}^{{+}}$        & 350(10)  & 149.1(4) & 176.8(2)  & 30 & C$_{{2}}$D$_{{3}}^{{+}}$        & 282.1(2) \\
29 & C$_{{2}}$H$_{{5}}^{{+}}$        & 440(10)  & 212.2(4) & 218.2(2)  & 34 & C$_{{2}}$D$_{{5}}^{{+}}$        & 474.5(2) \\
31 & CH$_{{2}}$OH$^{{+}}$            & 930(20)  & 498.1(8) & 437.4(8)  &    & CD$_{{2}}$OD$^{{+}}$            & ``''     \\
43 & C$_{{2}}$H$_{{3}}$O$^{{+}}$     & 63(1)    & 44.1(1)  & 45.62(8)  & 46 & C$_{{2}}$D$_{{3}}$O$^{{+}}$     & 35.84(4) \\
45 & CH$_{{3}}$CH$_{{2}}$O$^{{+}}$   & 280(6)   & 95.5(4)  & 89.3(2)   & 50 & CD$_{{3}}$CD$_{{2}}$O$^{{+}}$   & 75.83(5) \\
46 & CH$_{{3}}$CH$_{{2}}$OH$^{{+}}$  & 100      & 100      & 100       & 52 & CD$_{{3}}$CD$_{{2}}$OD$^{{+}}$  & 100   \\
    \end{tabular}
\end{table}

\begin{figure}[ht]
    \centering
    \includegraphics[width=\linewidth]{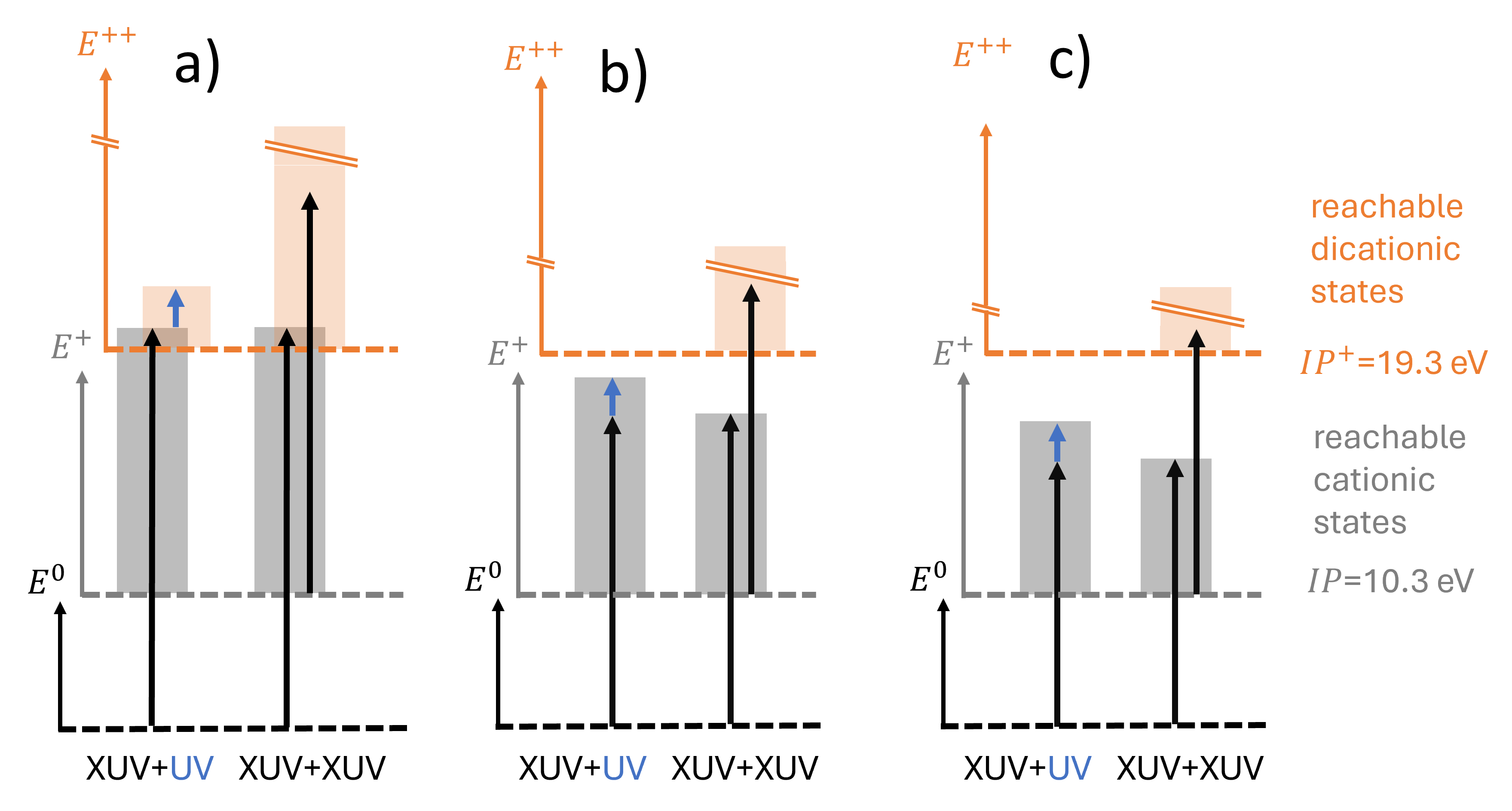}
    \caption[]{\label{fig:state_picture} Schematic representation of possible single- and double-ionization pathways using a) 31.7\,eV b) 24.7\,eV or c) 21.1\,eV XUV pump photons (black vertical arrows) and 391\,nm (blue vertical arrows) UV probe photons. Grey regions represent reachable cationic states; orange regions represent reachable dicationic states.}
\end{figure}

In accordance with other experiments\cite{Hoshina_2008_JChemPhys}, the formation of H$_{3}^{+}$ from ethanol dications is a minor channel, with an experimental H$_3^+$ signal amounting to about 5\% of the signal for intact ethanol cations (see Table~\ref{tab:fragment_abundances}). 
Despite the low yield of H$_3^+$, the data clearly show that H$_3^+$ is formed after interaction with XUV radiation above (31.7\,eV) as well as below (24.7 and 21.1\,eV) the double ionization threshold (29.6\,eV).
This indifference to photon energy indicates a predominant two-photon double ionization process, as opposed to one-photon double ionization (see Fig.~\ref{fig:state_picture}a).
This is possible due to the high XUV intensity generated by the FEL FERMI and is corroborated by the non-linear dependence of the H$_3^+$, H$_2^+$ and H$^+$ ion yields on the XUV intensity (see supplementary material).
Although the XUV energy of 21.1\,eV also allows for two-photon double ionization, we see a significantly reduced H$_{3}^{+}$ yield, compared to 24.7 and 31.7\,eV.
This is likely due to the lower XUV intensity available at 21.1\,eV and not necessarily due to the lower photon energy.
The reduction of the FEL intensity by about a factor of three (see supplementary material) makes the two-photon double-ionization process less efficient and reduces the yields of H$^{+}$, H$_{2}^{+}$, H$_{3}^{+}$ considerably.
Without ion-ion coincidence detection or a measurement of ion kinetic energy release, it is not straightforward to assign fragment ions to either a single or double ionization process, as double ionization produces identical fragments to single ionization\cite{Gope_2023_PhysChemChemPhys}.

Mass spectra obtained analogously for 2-aminoethanol, also for two different photon energies (31.7\,eV and 21.1\,eV) are shown in the bottom panel of Figure~\ref{fig:ethanol_mass_spectrum}.
To our knowledge, the double ionization threshold of 2-aminoethanol has not been experimentally determined, but a coarse estimation based on the single ionization potential\cite{Eland_1980} yields a value of $\approx$25.1 eV.
Other similar molecules (ethanol\cite{vonNiessen_1980_JElectronSpectrosc}, 1,2-ethanediol\cite{vonNiessen_1980_JElectronSpectrosc}, ethylamine\cite{Andrews_1995_JMassSpectrom}) have similar single and double ionization thresholds within 10–11\,eV and 26–28\,eV, respectively.
From this comparison we assume the two chosen photon energies to lie above and below the double ionization energy of 2-aminoethanol, respectively.
For 2-aminoethanol, only the measurement at 31.7\,eV shows H$_3^+$ ion signal in the mass spectrum.
At photon energies of 21.1\,eV, H$_3^+$ could not be detected, again likely due to the lower XUV intensity available, analogous to the measurement with ethanol.
Due to the low absolute signal of H$_{3}^{+}$ in 2-aminoethanol, we were not able to resolve the dynamics of its formation within reasonable data acquisition time, and we will focus solely on ethanol from this point onward.

\begin{figure}[ht]
    \centering
    \includegraphics[width=0.45\linewidth]{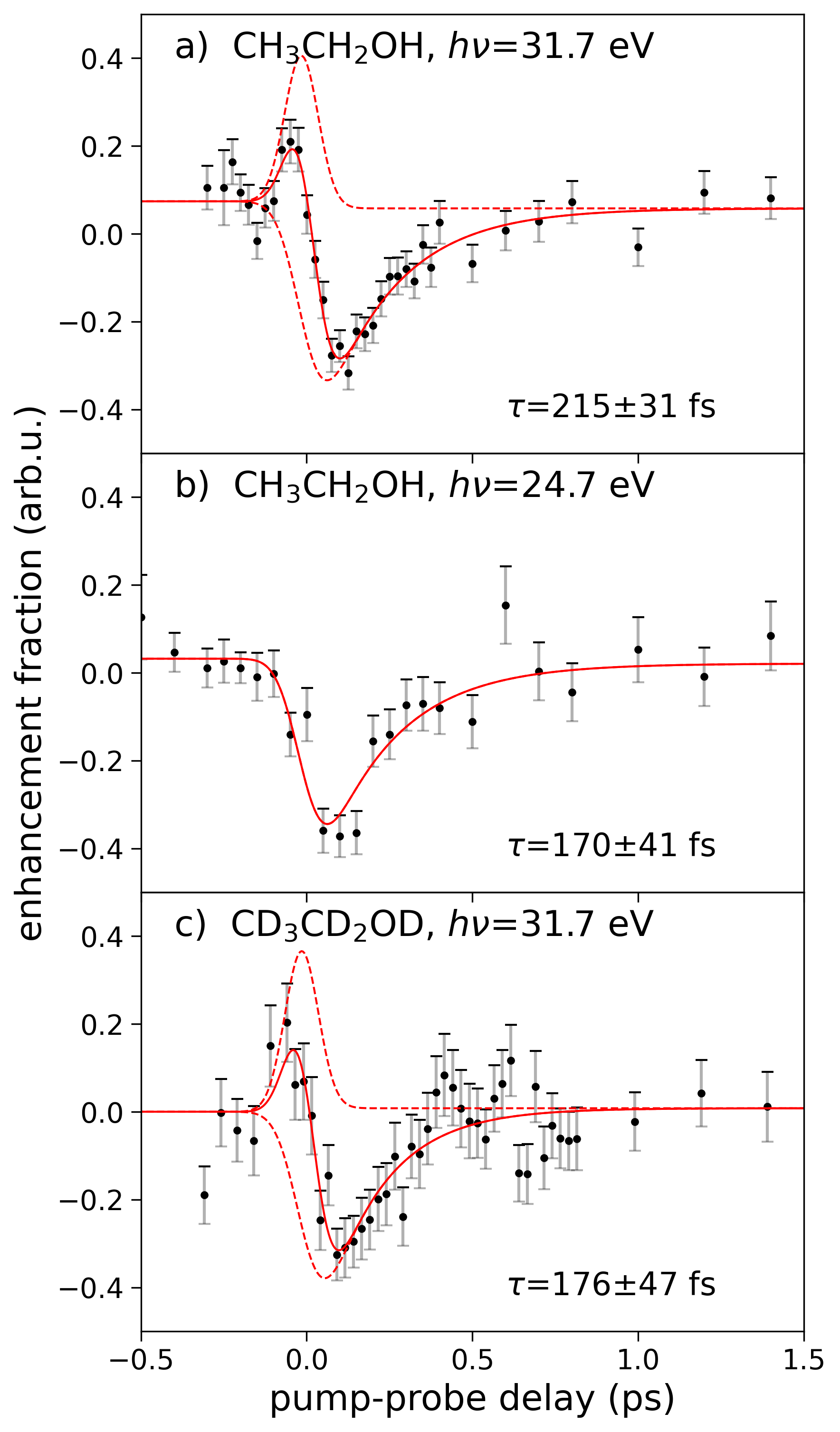}
    \caption[]{\label{fig:dynamics} H$_{3}^{+}$/D$_{3}^{+}$ ion yields as a function of pump-probe delay. Comparison of: a) non-deuterated ethanol at $h\nu=31.7$\,eV, b) non-deuterated ethanol at $h\nu=24.7$\,eV, and c) fully-deuterated ethanol at $h\nu=31.7$\,eV. The solid red line shows the fitted curve composed of signal depletion and enhancement contributions shown as red dashed lines.
    }
\end{figure}

To elucidate the dynamics of the fragmentation of ethanol after single- and double ionization, we performed a time-resolved measurement, in which a UV probe pulse disruptively probes the fragmentation dynamics.
These experiments were performed on both non-deuterated (C$_{2}$H$_{6}$O) and fully-deuterated ethanol (C$_{2}$D$_{6}$O).
Fig.~\ref{fig:dynamics} shows the dependence of the H$_{3}^{+}$ (D$_{3}^{+}$) yield on the delay between the XUV pump and the UV probe pulse.
At short positive time delays, the UV pulse depletes the H$_3^+$ signal, as was previously observed for intense IR pulses~\cite{Ekanayake_2018_JChemPhys,Ekanayake_2018_NatComm,Ekanayake_2017_SciRep,Livshits_2020_CommunChem} and XUV pulses\cite{Wang_2023_JPhysChemLett}.
Qualitatively, the data recorded at 31.7\,eV differ from the data recorded at 24.7\,eV by net increase in H$_3^+$ (D$_3^+$) yield around zero delay times before the yield is depleted at slightly positive delays.
A similar enhancement feature has been observed in single-color IR pump-probe studies of H$_2$-roaming reactions in alcohols\cite{Ekanayake_2018_JChemPhys,Ekanayake_2018_NatComm,Ekanayake_2017_SciRep}, where the behaviour was attributed to the increase of IR pulse intensity during the overlap of the pump and probe pulses.
In a previous experimental study where methanol was doubly-ionized by a broadband XUV pulse and probed disruptively by a 800 nm NIR pulse\cite{Livshits_2020_CommunChem}, no enhancement was observed and the data resembles our data obtained at 24.7\,eV.
In our experiment, the enhancement feature can be explained by a two-photon double ionization process with one photon of the 31.7\,eV XUV pulse and one photon of the 3.16\,eV UV pulse (see Figure 2b).
In contrast, the energy of one photon of the 24.7\,eV XUV pulse and one photon of the UV pulse is not sufficient for double ionization, explaining the absence of the enhancement at the lower XUV energy.

To quantitatively assess the roaming dynamics displayed in Fig.~\ref{fig:dynamics}, we fitted the data with a model taking into account the enhancement of the H$_3^+$ ion yield during the pulse overlap as well as its depletion (see supplementary material for details).
The positive enhancement feature is described by the Gaussian cross-correlation between the XUV and UV pulses, and the exponentially decaying depletion feature takes into account the cross-correlation corresponding to the two XUV photons and one UV photon involved in the process.
An additional step function accounting for small differences in the H$_3^+$ yield at long positive and negative delays is included as well. A slightly reduced yield of H$_3^+$ ions at long positive delays was also previously reported by Livshits \textit{et al.} \cite{Livshits_2020_CommunChem} in similar experiments on methanol, and was attributed to the dissociation of H$_3^+$ induced by the probe laser\cite{cosby_1988}.
This approach yields a more complete description of the dynamics than fitting a single exponential to the rising flank of the signal.
In particular, this takes into account the significant overlap between the positive enhancement feature and the exponential depletion feature caused by the cross-correlation of $119\pm 5$\,fs FWHM (full-width at half-maximum) between the XUV and UV pulses. 
In the fitting procedure, the cross-correlation was kept fixed to the experimentally-determined value for 31.7\,eV (119\,fs), determined by analyzing sidebands in the photoelectron signals of Ne upon ionization with the XUV and UV pulses.
The cross-correlation for 24.7\,eV is estimated to be $\sim$135\,fs, assuming the pulse duration scales with the harmonic order of the FEL photon energy\cite{Finetti_2017_PhysRevX}. 
A small correction to the experimentally-determined zero delay was fitted to the data set in Figure~\ref{fig:dynamics}a) and kept constant for the other data sets.

The time constants extracted from the fits for all three data sets are summarized in Fig.~\ref{fig:dynamics} and compared to the results from previous experiments\cite{Ekanayake_2017_SciRep, Ekanayake_2018_NatComm, Ekanayake_2018_JChemPhys, Wang_2023_JPhysChemLett} in Table~\ref{tab:other_experiments}.
Within their uncertainties, retrieved from the fitting procedure, the time constants for the different measurements are very similar with an average slightly below $\sim$200\,fs. The obtained uncertainties are around half the width of the Gaussian cross-correlation, which appears reasonable.
Not all our datasets are in agreement with previous values reported by Ekanayake \textit{et al.} \cite{Ekanayake_2017_SciRep,Ekanayake_2018_NatComm} and Wang \textit{et al.}\cite{Wang_2023_JPhysChemLett}, and rather point to a slightly shorter time constant.
However, within uncertainties, these differences seem barely significant.
Surprisingly, our measured time constants, for both XUV photon energies, as well as with and without deuteration, are indistinguishable within their error bars.
Such an indifference of the roaming dynamics with respect to the XUV photon energy (Figs.~\ref{fig:dynamics}a and \ref{fig:dynamics}b) is expected if only a few of the lowest-lying electronic states of the dication contribute to the roaming reaction, where these states are populated by both XUV photon energies (see Fig.~\ref{fig:state_picture}).
In the case of methanol, only the lowest dicationic states contribute to the roaming reaction\cite{Luzon_2019_JPhysChemLett}, and a similar scenario for ethanol would explain the observed behaviour.
Furthermore, this explanation is in agreement with the enhancement of H$_3^+$ production during the overlap of the XUV and UV pulses for 31.7\,eV photons as a two-color absorption, and the absence thereof for 24.7\,eV. 
Using this information and the double ionization potential of ethanol\cite{Linusson_2009_PhysRevA}, we can locate the excitation energies of the electronic state(s) producing H$_3^+$ ions between 29.6 and 34.9\,eV. 
The low number of states contributing to the roaming reaction also explains why different experiments inducing roaming reactions by strong-field double ionization\cite{Ekanayake_2017_SciRep,Ekanayake_2018_NatComm}, single-photon double ionization by XUV pulses\cite{Wang_2023_JPhysChemLett}, or XUV two-photon double ionization in the present study all show the same dynamics.

The indifference of the roaming dynamics to either deuterated or non-deuterated ethanol (see Figures~\ref{fig:dynamics}a, \ref{fig:dynamics}c and Table~\ref{tab:other_experiments}) is unexpected.
A recent study by Ekanayake \textit{et al.}\cite{Ekanayake_2017_SciRep} using intense IR pulses on methanol isotopomers reported on the dynamics of the production of H$_3^+$, H$_2$D$^+$ and D$_3^+$. 
Differences in the formation times of H$_3^+$ and H$_2$D$^+$ from doubly ionized CH$_3$OD were attributed to two different reaction pathways. The pathway involving proton transfer from the carbon atom was observed to be significantly faster than the pathway including proton transfer from the hydroxyl group, in agreement with the difference in distance, which the neutral roaming H$_2$ moiety has to travel. 
The comparison between the formation of H$_3^+$ from CH$_3$OH and D$_3^+$ from CD$_3$OD showed a roughly 35 \% increase in formation time. 
This increase can either be caused by generally lower reaction rates for deuterons as compared to protons, or by a shift in the branching ratios for the different pathways, induced by the isotope exchange.
The interplay between different reaction pathways and the possibility to have reduced dissociation and transfer rates for heavier isotopes creates ambiguity in the interpretation.
The results on the H$_2$D$^+$ formation were confirmed by Gope \textit{et al.}\cite{Gope_2021_NatSci}, who also performed ab initio molecular dynamics simulations revealing differences in the ion kinetic energy releases observed for different isotopomers. The latter result indicates a complex influence of the isotope exchange.

\begin{table}
    \centering
    \caption[]{\label{tab:other_experiments} Comparison of previous work on the H$_{2}$-roaming dynamics of photoionizated ethanol with the results from this study.}
    \begin{tabular}{l | c | c | r c l}
        \hline\hline
        reference & ethanol species & pump-probe scheme & \multicolumn{3}{c}{time constant} \\
        \hline\hline
        Ekanayake \textit{et al}. (2018) \cite{Ekanayake_2018_NatComm, Ekanayake_2018_JChemPhys} & CH$_{3}$CH$_{2}$OH & IR-IR (1.55\,eV) & 235 & $\pm$ & 10\,fs \\
        & CH$_{3}$CH$_{2}$OH &  & 220 & $\pm$ & 6\,fs \\
        \hline
        Wang \textit{et al}. (2023) \cite{Wang_2023_JPhysChemLett} & CH$_{3}$CH$_{2}$OH & XUV-XUV (28, 28\,eV) & 296 & $\pm$ & 87\,fs \\
        \hline
        This work & CH$_{3}$CH$_{2}$OH & XUV-UV (31.7, 3.16\,eV) & 215 & $\pm$ & 31\,fs \\
        & CH$_{3}$CH$_{2}$OH & XUV-UV (24.7, 3.16\,eV) & 170 & $\pm$ & 41\,fs \\
        & CD$_{3}$CD$_{2}$OD  & XUV-UV (31.7, 3.16\,eV) & 176 & $\pm$ & 47\,fs \\
    \end{tabular}
\end{table}

In ethanol, the situation is even more complex, with at least five possible reaction pathways (see Table~\ref{tab:channels}). 
Of these possible pathways, at least four have been observed experimentally\cite{Ekanayake_2018_NatComm}, although the pathways involving formation of H$_2$ from the $\alpha$-carbon dominate.
Our experiment cannot distinguish these pathways, and the dynamics we observe include all contributing pathways.
Intuitively, all involved individual steps, i.e neutral H$_2$ dissociation, H$_2$ roaming, and final proton transfer, should proceed more slowly if protons are substituted by deuterons.
Therefore the observations reported for methanol\cite{Ekanayake_2017_SciRep} are in agreement with expectation.
The similarity of H$_3^+$ and D$_3^+$ formation times that we observe in ethanol, contradict this intuition.
It remains unclear, whether all reaction pathways are unaffected by the isotope exchange, or there is coincidental cancellation of isotope effects on the reaction dynamics and the branching ratios of different pathways.
Furthermore, the isotope exchange could result in spatially shorter roaming trajectories, possibly counteracting the expected slower motion of the heavier isotopes.
The absence of kinetic isotope effects could be an indication that tunneling processes are not involved in the formation of H$_3^+$.
Additional time-resolved measurements on different ethanol isotopomers could help to clarify potential differences and isotope effects for individual roaming pathways. 
Small kinetic isotope effects on the dynamics could furthermore be concealed by the significant uncertainties of our measurements.
Note, that while we do not observe a kinetic isotope effect on the dynamics of H$_3^+$ (D$_3^+$) formation, we do see an effect in the relative ion yields of most fragment ions (Table~\ref{tab:fragment_abundances}). Relative to the respective parent ion, deuterated ethanol shows less intensity in all isolated fragments, indicating that a larger portion of the photoions remain intact.

\begin{table}
\centering
\caption{\label{tab:channels}: Different possible roaming reaction pathways for the formation of H$_3^+$ from doubly ionized ethanol, assuming no rearrangement of H atoms.}
\begin{tabular}{c||c||c}
Initial dication & \multicolumn{2}{|c}{CH$_3$-CH$_2$-OH$^{2+}$} \\ \hline
Transient isomer & C\textcolor{cyan}{H}-C\textcolor{blue}{H}$_2$-O\textcolor{red}{H}$^{2+}$ + H$_2$ & C\textcolor{cyan}{H}$_3$-C-O\textcolor{red}{H}$^{2+}$ + H$_2$  \\ \hline
Product isomers   & C\textcolor{cyan}{H}-C\textcolor{blue}{H}$_2$-O$^{+}$ + H$_2$\textcolor{red}{H}$^+$ & C\textcolor{cyan}{H}$_3$-C-O$^{+}$ + H$_2$\textcolor{red}{H}$^+$  \\
                 & C\textcolor{cyan}{H}-C\textcolor{blue}{H}-O\textcolor{red}{H}$^{+}$ +  H$_2$\textcolor{blue}{H}$^+$   & C\textcolor{cyan}{H}$_2$-C-O\textcolor{red}{H}$^{+}$ + H$_2$\textcolor{cyan}{H}$^+$ \\
                 & C-C\textcolor{blue}{H}$_2$-O\textcolor{red}{H}$^{+}$ + H$_2$\textcolor{cyan}{H}$^+$   &                           \\
\end{tabular}
\end{table}

It is also surprising that largely different probe wavelengths (IR, XUV or UV) do not affect the time scale of the observed dynamics.
The description as disruptive probe pulses does not address the physics of the probe interactions.
Generally, disruptive probing is intended to deplete population from the roaming state. 
Once the roaming process is completed, after either proton or electron transfer has occurred and the two singly-charged fragments undergo Coulomb explosion, the disruptive probe pulse would no longer be able to affect the formed ionic fragments.
A strong IR pulse can in principle be used to deplete a certain electronic state and disrupt its dynamics by further ionizing the system or by driving resonant vibronic transitions.
The states populated by the probe pulse would then follow a different fragmentation pathway, leading to the observed depletion of a certain product, and the enhancement of another.
For highly energetic XUV photons, depletion of a reaction pathway would rather occur by additional ionization of the dicationic fragment or of a neutral roaming H$_2$ moiety.
A possible complication can arise in such cases, if the probe pulse can also affect the final ionic fragments via further ionization or photodissociation. 
In this case, one would observe a significant depletion of the reaction products also at long positive delays, after the reaction dynamics. 
We do not observe such an effect in our study, nor has this been a limiting factor in previous studies using XUV or IR probe pulses\cite{Ekanayake_2017_SciRep,Ekanayake_2018_NatComm,Ekanayake_2018_JChemPhys,Livshits_2020_CommunChem,Wang_2023_JPhysChemLett}.
The UV pulses, used in the scope of this work to disruptively probe the roaming process and formation of H$_3^+$, had intensities of up to 1.6$\times$10$^{13}$\,W~cm$^{-2}$. 
These pulses induce a maximum reduction of the H$_3^+$ yield by about 30\% (see Figure~\ref{fig:dynamics}).
In comparison, previous studies using disruptive NIR pulses for methanol observed slightly lower maximum reduction of the H$_3^+$ yield around and below 10 \%\cite{Ekanayake_2017_SciRep,Livshits_2020_CommunChem}.
Data on ethanol is only available without quantitative information on the relative depletion\cite{Ekanayake_2018_NatComm,Ekanayake_2018_JChemPhys}.
The stronger depletion with lower probe intensities (176\,$\mu$J for CH$_{3}$CH$_{2}$OH, ~88\,$\mu$J for CD$_{3}$CD$_{2}$OD) in our experiment may indicate a different probe mechanism, although the observed differences can also be affected by differences in the experimental spatial overlap of pump and probe pulses, making a quantitative comparison difficult.
While we cannot rule out multi-photon ionization processes by the probe pulse, it is more likely that the probe pulse depletes the roaming states by driving resonant or near-resonant electronic transitions, given that the roaming states are likely energetically low-lying, as explained previously for methanol\cite{Luzon_2019_JPhysChemLett}.

In summary, the relative indifference of the experimentally observed time constants to the probing scheme shows that all different probe wavelengths can disrupt the roaming process over essentially the same time span, which is therefore likely to be the complete time span of the roaming process. 

\paragraph{Other masses}

\begin{figure}[ht]
    \centering
    \includegraphics[width=\linewidth*4/4]{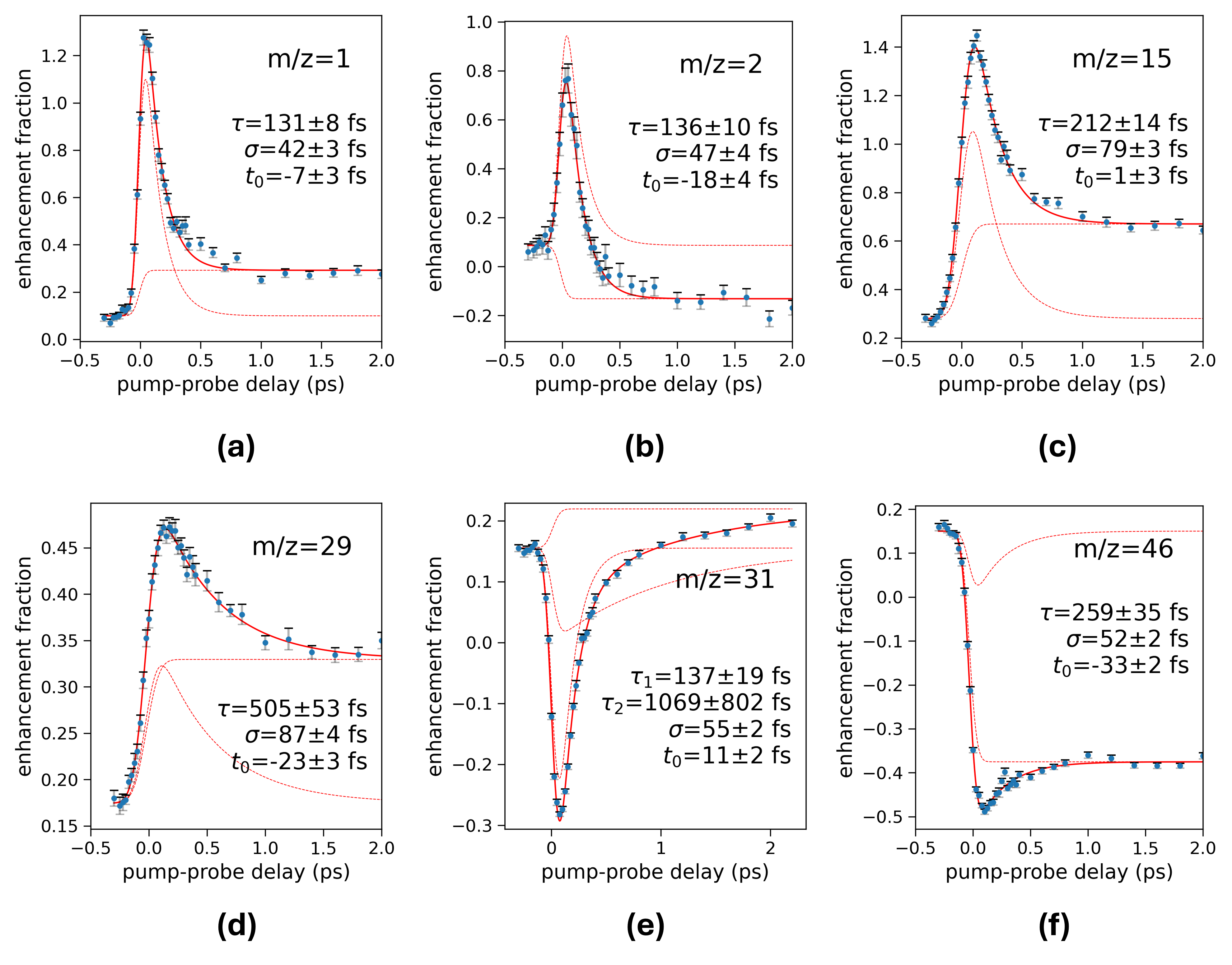}
    \caption[]{\label{fig:other_dynamics} Ion yields for (a) $m/z=1$ (b) $m/z=2$ (c) $m/z=15$ (d) $m/z=29$ (e) $m/z=31$ (f) $m/z=46$ as a function of the pump-probe delay between the 31.7\,eV XUV pulse and a 392 nm UV pulse. Solid red lines show fitted curves with the individual contributions shown by red dotted lines. All ion yields except $m/z$=31 (panel (e)) are fitted by single exponential decays and step functions, convoluted with Gaussian cross-correlations. The fit of $m/z$=31 contains two exponentially decaying components, see text.}
\end{figure}

While the dynamics of the exotic H$_{2}$-roaming mechanism leading to H$_3^+$ formation from small doubly-ionized organic molecules has received considerable attention, the rich dynamics observable for other, much more abundant fragments, mostly formed by single ionization, has been somewhat neglected.
We show a selection of the main fragment ion yields from our measurements on non-deuterated ethanol with photon energies of 31.7\,eV in Figure~\ref{fig:other_dynamics}, and we give the relative ion yields in Table~\ref{tab:fragment_abundances}.
Data on further fragment ions, as well as for the fragment ions of 2-aminoethanol, are given in the supplementary material.
It is evident that the different fragmentation reactions occur on various sub-ps timescales with their yields either being enhanced or depleted by the probe pulse.
In general, we observe that larger fragments are rather depleted by the probe pulse, while smaller fragments are rather enhanced (with the notable exception of H$_{3}^{+}$). This is in agreement with the general intuition, that additional energy in the system induces further fragmentation.
Many fragments are also enhanced or depleted at long positive delays (>1\,ps), indicating that after the end of ultrafast dynamic rearrangement, some (meta-)stable fragment ions can be further fragmented by the probe pulse.
Detailed knowledge of the involved potential surfaces on which the fragmentation reaction occur would allow to quantitatively model the different observed dynamics.
A theoretical treatment of the complex network of possible fragmentation pathways, involving highly-excited cationic states is currently not feasible, and therefore beyond the scope of this particular study.

To obtain a semi-quantitative description of the observed dynamics, we use a fit model similar to the one used above for the description of H$_3^+$ formation.
This model includes a single exponentially decaying enhancement or depletion feature and a step function broadened by the cross-correlation between the XUV pump and UV probe pulses.
The latter accounts for different signal levels at long positive and negative delays.
We freely adjust all parameters in the fitting procedure.
Some of the resulting cross-correlation values differ from the experimentally measured values, as do some of the effective zero-delay positions between the XUV and UV pulses ($t_0$).
Also, the dynamics of some of the product masses are fit poorly by a simple exponential model, implying more complex fragmentation dynamics.
For example, the CH$_2$OH$^+$ ion ($m/z$=31, Fig.~\ref{fig:other_dynamics}e) is described exceptionally poorly by a simple exponential model (see Supplementary Figure 4(f)). We thus fit the ion yield of $m/z$=31 with a superposition of two exponential decay curves with individual time constants, which captures the dynamics well.

Comparing the time constants retrieved by this approach can in principle provide some insight into the changes the probe pulse introduces to the fragmentation pathways. 
For pump-probe delays larger than the cross correlation, the total ion yield, distributed over all ion masses, can be assumed to be conserved, such that the depletion of one fragment enhances another. 
For example, it is noteworthy that the smallest ions H$^+$ and H$_2^+$ show an enhancement decaying with a very short time constant even shorter than the time constant of the H$_3^+$ depletion. It seems intuitive that the H$_3^+$ formation, requiring the motion of a neutral molecular H$_2$ fragment over a significant distance, requires more time than the direct dissociation of light fragments. One might suggest that the mechanism behind the depletion of H$_3^+$ signals proceeds via the multiphoton ionization or dissociative ionization of the roaming neutral H$_2$. If so, the dynamics of the H$^+$ and H$_2^+$ would be complementary to the dynamics of H$_3^+$. However, the absolute signal changes in the H$^+$ and H$_2^+$ ion yields are much larger than in the H$_3^+$ depletion.
Such a discrepancy in the absolute signal changes could be explained by a large portion of neutral roaming H$_2$ that does not finally lead to the production of H$_3^+$ but results in dissociation.
In this case, however, we would expect a large enhancement of the H$^+$ and H$_2^+$ signals also at large delays, created by the ionization and dissociative ionization of the large amount of produced isolated neutral H$_2$.
Thus, we cannot confirm the direct connection between the ion yields of H$^+$, H$_2^+$ and H$_3^+$.

The most abundant ion in the mass spectrum of ethanol is CH$_2$OH$^+$ ($m/z=31$), and we note that this is the only fragment that noticeably disagrees with a fit including a single exponential time constant. We thus employ a bi-exponential fit model for $m/z=31$, that yields a fast and a slow time constant, albeit the latter has a large uncertainty. 
The faster timescale and absolute signal change of the depletion of the fragment of $m/z=31$ (CH$_2$OH$^+$) and the enhancement of $m/z=15$ (CH$_3^+$) are similar. Ion pairs of these masses, adding up to the mass of the intact ethanol parent ion, are a dominant product of the double ionization of ethanol\cite{Gope_2023_PhysChemChemPhys}. 
Nevertheless, we assume that most of the respective ions in our mass spectra are created by single ionization, followed by C-C bond rupture.
From photoelectron-photoion coincidence experiments\cite{Niwa_1982_IntJMassSpectrom}, it is known that C-C bond rupture predominantly produces CH$_2$OH$^+$ and neutral CH$_3$ radicals, and only at large excess energies ($\sim$6\,eV above the ionization threshold of ethanol) can the less favorable combination of CH$_3^+$ cations and neutral CH$_2$OH radicals be formed.
In our experiment, the electronic excitation of the ethanol cation by the UV probe pulse may induce this more energetic charge distribution, leading to the enhancement of CH$_3^+$ ions and the depletion of CH$_2$OH$^+$.
Nonetheless, we cannot make a final assignment with the current data quality and temporal resolution without an extensive theoretical treatment.

The ion signal corresponding to $m/z=29$ shows an enhancement by the probe pulse, which decays on a remarkably long time scale of $\sim$500\,fs. This signal can contain contributions from COH$^+$ (or HCO$^+$) as well as from C$_2$H$_5^+$. The comparison of the ion yields (see Table~\ref{tab:fragment_abundances}) of $m/z=31$ (CH$_2$OH$^+$) and $m/z=27$ (C$_2$H$_3^+$) with the ion yields of $m/z=34$ (C$_2$D$_5^+$ and CD$_2$OD$^+$) and $m/z=30$ (CDO$^+$ and C$_2$D$_3^+$), allows to estimate the relative contributions between HCO$^+$ and C$_2$H$_5^+$ to the signal of $m/z=29$.
Assuming that otherwise the isotope exchange does not affect the branching ratios, we infer that COH$^+$ (or HCO$^+$) dominates the signal of $m/z=29$.

The enhancement of $m/z=29$ shows significantly slower dynamics than any other fragment. The only comparable time constant is observed for the slow component of $m/z=31$.
This suggests that UV-induced photodissociation of CH$_2$OH$^+$\,$\rightarrow$\,COH$^{+}+$H$_{2}$ may cause the enhancement of $m/z=29$. 
An ongoing relaxation process of CH$_2$OH$^+$ after C-C bond rupture possibly reduces the cross section of this photodissociation process, thus causing the slow dynamics in the respective ion signals.

Since we cannot unambiguously separate a possibly small contribution of C$_2$H$_5^+$ from the  COH$^+$ signal, C$_2$H$_3$ ($m/z=27$) is the only clearly observable fragment formed after C-O bond rupture. The formation is enhanced by the UV pulse, with a fast component that decays on a timescale of $\sim$200\,fs (see supplementary material), and a remaining constant enhancement at long positive delays.
The latter can realistically only be created by the UV action on intact ethanol cations, that are photodissociated. 
This is in agreement with photoelectron-photoion coincidence studies, which find a threshold for the formation of C$_2$H$_3^+$ at about 14\,eV, within one probe photon energy (3.16\,eV) above the cationic ground state (10.5-11\,eV), producing intact ethanol cations.
Which other fragment is depleted on fast timescales, complementary to the enhancement of C$_2$H$_3$, cannot be determined without a full quantitative model of the fragmentation processes.
Coincidence measurements, as well as measurements of ion kinetic energy releases could help to elucidate the different fragmentation mechanisms and differentiate between fragmentation occurring in doubly-ionized and singly-ionized species.

\section*{Conclusion}

In conclusion, we have presented a study of the H$_2$ roaming reaction leading to the production of H$_3^+$ ions from doubly-ionized ethanol and 2-aminoethanol.
The experimental scheme, using two-photon double ionization by intense XUV pulses above and below the double ionization threshold, and a disruptive UV probe, yields similar timescales for H$_3^+$ formation as previous one-color strong-field IR-IR\cite{Ekanayake_2018_NatComm,Ekanayake_2018_JChemPhys,Ekanayake_2017_SciRep} or XUV-XUV\cite{Wang_2023_JPhysChemLett} experiments on ethanol.
The indifference of the dynamics with respect to the probe wavelength indicates that the disruptive probe pulses depletes the population of the H$_3^+$ formation pathway, although likely via differing physical processes.
Most strikingly, we observe the absence of an isotope effect in the D$_3^+$ formation dynamics following the double ionization of fully-deuterated ethanol, which is in contrast to an earlier experiment on methanol which showed an increase of the D$_3^+$ formation time of 35 \% in D$_4$-methanol\cite{Ekanayake_2017_SciRep}.
The origin of this difference between methanol and ethanol remains unclear, but may involve the larger number of possible reaction pathways as well as fundamental differences in the reaction mechanisms.
More experiments and theoretical modelling in the future will be necessary to elucidate the possible different pathways for the formation of H$_3^+$ and how their dynamics and relative branching ratios are influenced by isotope exchange.
In addition, we report on the dynamics of various ion yields, most likely dominated by singly-ionized ethanol cations, which feature dynamics on various timescales.
The study of these dynamics provide initial insights into the action of the disruptive probe pulse in the UV range, which seems to induce mainly electronic excitation of the cations.
This excitation depletes the yields of certain ions in favor of alternative ionic products.
More information could be gained in the future from experiments providing ion-ion coincidence detection and ion kinetic energy release data to disentangle products from single and double ionization.

\section*{Methods}

The experiments were performed at the Low-Density Matter (LDM) endstation of the FERMI Free-Electron Laser at the Elettra synchrotron facility in Trieste.
A molecular beam was produced by expanding ethanol and 2-aminoethanol, respectively, seeded in neon carrier gas through a commercially-available pulsed solenoid valve (Parker Hannifin, Series 9).
The gas mixture was produced by passing the neon carrier gas through a bubbler filled with liquid ethanol cooled to 0$^\circ$C (or 2-aminoethanol heated to $\sim$80$^\circ$C), before reaching the pulsed valve heated to $\sim$80$^\circ$C.
Backing pressure (typically 0.7\,bar for both ethanol and 2-aminoethanol) and nozzle temperature ($\sim$80$^\circ$C) were adjusted to minimize cluster formation.
After passing a skimmer, the molecular beam was intersected at right angles by the XUV beam from the FEL and the UV beam.
The latter was used as the disruptive probe.
In the case of ethanol, the XUV pump pulse used photon energies of either 24.7\,eV or 31.7\,eV, produced as either the 7th or 9th harmonic of the 352.3\,nm seed laser (3.52\,eV) respectively.
Higher harmonics of the FEL photon energy were filtered out using an aluminium-magnesium metal filter.
The 392\,nm UV probe pulse was produced by a frequency-doubled Ti:Sapphire laser.
XUV pulse energies up to 62\,$\mu$J with spot sizes ranging from 50 to 150\,$\mu$m FWHM were used, corresponding to peak intensities up to $2.2\times 10^{13}$\,W~cm$^{-2}$.
UV pulse energies of up to $\sim$200\,$\mu$J were used with spot sizes up to 100\,$\mu$m FWHM, corresponding to peak intensities up to $1.6\times 10^{13}$\,W~cm$^{-2}$. Specific values corresponding to Fig.~\ref{fig:dynamics} are provided in the supplementary material.
The XUV-UV cross-correlation was measured as $119\pm 5$\,fs FWHM (51\,fs standard deviation) by observing sidebands created by the UV pulses in the XUV photoelectron signals.
The produced ions and electrons were extracted perpendicular to the laser beams and the molecular beam towards a magnetic bottle electron spectrometer and an ion time-of-flight mass spectrometer mounted in in-line tandem configuration~\cite{Squibb_2018_NatComm}.
To achieve maximum ion mass resolution, a retarding potential was applied to the magnet of the magnetic bottle spectrometer for most measurements, rendering the electron signals unusable.  

\section*{Acknowledgements}

We thank the Deutsche Forschungsgemeinschaft (DFG) for funding in the framework of the research training group DynCAM (RTG 2717) and grant number STI 125/19-2.
Furthermore, we acknowledge support by the European Cooperation in Science \& Technology (COST) Action CA21101 – Confined Molecular Systems: From a New Generation of Materials to the Stars (COSY).
This work has also been financially supported by the Swedish Research Council (VR) (grant numbers 2018-03731 and 2023-03464) and the Knut and Alice Wallenberg Foundation (grant number 2017.0104), Sweden.
We thank Professor Daniel Strasser for sharing unpublished results and helpful discussions.

\section*{Author contributions statement}

K.D. conceived the experiment, S.H., K.D. and M.M. planned and prepared the beamtime in coordination with O.P..
M.D.F., O.P, N.P. M.B. and C.C. operated the LDM endstation and provided input regarding experimental conditions. M.D. managed the seed laser for the FEL and L.R. was in charge of the FEL beam transportation and alignment. 
All authors were involved in the experimental measurements during the beamtime at the LDM endstation or contributed to valuable discussions during the beamtime.
A.N. analyzed the data under supervision of S.H. and F.S.
A.N. and S.H. wrote the original draft and all authors read, commented on, and approved the manuscript.

\section*{Additional information}

The authors have no conflicts of interest to disclose.

\section*{Data availability}

The datasets used and analyzed during the current study are available from the corresponding author upon reasonable request. 

\section*{Supplementary Material}

Additional experimental results on 2-aminoethanol, dynamics of additional fragmentation reactions.

\bibliography{references}

\end{document}


\flushbottom
\maketitle

\section{Experimental parameters}

XUV pulse energies from 0 to 62\ $\mu$J, after beam-transport-induced transmission losses, were used, with spot sizes ranging from 50 to 150\ $\mu$m FWHM.
UV pulse energies ranged from 50-200\ $\mu$J. The specific pulse energies used for each measurement are summarized in Supplementary Table\ \ref{tab:pulse_parameters}.

\begin{table}
    \centering
    \caption[]{\label{tab:pulse_parameters}XUV and UV pulse parameters used for datasets shown in the main text and in this supplementary material.}
    \begin{tabular}{c | c | c | c }
        \hline\hline
        Species & XUV photon energy (eV) & XUV pulse energy ($\mu$J) & UV pulse energy ($\mu$J) \\
        \hline\hline
        CH$_{3}$CH$_{2}$OH         & 31.7 & $\sim$47 & $\sim$176 \\
        CH$_{3}$CH$_{2}$OH         & 24.7 & $\sim$30 & $\sim$177 \\
        CH$_{3}$CH$_{2}$OH         & 21.1 & $\sim$14 & — \\
        \hline
        CD$_{3}$CD$_{2}$OD         & 31.7 & $\sim$44 & $\sim$88 \\
        \hline
        NH$_{2}$CH$_{2}$CH$_{2}$OH & 31.7 & $\sim$50 & $\sim$120 \\
        NH$_{2}$CH$_{2}$CH$_{2}$OH & 21.1 & $\sim$14 & $\sim$130 \\
        \hline
    \end{tabular}
\end{table}

\section{Data analysis}

Raw data were taken by all combinations of having the pump and probe pulses blocked and unblocked. 
"Foreground" data is taken as the difference between pump ON probe ON and pump OFF probe ON.
"Background" data is taken as the difference between pump ON probe OFF and pump OFF probe OFF.
Enhancement of ion yield is defined as ("Foreground"/"Background" - 1), e.g., an enhancement=0\% denotes no difference between foreground and background, and enhancement=100\% denotes a foreground signal twice as large as the background signal.

To obtain averages and errors of overlapping mass peaks, we first fit the TOF spectrum with multiple Gaussians.
From the fitted Gaussian, we then create a masking filter for each mass peak, that describes which portion of the ion signal can be ascribed to a certain ion mass (see Supplementary Figure~\ref{fig:masking}).
We then multiply the raw data with the masking filter, which yields the isolated mass peak, over which we integrate to obtain the peak signal.
The advantage of this slightly complicated procedure lies in the accurate description of bands that may deviate from ideal Gaussians due to ringing effects and other distortions (see Supplementary Figure~\ref{fig:masking}).
We perform this procedure on multiple TOF spectra to obtain the average peak signal and its corresponding error.

\begin{figure}[ht]
    \begin{subfigure}{0.32\linewidth}
        \includegraphics[width=\linewidth]{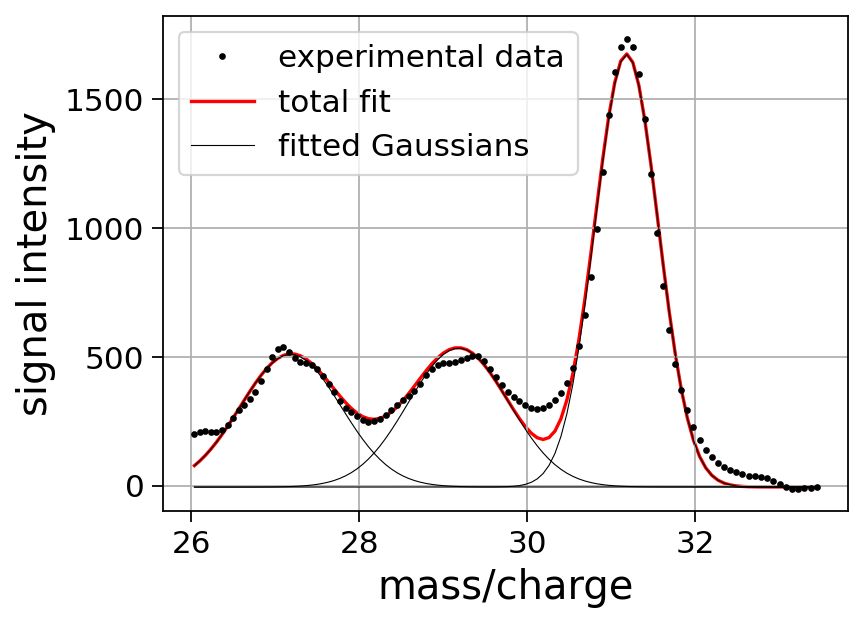}
        \caption[]{\label{subfig:masking_fit}}
    \end{subfigure}
    ~
    \begin{subfigure}{0.32\linewidth}
        \includegraphics[width=\linewidth]{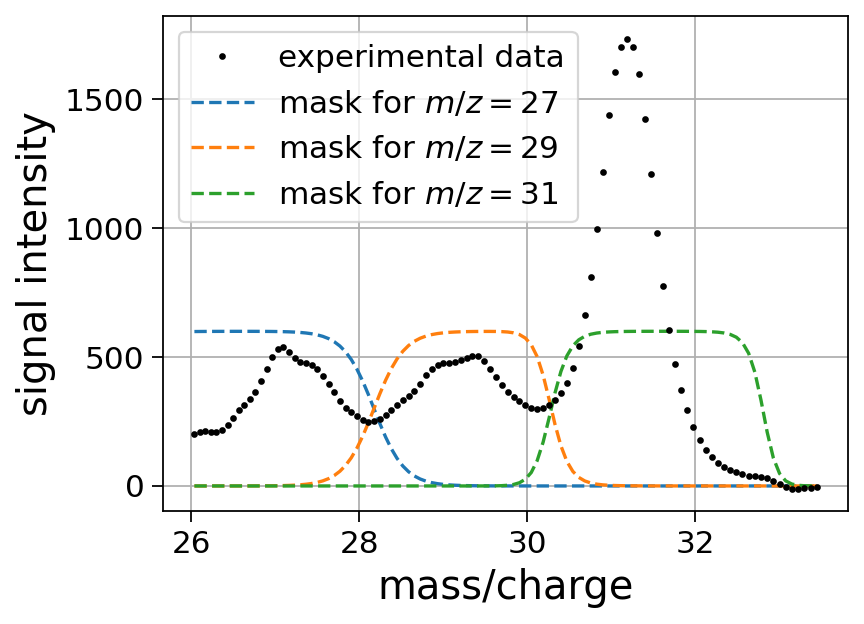}
        \caption[]{\label{subfig:masking_attenuate}}
    \end{subfigure}
    ~
    \begin{subfigure}{0.32\linewidth}
        \includegraphics[width=\linewidth]{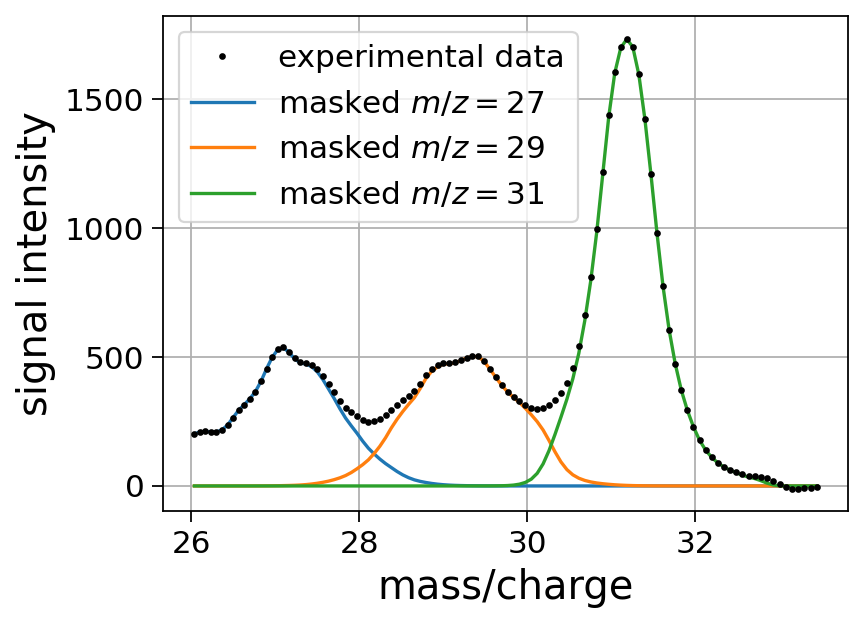}
        \caption[]{\label{subfig:masking_peaks}}
    \end{subfigure}
    
    \caption[]{\label{fig:masking} Example of separating mass peaks $m/z=27,29,31$ through masking. (\subref{subfig:masking_fit}) Simultaneous fit of three Gaussians on the TOF spectrum (\subref{subfig:masking_attenuate}) Individual masks for each mass peak (\subref{subfig:masking_peaks}) Resulting masked peaks.}
\end{figure}

\section{XUV intensity dependence}

\paragraph{Data normalization} The peak intensities from the XUV pulse are calculated assuming a Gaussian beam with a measured spot size FWHM of 52\ $\mu$m\ $\times$\ 61\ $\mu$m, an XUV-UV temporal cross-correlation measurement of 119\,fs, and an assumed temporal FWHM of the UV pulse of 100\,fs, which yields a temporal FWHM of the XUV of 54\,fs.

\begin{figure}[ht]
    \begin{subfigure}{0.45\linewidth}
        \includegraphics[width=\linewidth]{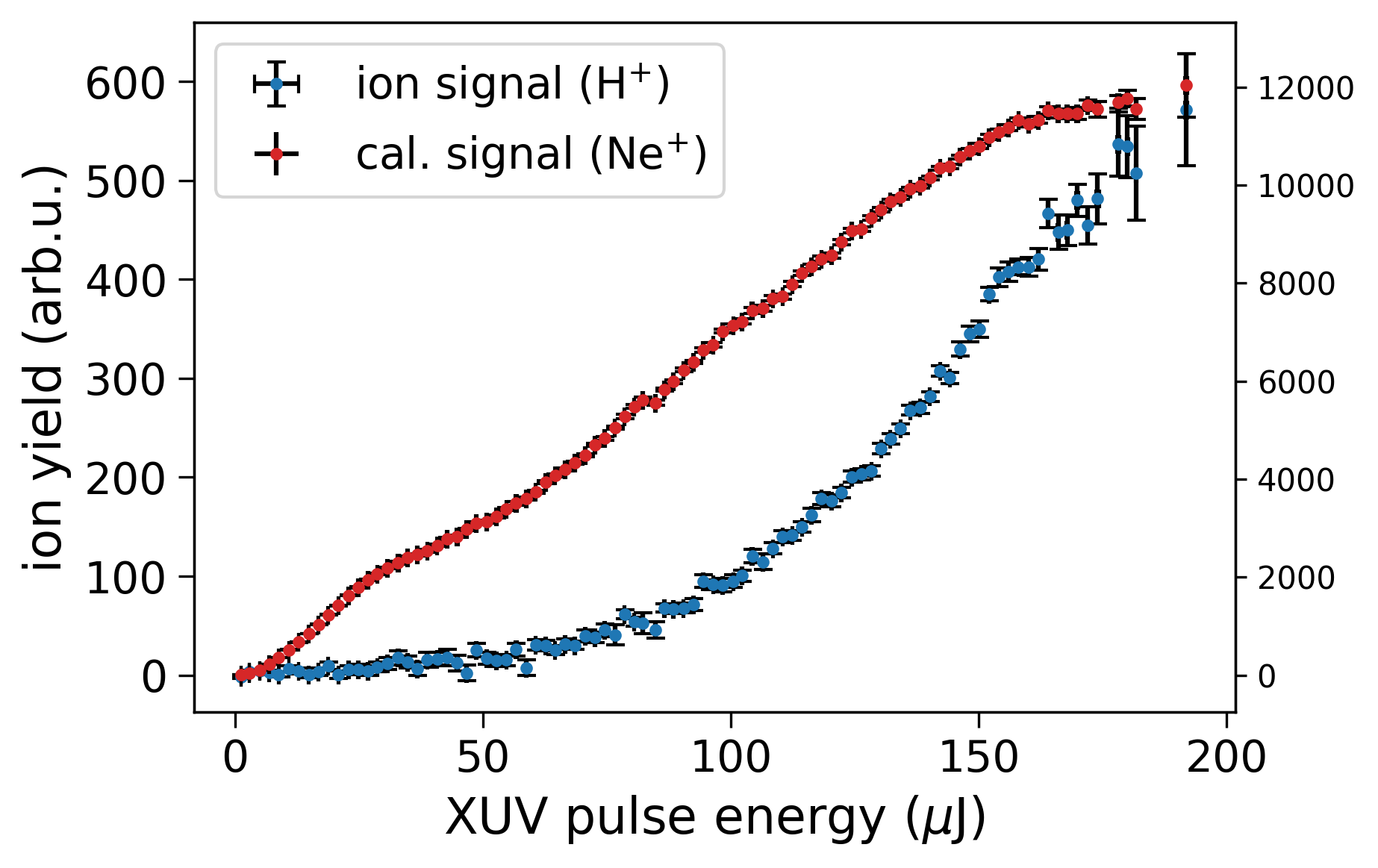}
        \caption[]{\label{subfig:RawIntensityH+}}
    \end{subfigure}
    ~
    \begin{subfigure}{0.45\linewidth}
        \includegraphics[width=\linewidth]{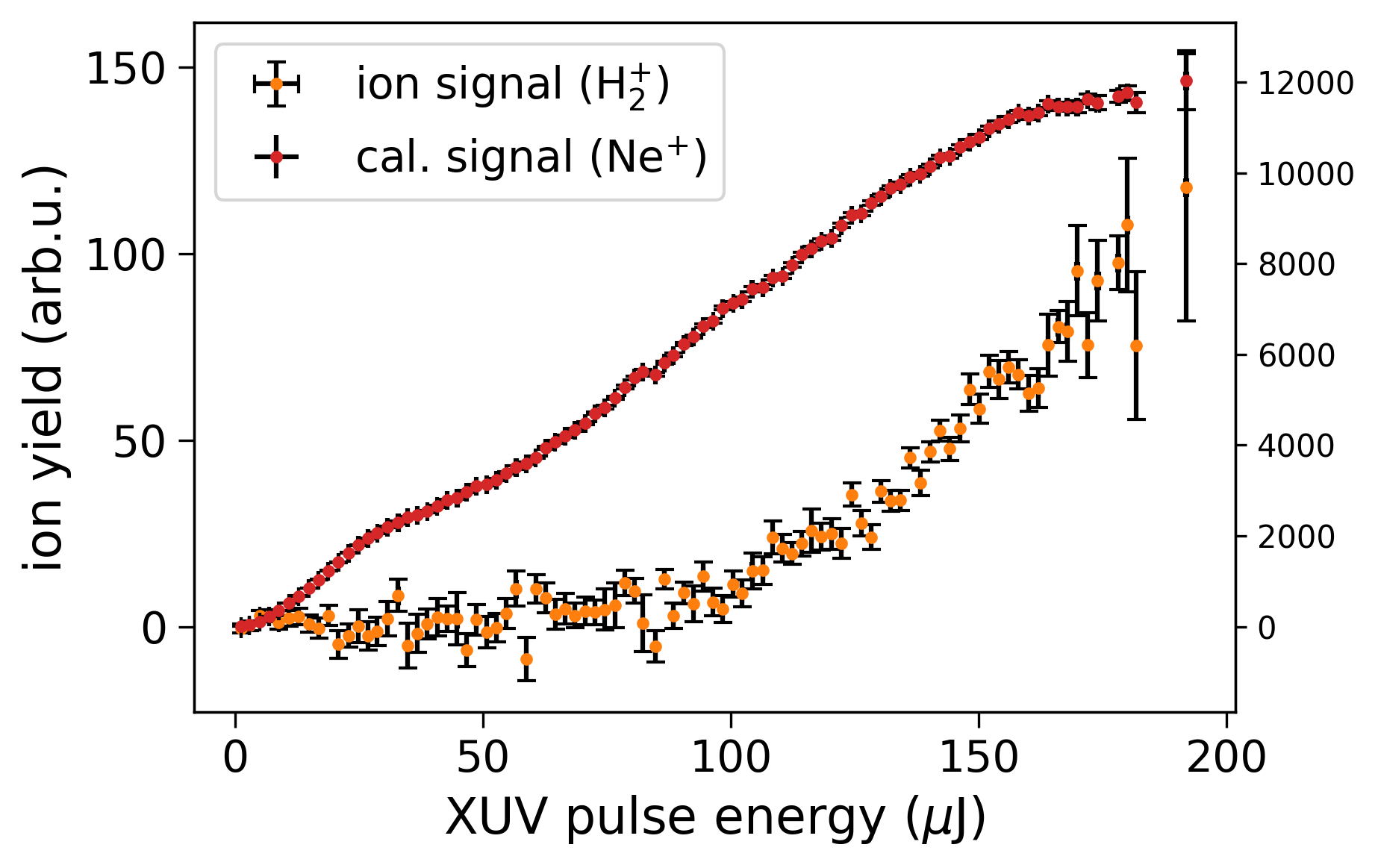}
        \caption[]{\label{subfig:RawIntensityH2+}}
    \end{subfigure}
    
    \begin{subfigure}{0.45\linewidth}
        \includegraphics[width=\linewidth]{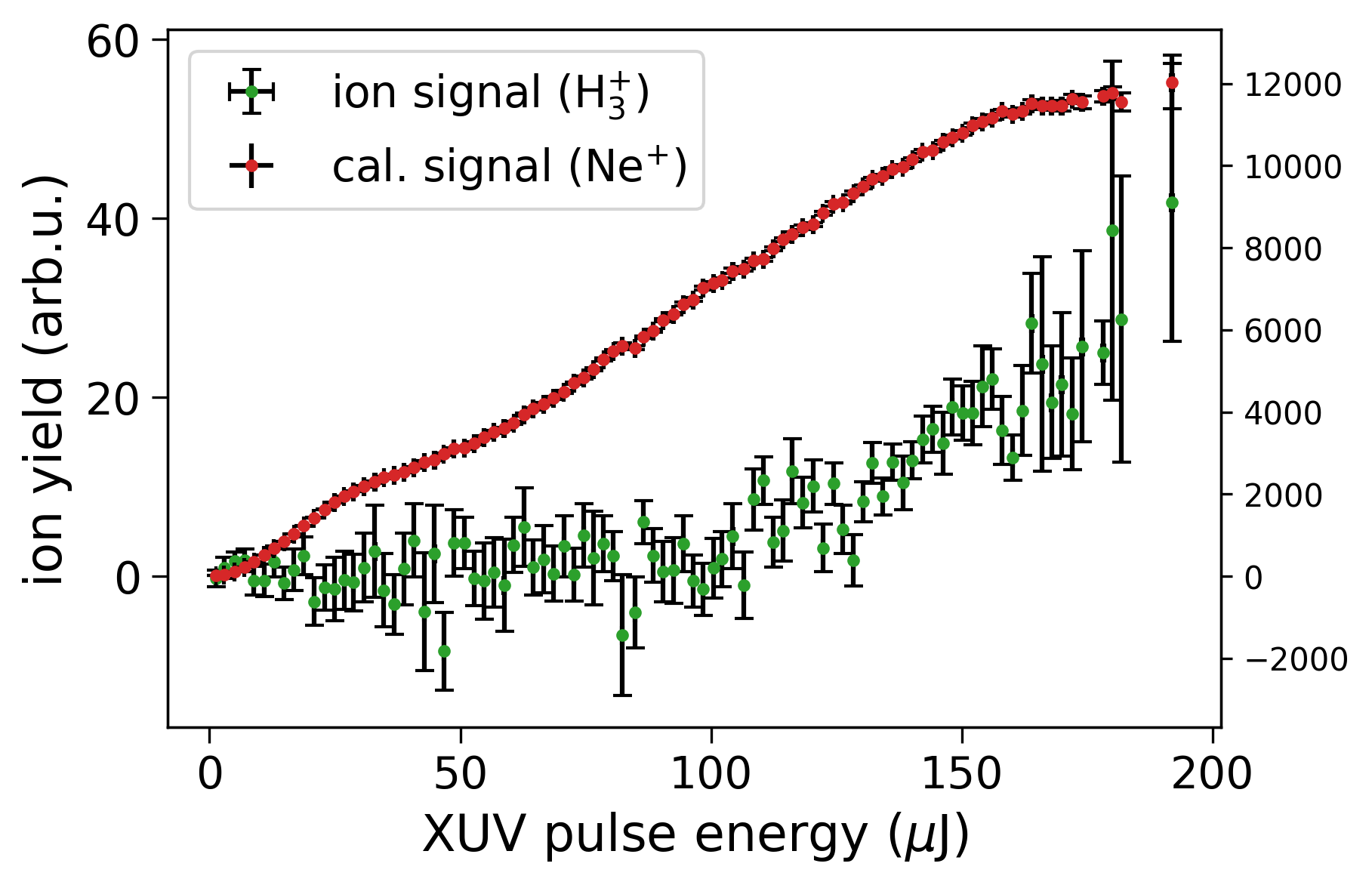}
        \caption[]{\label{subfig:RawIntensityH3+}}
    \end{subfigure}
    ~
    \begin{subfigure}{0.45\linewidth}
        \includegraphics[width=\linewidth]{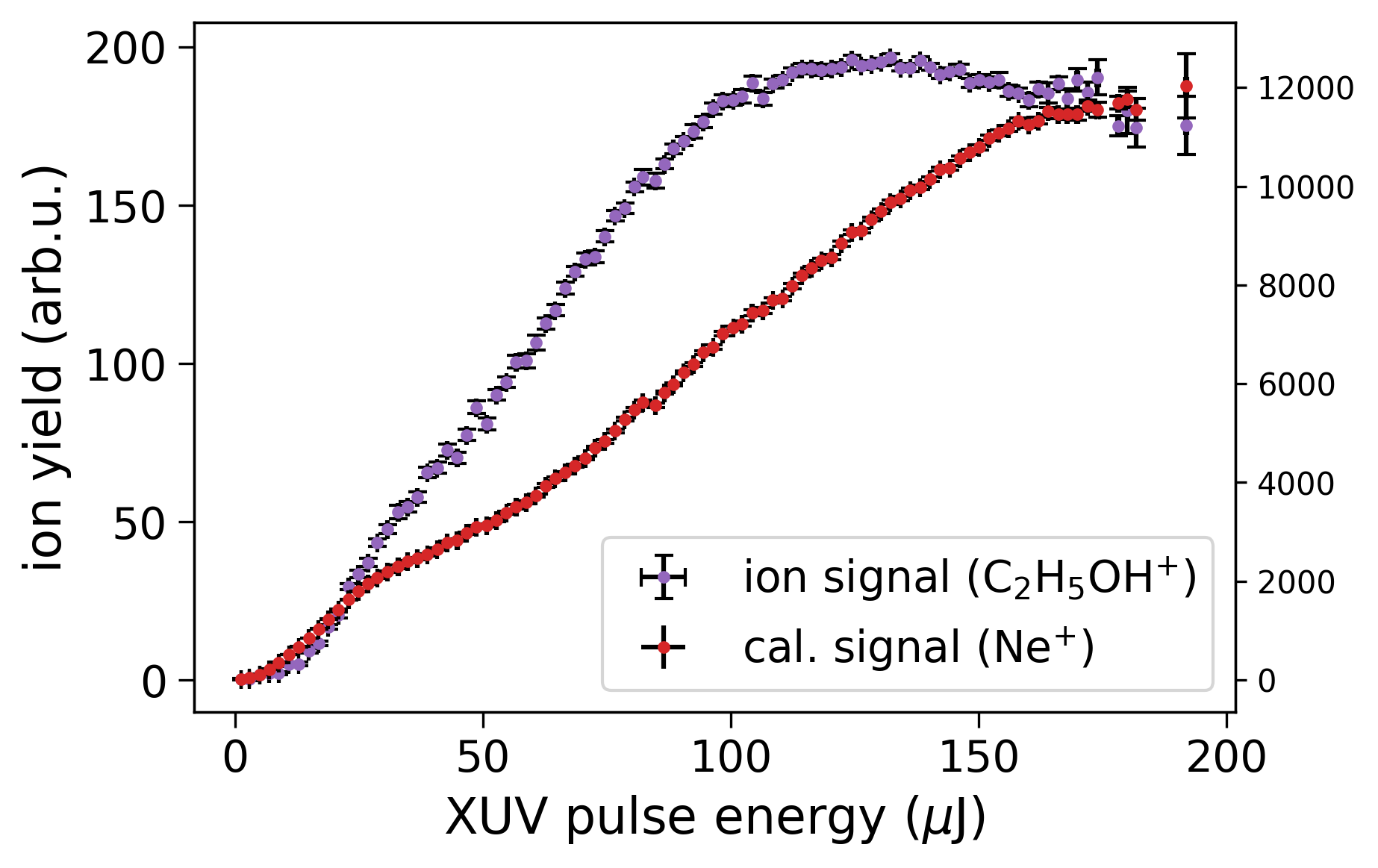}
        \caption[]{\label{subfig:RawIntensityParent+}}
    \end{subfigure}
    \caption[]{\label{fig:raw_pow_dep} Raw intensity dependence of the (\subref{subfig:RawIntensityH+}) H$^+$ (\subref{subfig:RawIntensityH2+}) H$_2^+$ (\subref{subfig:RawIntensityH3+}) H$_3^+$ and (\subref{subfig:RawIntensityParent+}) parent C$_2$H$_5$OH$^+$ ion yields on the XUV (31.7\ eV) intensity.}
\end{figure}

The raw data for the Ne$^{+}$ photoion intensity-dependence measurements show deviations from ideal linear behaviour, and the other ions show similarly correlated deviations as well (Supplementary Figure~\ref{fig:raw_pow_dep}). These deviations do not seem to originate from the detected ion signals, but rather from the intensity monitor which measures the pulse energy of the FEL radiation. To correct for this, we normalize the intensity-dependent ion data by assuming perfectly linear behaviour for the Ne$^{+}$ with respect to the XUV pulse energy. We use this to correct the XUV pulse energies of the other ions.

\begin{figure}[ht]
    \begin{subfigure}{0.45\linewidth}
        \includegraphics[width=\linewidth]{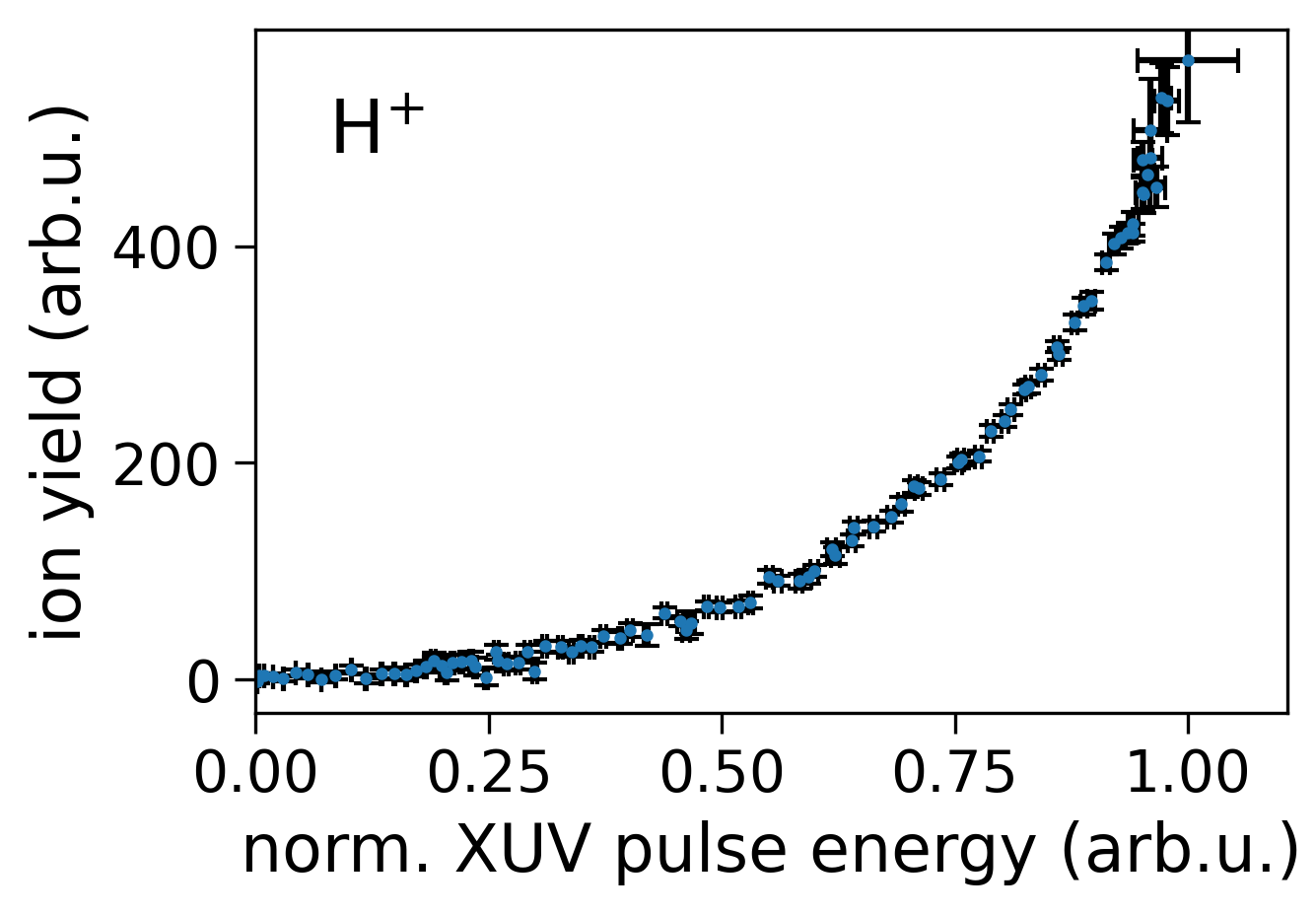}
        \caption[]{\label{subfig:IntensityH+}}
    \end{subfigure}
    ~
    \begin{subfigure}{0.45\linewidth}
        \includegraphics[width=\linewidth]{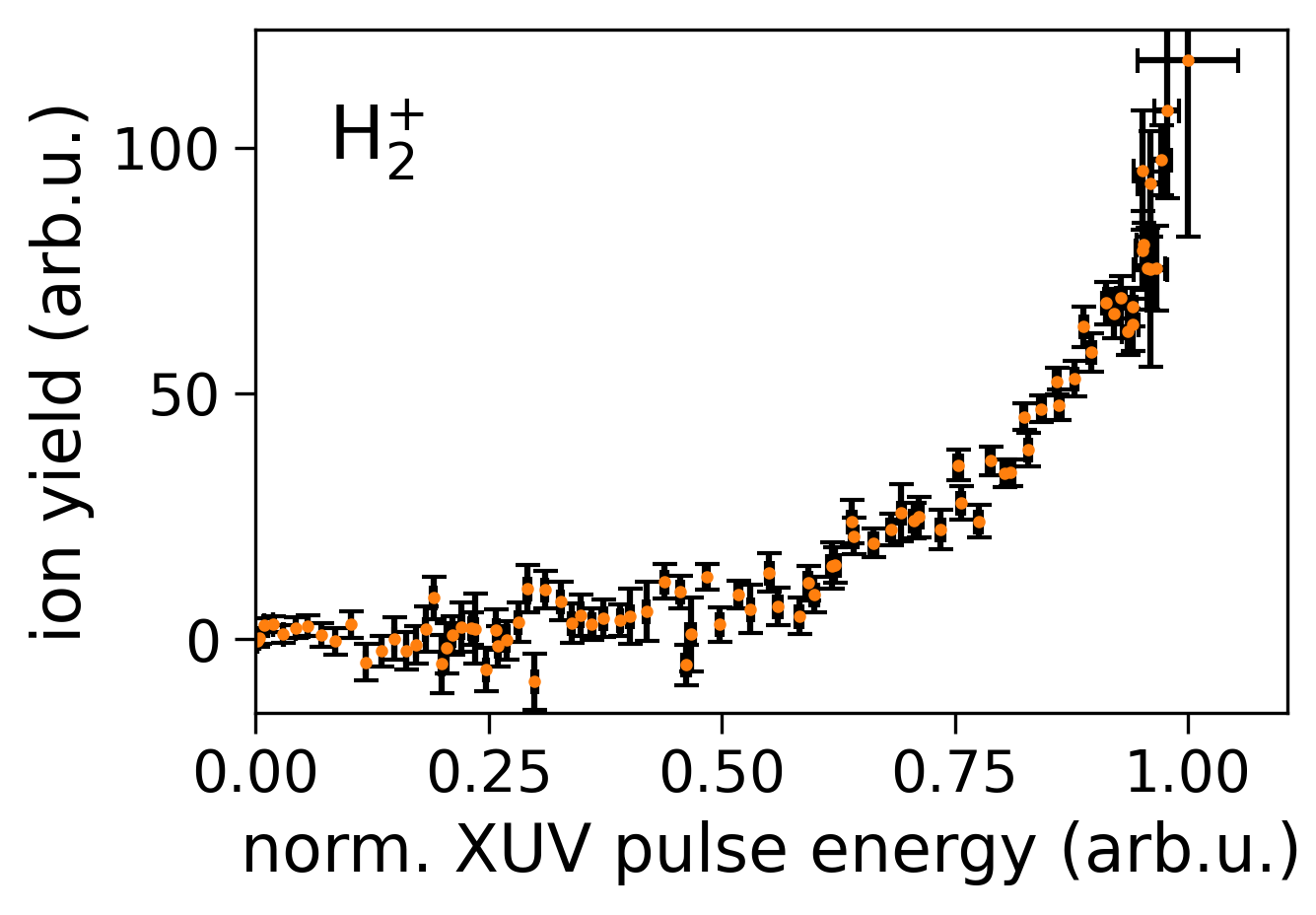}
        \caption[]{\label{subfig:IntensityH2+}}
    \end{subfigure}
    
    \begin{subfigure}{0.45\linewidth}
        \includegraphics[width=\linewidth]{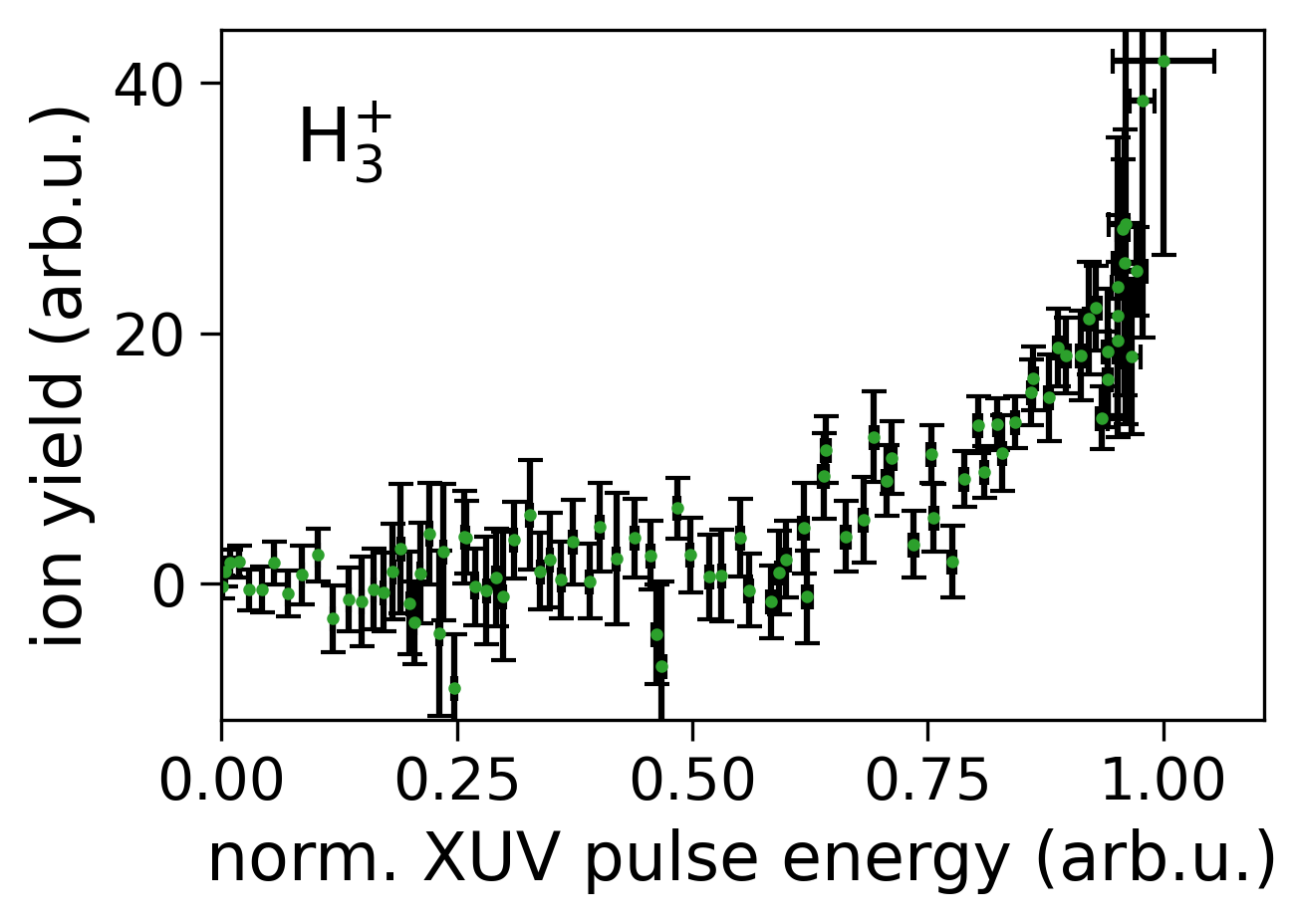}
        \caption[]{\label{subfig:IntensityH3+}}
    \end{subfigure}
    ~
    \begin{subfigure}{0.45\linewidth}
        \includegraphics[width=\linewidth]{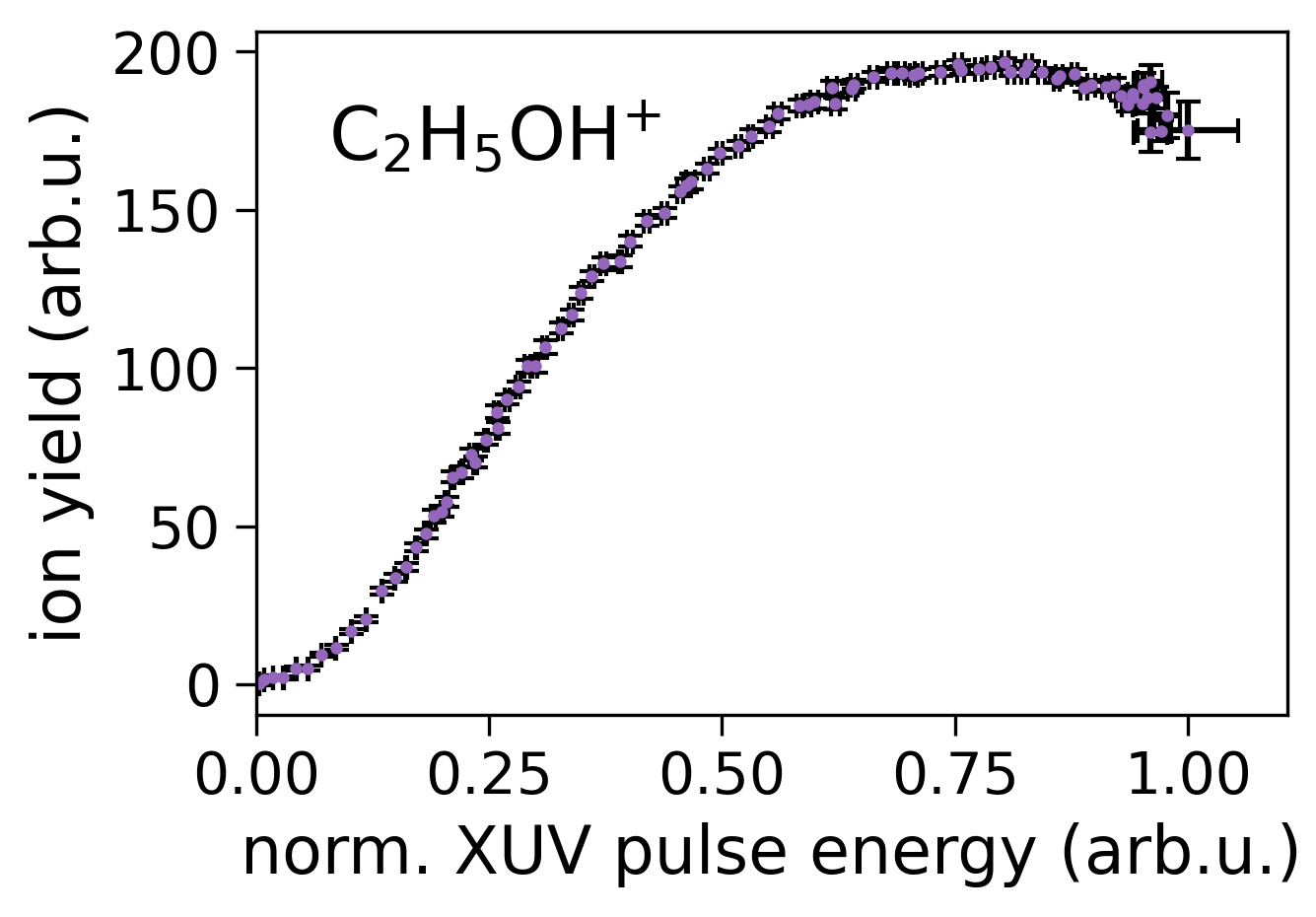}
        \caption[]{\label{subfig:IntensityParent+}}
    \end{subfigure}
    \caption[]{\label{fig:pow_dep} Intensity dependence of the (\subref{subfig:IntensityH+}) H$^+$ (\subref{subfig:IntensityH2+}) H$_2^+$ (\subref{subfig:IntensityH3+}) H$_3^+$ and (\subref{subfig:IntensityParent+}) parent C$_2$H$_5$OH$^+$ ion yields on the XUV (31.7\ eV) intensity. XUV pulse energy normalized to the Ne$^{+}$ signal level, assuming linear dependence of Ne$^{+}$ photoionization at 31.7\ eV up to $3 \cdot 10^{13}$ W/cm$^{2}$. Global error of $\sim$10\% for the XUV pulse energy.}
\end{figure}

\paragraph{Deviations from expected behaviour} The XUV pulse energy dependence of various ion yields are shown in Supplementary Figure~\ref{fig:pow_dep}. 
H$_{1,2,3}^{+}$ are presumably purely dicationic products, and are expected to show a quadratic behaviour with respect to the XUV pulse energy, assuming two-photon double ionization is the dominating process.
We notice that, at low XUV pulse energy, the parent cation exhibits non-linear behaviour (Supplementary Figure~\ref{subfig:IntensityParent+}), whereas it is expected to be linear as a one-photon ionization process.
We attribute this to the minor presence of ethanol clusters in this intensity-dependence measurement.
At higher XUV pulse energies, we see that the parent cation yield saturates, while the H$_{1,2,3}^{+}$ ion signals show a significant increase.
This strong increase is attributed to the non-linearity of the two-photon double-ionization process of ethanol.
We note that in the main text, the datasets presented do not have any contribution from clusters.

\section{Fitting of pump-probe delay-dependent ion yields}

The fit model employed for the pump-probe delay-dependent ($t_{P}$) yield of H$_3^+$ or D$_3^+$ ions consists of three contributions, as discussed in the main text. 
The positive enhancement around zero delay was fitted as a Gaussian cross-correlation feature
\begin{equation}
\label{eq:cross_contribution}
    y_{C}(t_P; \sigma) \coloneqq a_C e^{\frac{1}{2} \left(\frac{t_P-t_0}{\sigma_C}\right)^2},
\end{equation}
where $\sigma_C \coloneqq \sqrt{\sigma_{X}^2 + \sigma_{U}^2}=$119\,fs FWHM corresponds to the experimentally measured cross-correlation, $t_{0}$ is the zero-delay position between the two pulses, and $a_C$ is a fitted amplitude.

The additional step function, which is used to account for signal differences at long negative and long positive delays ($a_{\pm\infty}$), has the form,
\begin{equation}
\label{eq:offset_contribution}
    y_{O}(t_P) \coloneqq   \frac{a_{\infty} - a_{-\infty}}{2} \erf\left\{\frac{t_P}{\sigma_C\sqrt{2}}\right\} - a_{-\infty}.
\end{equation}

The exponentially decaying depletion feature $y_{D}$ is a convolution between a Gaussian and a step function with an exponential decay.
For example, the depletion of the H$_{3}^{+}$ ion, originating from double photoionization is found as
\begin{equation}
    y_{D}(t_P; \sigma) \approx s(t;\sigma) * \theta(t)e^{-R t}
\end{equation}
where $\theta(t)$ is a step function, and ``$*$'' denotes convolution, $s(t;\sigma_D)$ is a Gaussian resulting from convoluting three other Gaussians (two from XUV corresponding to two-photon double ionization, and one from UV), with a corresponding width fixed as $\sigma_D=\sqrt{2\sigma_{X}^2 + \sigma_{U}^2}=$135\,fs FWHM.
The convolution evaluates as
\begin{equation}
\label{eq:depletion_approximation}
    y_D(t_P; \sigma) \approx 1+\frac{1}{2}(C-1)e^{\frac{1}{2}(R \sigma)^2}e^{-R t_P} \erfc\left\{\frac{R \sigma^2-t_P}{\sqrt{2}\sigma}\right\},
\end{equation}
where $C$ is interpreted as the depletion strength (e.g. $C=1$ is no depletion, $C=0$ is total depletion, $C=2$ is a two-fold enhancement), and $\erfc$ is the complementary error function.
This ``simple'' form holds as long as the depletion is small and probe pulses are short.
In the case of strong depletion by long probe pulses, saturation of the probe effects may lead to failure of this approximation. Nevertheless, only the apparent time-zero $t_{0}$ and convolution width $\sigma$ would be affected; the rate constant $R$ remains unchanged.

The overall fit for H$_{3}^{+}$ is then
\begin{equation}\label{eq:depletion_fit}
    y(t_P)=y_O(t_P; \sigma_C) + y_C(t_P; \sigma_C)+y_D(t_P; \sigma_D).
\end{equation}
Mass fragments other than H$_{3}^{+}$ were fitted with a similar form, but using the cross-correlation of XUV+UV for all contributions, rather than the XUV+XUV+UV cross-correlation of H$_{3}^{+}$,
\begin{equation}\label{eq:others_depletion_fit}
    y(t_P)=y_O(t_P; \sigma_C) + y_C(t_P; \sigma_C)+y_D(t_P; \sigma_C).
\end{equation}

For all fragments other than H$_3^+$ and D$_3^+$, which presumably originate dominantly from single photoionization, we let the width $\sigma_{D}$ fit freely.

\paragraph{Rate equation} In the case where the approximation fails, and precise values of $t_{0}$ and $\sigma$ are required, one must consider the system of rate equations whose approximation leads to Eq.~\eqref{eq:depletion_approximation}.
We will introduce this here.
The disruptive pathway is determined by the cross-section for the XUV-induced double-ionization $S$, the relaxation time-constant $R$ from the dication state to the roaming product, and the disruption strength $D$ of the UV probe on the roaming process.
To describe it in detail, we give the form of our XUV pump $x(t)$ and UV probe $u(t)$ pulses with a pump-probe delay $t_P$ between them,
\begin{equation}
\begin{aligned}
   x(t) &= \frac{P_x}{\sigma_{X}\sqrt{2\pi}} e^{-\frac{1}{2} \left(\frac{t}{\sigma_{X}}\right)^2} \\
   u(t;t_P) &= \frac{P_u}{\sigma_{U}\sqrt{2\pi}} e^{-\frac{1}{2} \left(\frac{t-t_P}{\sigma_{U}}\right)^2}, \\
\end{aligned}
\end{equation}
and now form a system of rate equations to describe the time-evolution as:
\begin{equation}
\label{eq:depletion_rate_equations}
\begin{aligned}
   y_{0}'(t) &= -S_1 x(t) y_{0}, \\ 
   y'_{1}(t) &= S_1 x(t) y_{0}(t) - S_2 x(t) y_{1}(t), \\ 
   y_{2}'(t) &= S_2 x(t) y_{1}(t) - D u(t;t_P) y_{2}(t) - R y_{2}(t), \\ 
   y_{D}'(t) &= R y_{2}(t), \\
\end{aligned}
\end{equation}
where we have the populations of: the ground state $y_{0}$, the cationic state $y_{1}$, the undissociated dicationic state $y_{2}$, and the H$_{3}^{+}$ product $y_{D}$; and the couplings: ionization rate $S_1$ of $y_{0}$ by $x(t)$, ionization rate $S_2$ of $y_{1}$ by $x(t)$, depletion rate $D$ of the $y_{2}$ by $u(t;t_P)$, and decay rate $R$ of $y_{2}$.
The ``depletion'' contribution is given as $y_{D} \coloneqq \lim_{t\rightarrow \infty} y_{D}(t)$,
\begin{equation}
\label{eq:depletion_contribution}
    y_{D} = R \int_{-\infty}^{\infty} dt \left\{ e^{-D u(t;t_P) - R t} \int_{-\infty}^{t} dt' \left\{e^{Du(t';t_P)+Rt'} B x(t') \left(\frac{1}{B-A} e^{-Ax(t')} - e^{-Bx(t')}\right)\right\}\right\}.
\end{equation}
A closed form for Eq.~\eqref{eq:depletion_contribution} does not exist, so it must either be numerically evaluated, or approximated (e.g. Eq.~\eqref{eq:depletion_approximation}).

\section{Ion yield data for 2-aminoethanol}

Analogous to ethanol in the main text, we provide relative ion yields from photoionization of 2-aminoethanol in Supplementary Table \ref{tab:ami_fragment_abundances}.

\begin{table}
    \centering
    \caption[]{\label{tab:ami_fragment_abundances} Relative ion yields for 2-aminoethanol (C$_{2}$H$_{7}$NO) upon XUV ionization.}
    \begin{tabular}{c | c | c | c }
        \hline\hline
        $m/z$ from & possible & relative yield at  & relative yield at \\
        C$_{2}$H$_{7}$NO & identity & $hv=21.1$\,eV & $hv=31.7$\,eV \\
        \hline\hline
1  & H$^{{+}}$                         & 55.7(4)  & 1550(10)  \\
2  & H$_{{2}}^{{+}}$                   & 22.0(3)  & 164(4)    \\
3  & H$_{{3}}^{{+}}$                   & 0.0(4)   & 13(2)     \\
13 & CH$^{{+}}$                        & 2.9(2)   & 154(2)    \\
14 & CH$_{{2}}^{{+}}$/N$^{{+}}$        & 6.7(1)   & 279(3)    \\
15 & CH$_{{3}}^{{+}}$/NH$^{{+}}$       & 265.8(8) & 831(6)    \\
17 & OH$^{{+}}$/NH$_{{3}}^{{+}}$       & 85.1(5)  & 373(4)    \\
18 & H$_{{2}}$O$^{{+}}$/NH$_{{4}}^{{+}}$ & 272(3)   & 664(5)    \\
19 & H$_{{3}}$O$^{{+}}$                & 58.9(3)  & 1073(8)   \\
28 & NH$_{{2}}$C$^{{+}}$               & 372(2)   & 595(4)    \\
30 & CH$_{{4}}$N$^{{+}}$               & 3173(8)  & 3150(20)  \\
42 & C$_{{2}}$H$_{{4}}$N$^{{+}}$       & 226.2(7) & 278(2)    \\
45 & CH$_{{3}}$CH$_{{2}}$O$^{{+}}$     & 28.3(3)  & 43.1(7)   \\
61 & C$_{{2}}$H$_{{7}}$NO$^{{+}}$      & 100      & 100       \\        
    \end{tabular}
\end{table}

\section{Depletion data for ethanol and 2-aminoethanol}

\begin{figure}[t]
    \centering
    \begin{subfigure}[t]{0.3\linewidth}
        \centering
        \includegraphics[width=\linewidth*4/4]{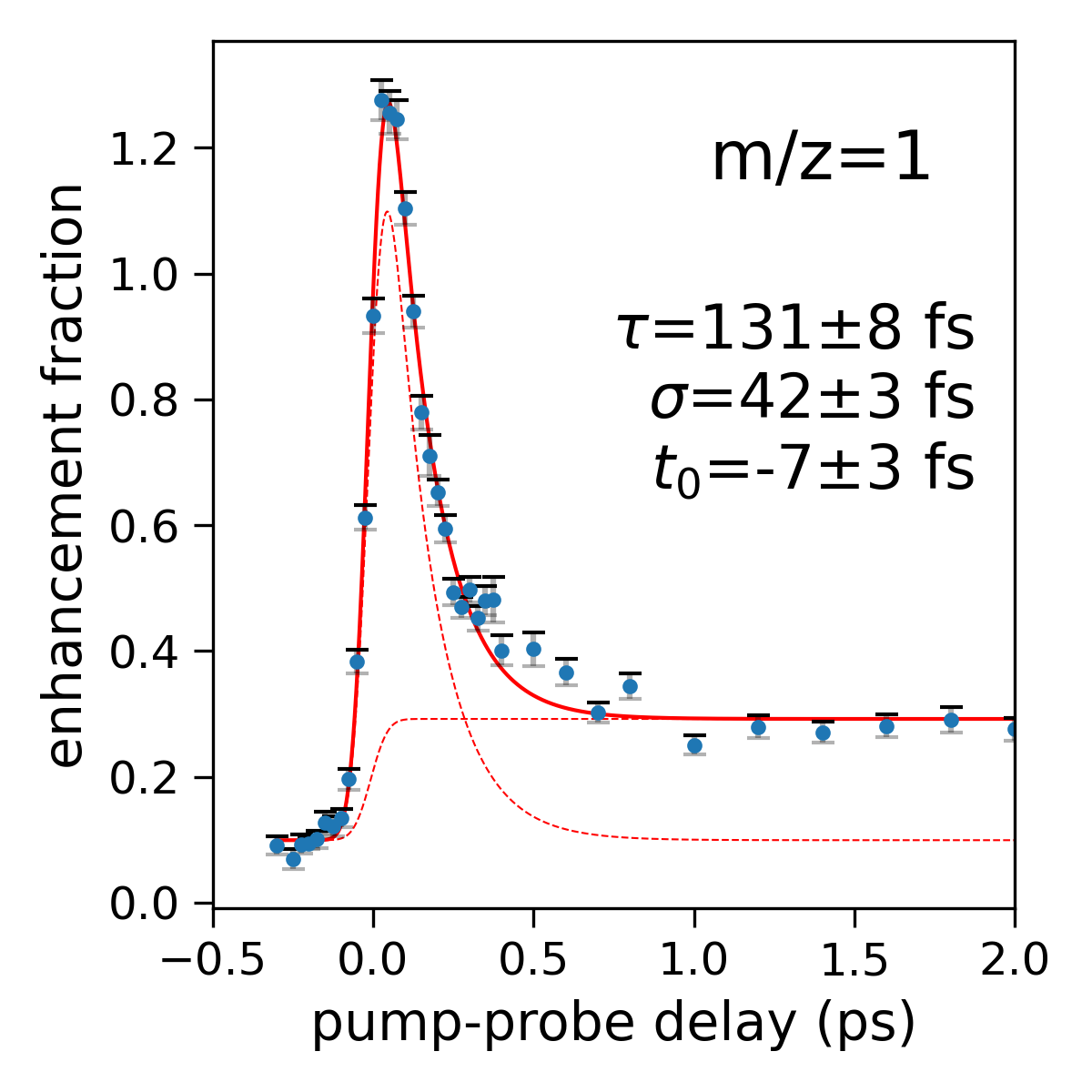}
        \caption[D0 H09 mq1]{\label{D0_H09_mq1} }
    \end{subfigure}%
    ~
    \begin{subfigure}[t]{0.3\linewidth}
        \centering
        \includegraphics[width=\linewidth*4/4]{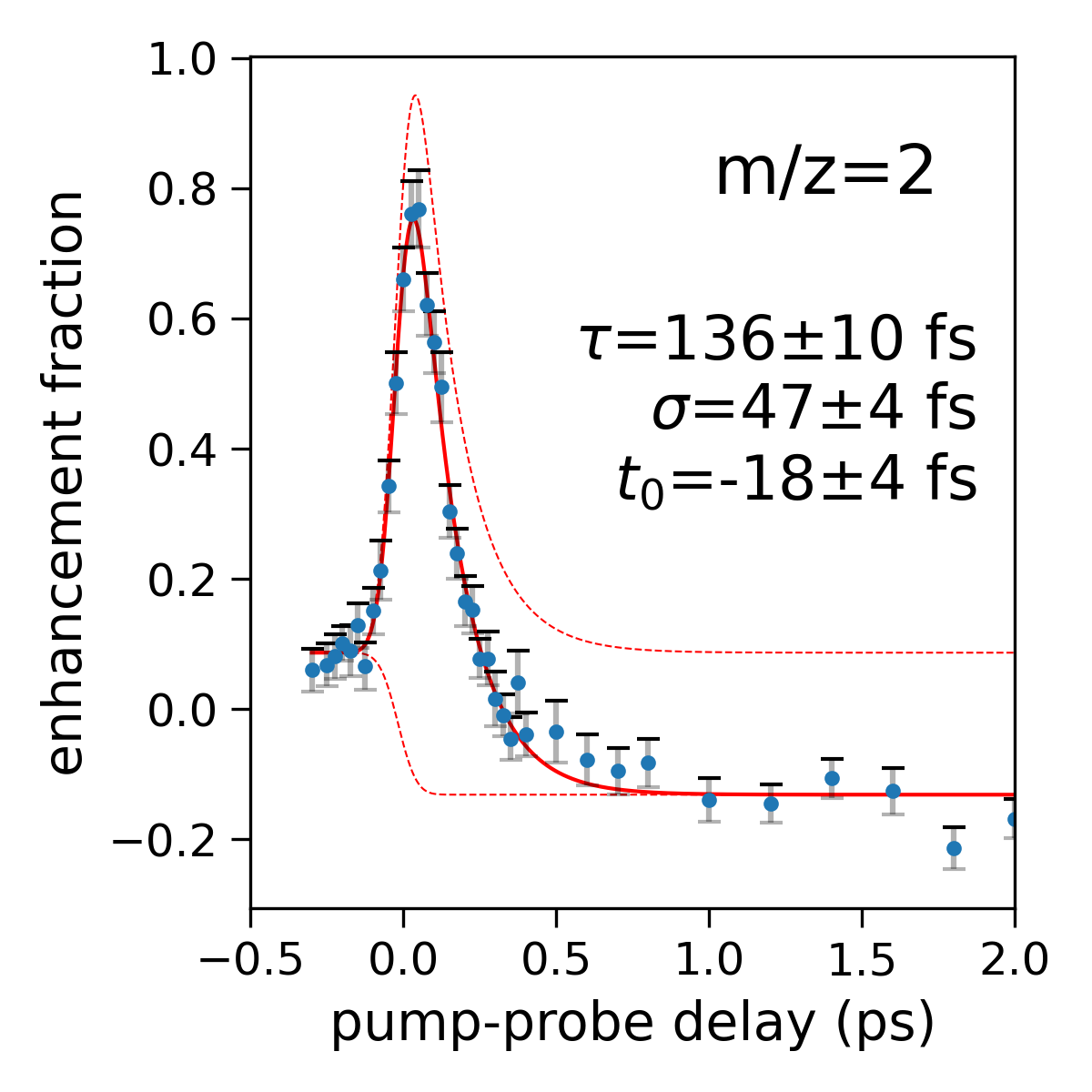}
        \caption[D0 H09 mq2]{\label{D0_H09_mq2} }
    \end{subfigure}%
    ~
    \begin{subfigure}[t]{0.3\linewidth}
        \centering
        \includegraphics[width=\linewidth*4/4]{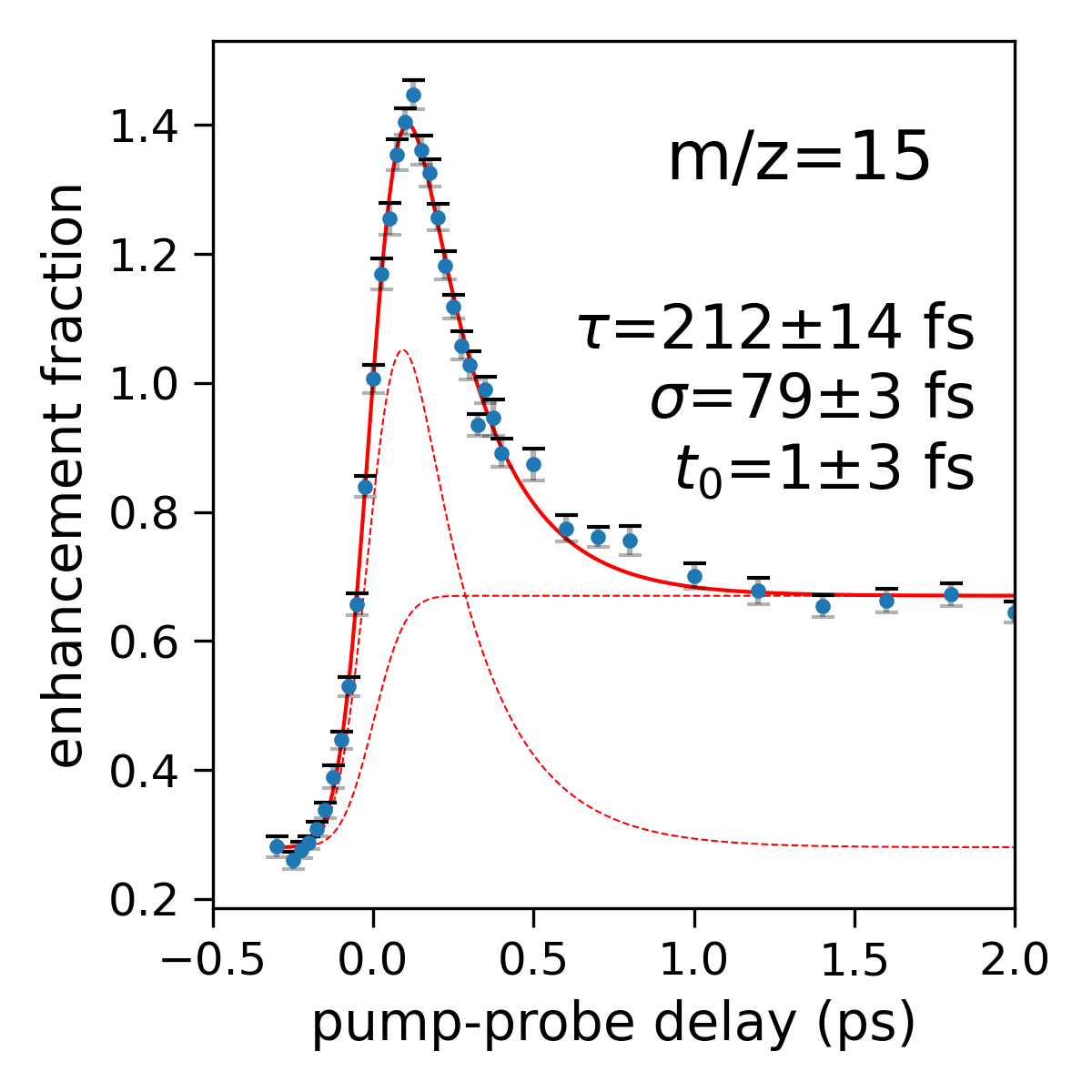}
        \caption[D0 H09 mq15]{\label{D0_H09_mq15} }
    \end{subfigure}%
    
    \begin{subfigure}[t]{0.3\linewidth}
        \centering
        \includegraphics[width=\linewidth*4/4]{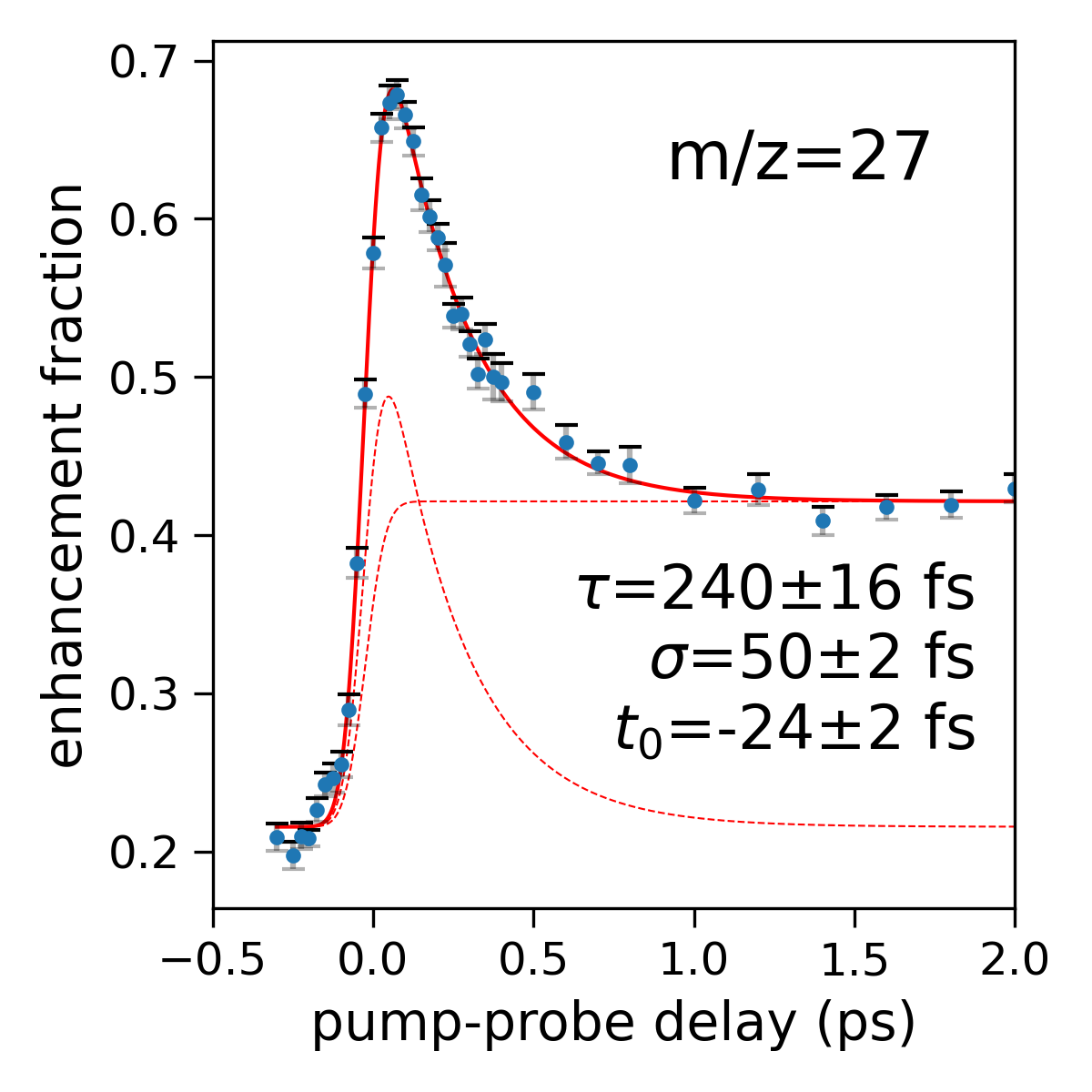}
        \caption[D0 H09 mq27]{\label{D0_H09_mq27} }
    \end{subfigure}%
    ~
    \begin{subfigure}[t]{0.3\linewidth}
        \centering
        \includegraphics[width=\linewidth*4/4]{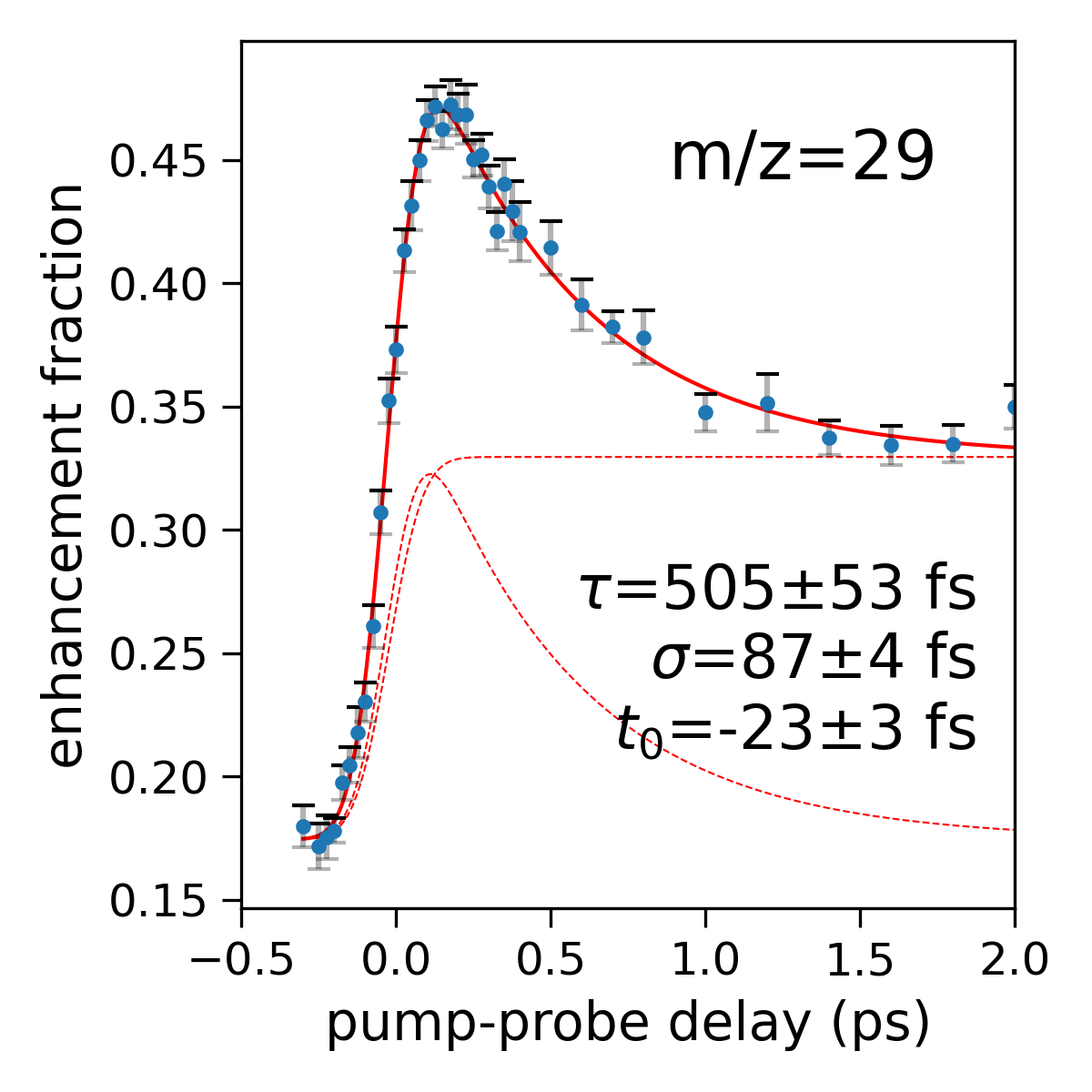}
        \caption[D0 H09 mq29]{\label{D0_H09_mq29} }
    \end{subfigure}%
    ~
    \begin{subfigure}[t]{0.3\linewidth}
        \centering
        \includegraphics[width=\linewidth*4/4]{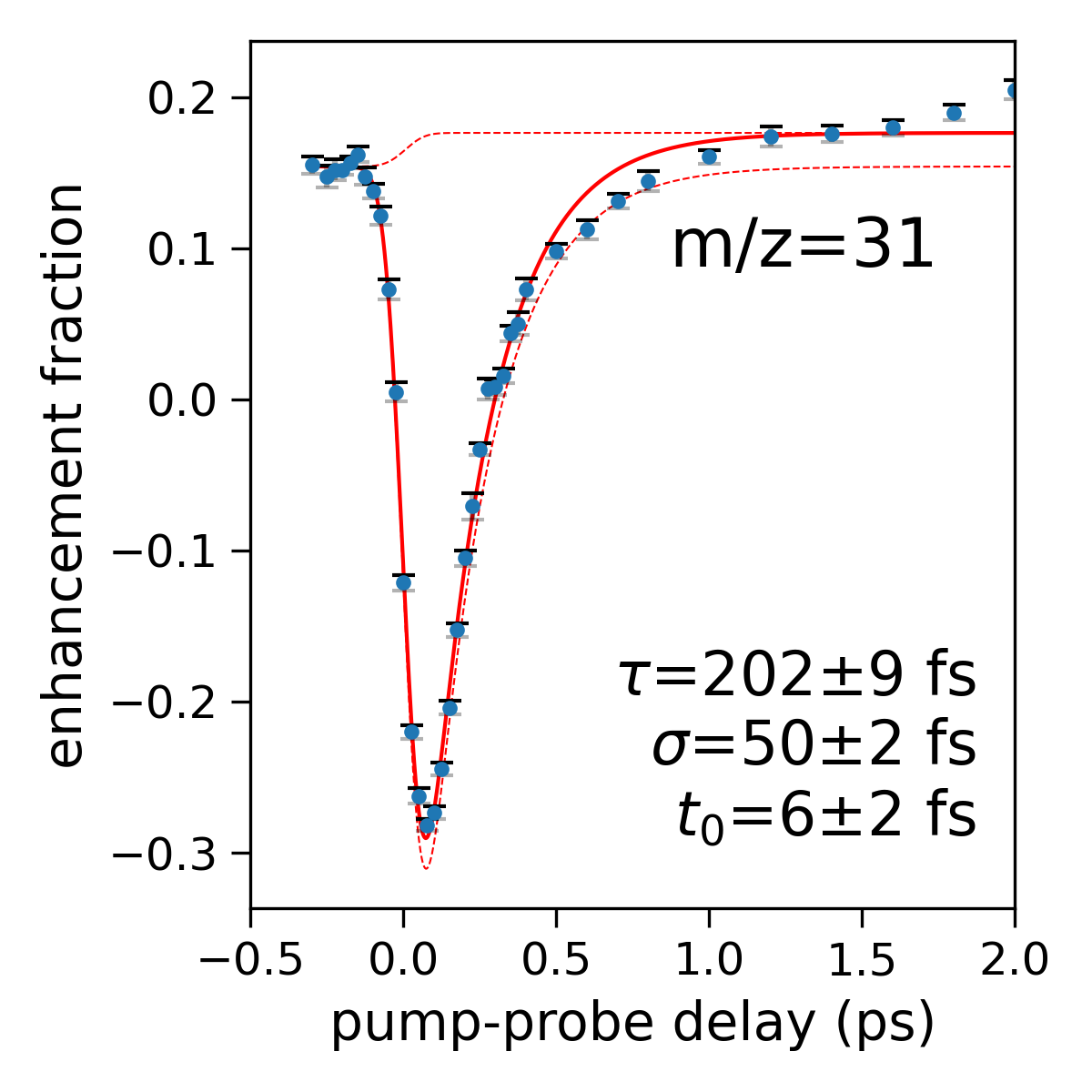}
        \caption[D0 H09 mq31]{\label{D0_H09_mq31} }
    \end{subfigure}%
    
    \begin{subfigure}[t]{0.3\linewidth}
        \centering
        \includegraphics[width=\linewidth*4/4]{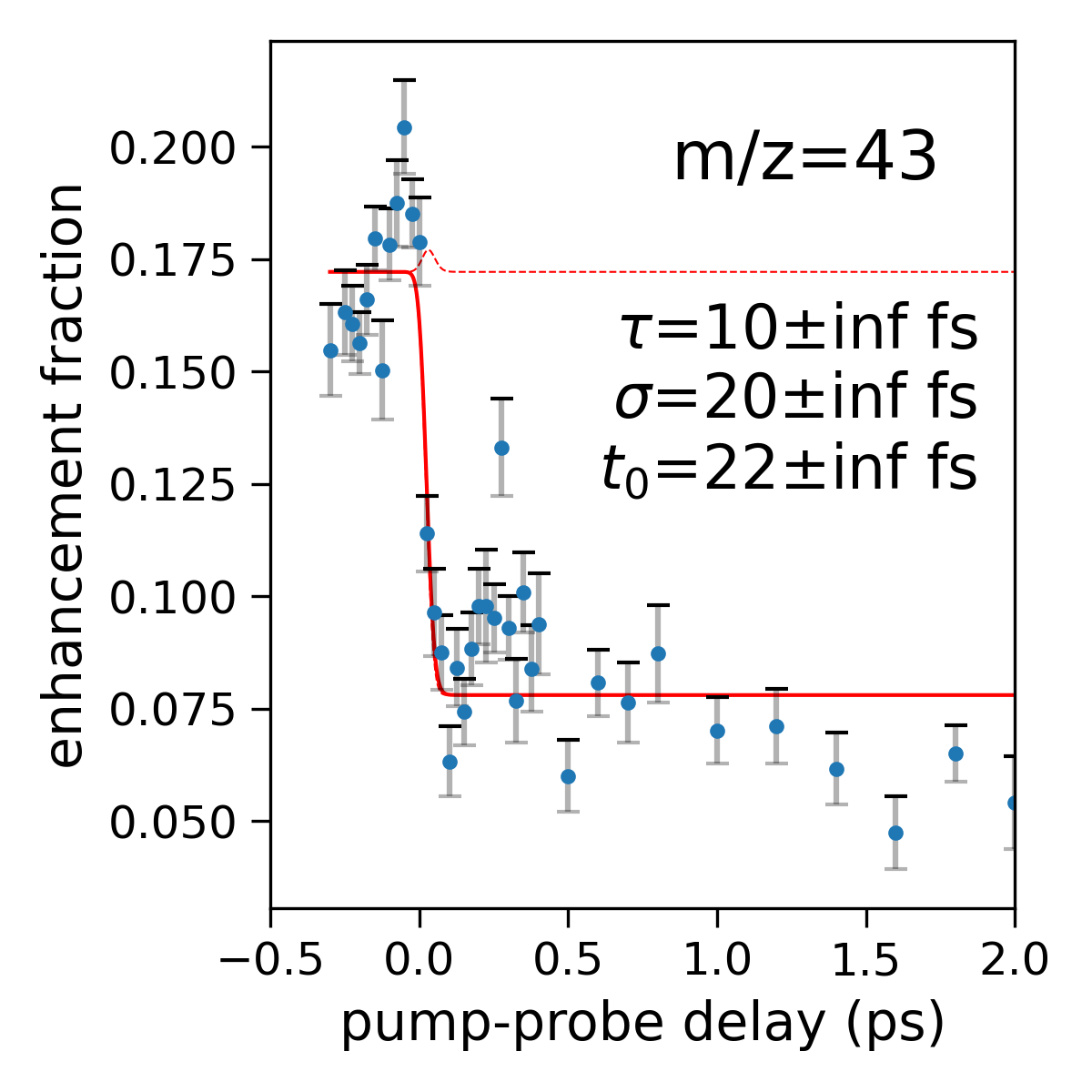}
        \caption[D0 H09 mq43]{\label{D0_H09_mq43} }
    \end{subfigure}%
    ~
    \begin{subfigure}[t]{0.3\linewidth}
        \centering
        \includegraphics[width=\linewidth*4/4]{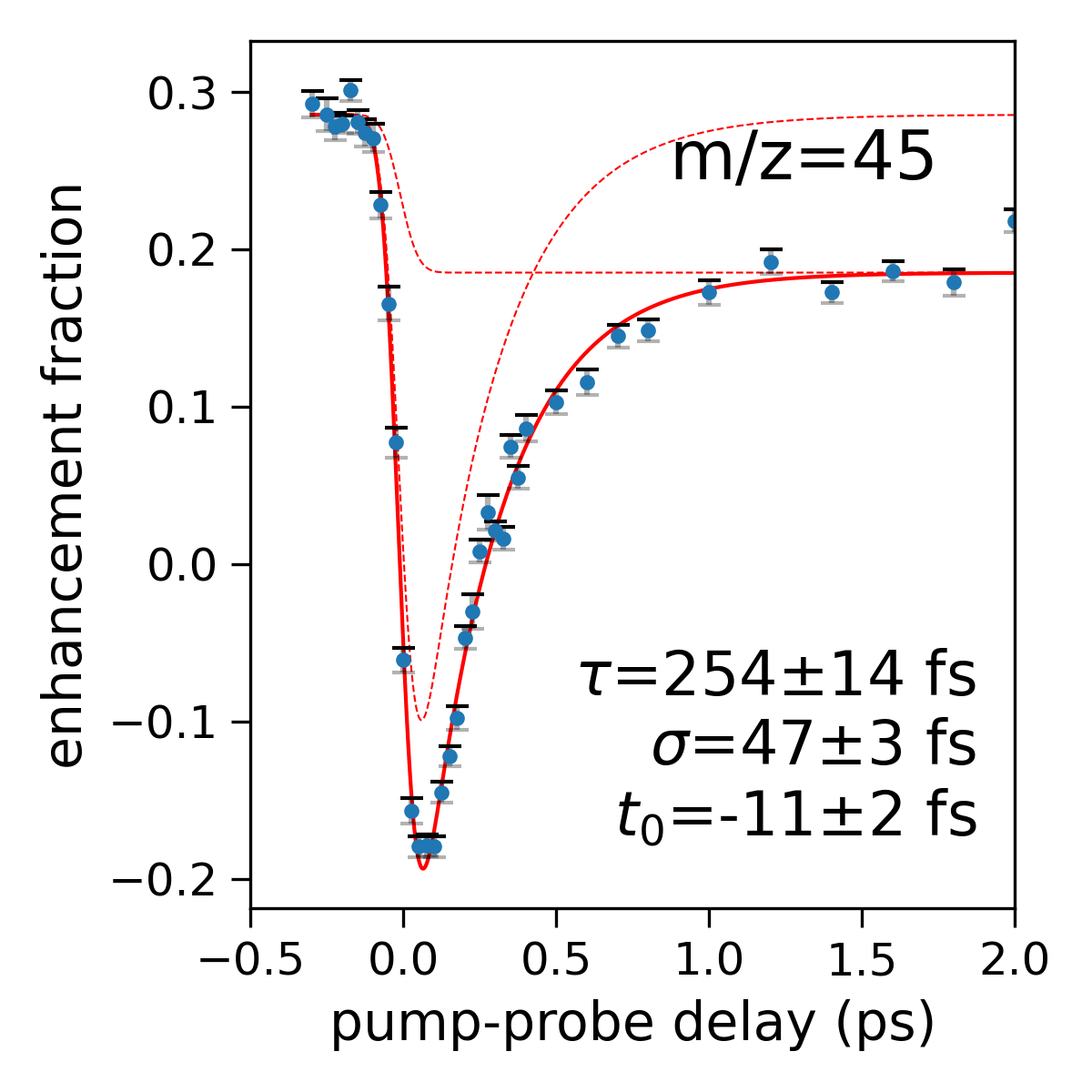}
        \caption[D0 H09 mq45]{\label{D0_H09_mq45} }
    \end{subfigure}%
    ~
    \begin{subfigure}[t]{0.3\linewidth}
        \centering
        \includegraphics[width=\linewidth*4/4]{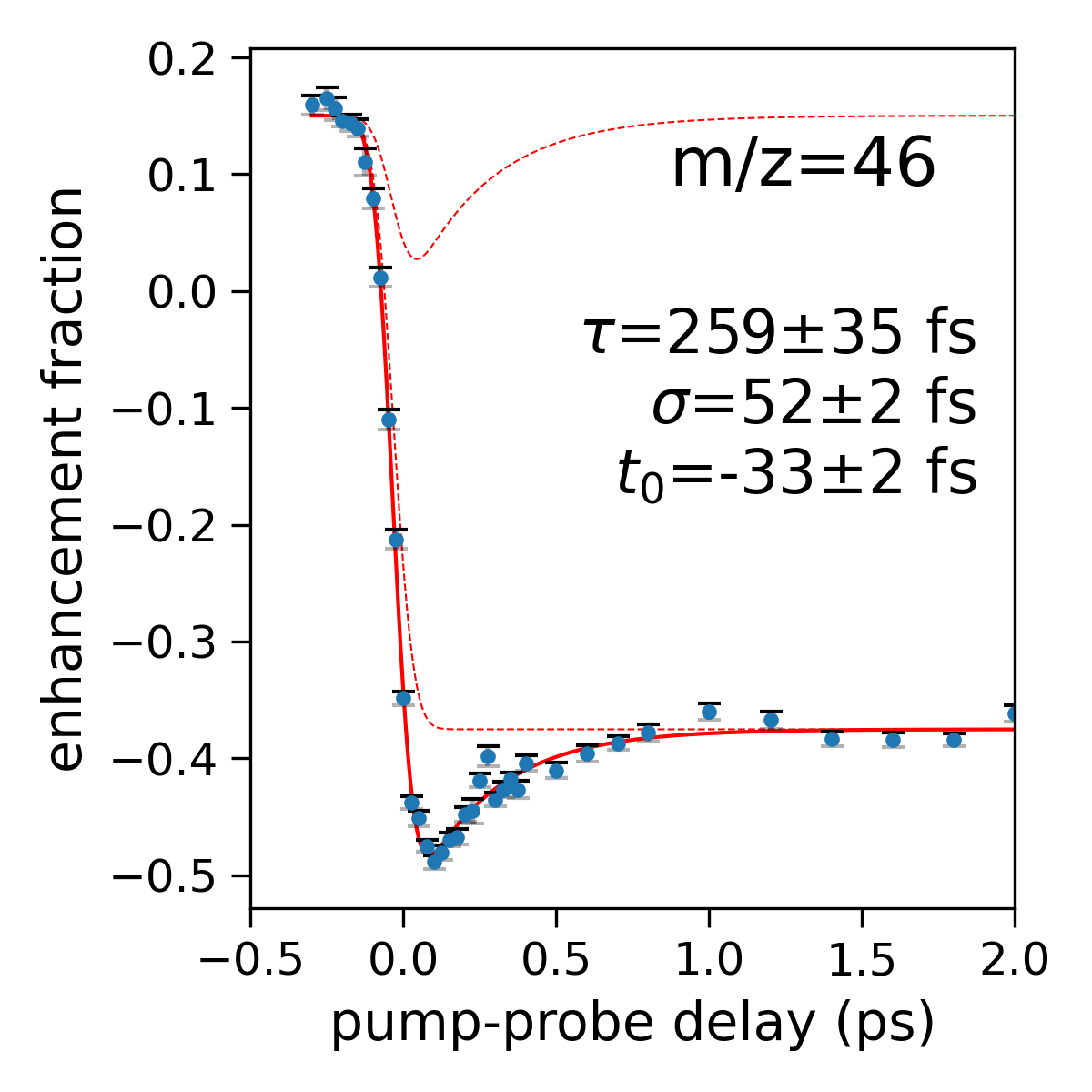}
        \caption[D0 H09 mq46]{\label{D0_H09_mq46} }
    \end{subfigure}%
		
    \caption[]{\label{fig:D0_H9} Pump-probe-delay-dependent ion yields from the photoionization of non-deuterated ethanol, at an XUV photon energy $hv$=31.7\,eV. The data was fit by the model in Eq.~\eqref{eq:depletion_fit}.}
\end{figure}

\begin{figure}[t]
    \centering
    \begin{subfigure}[t]{0.3\linewidth}
        \centering
        \includegraphics[width=\linewidth*4/4]{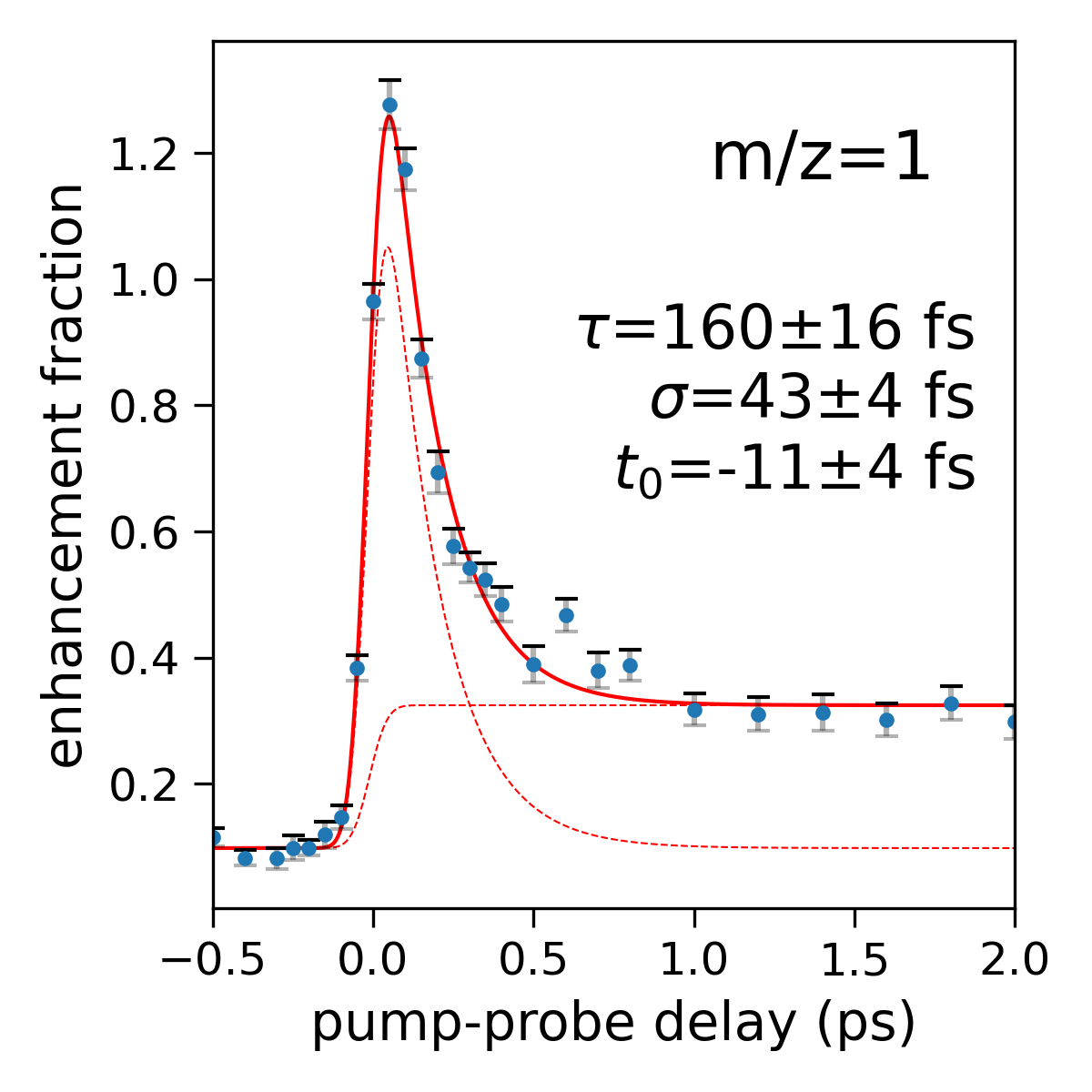}
        \caption[D0 H07 mq1]{\label{D0_H07_mq1} }
    \end{subfigure}%
    ~
    \begin{subfigure}[t]{0.3\linewidth}
        \centering
        \includegraphics[width=\linewidth*4/4]{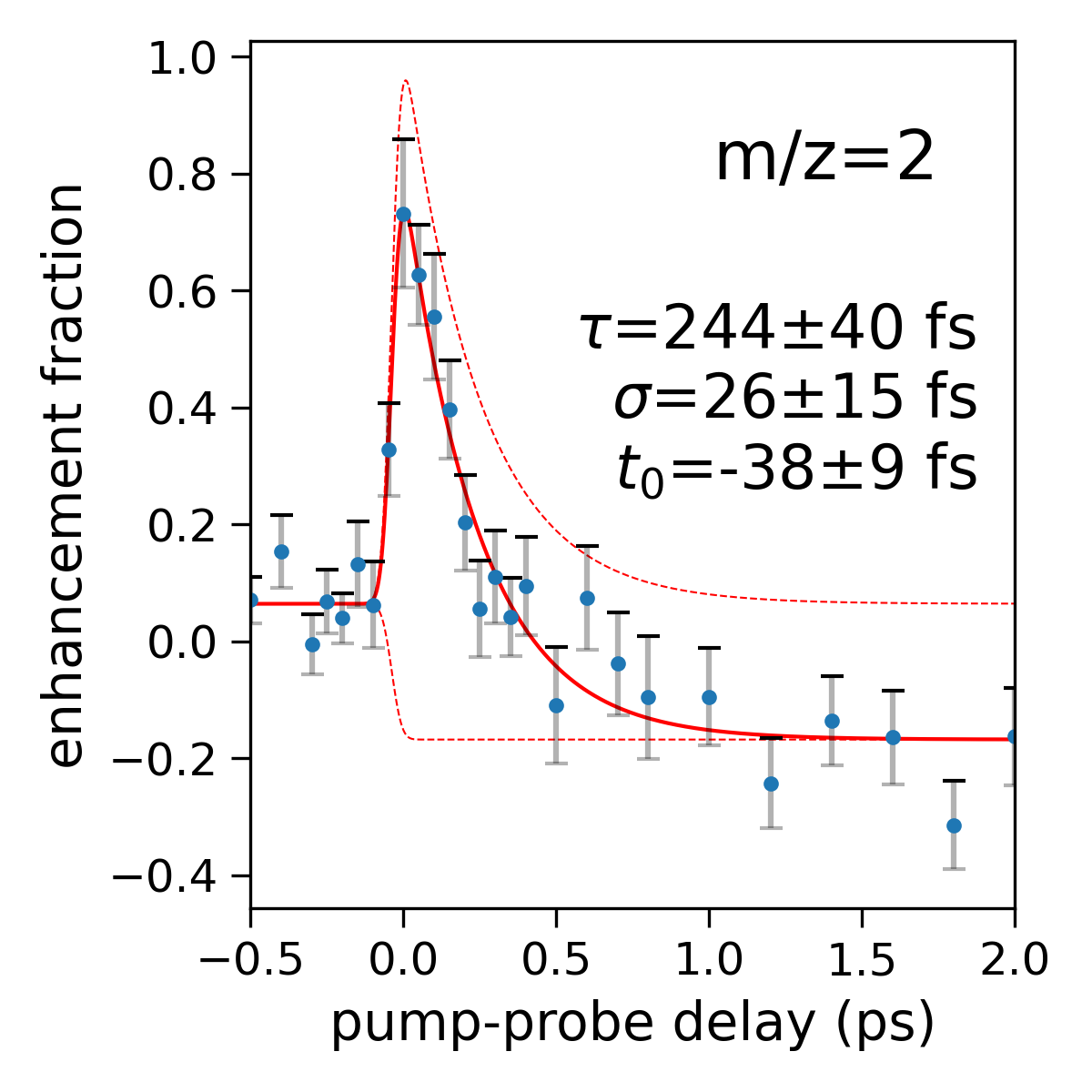}
        \caption[D0 H07 mq2]{\label{D0_H07_mq2} }
    \end{subfigure}%
    ~
    \begin{subfigure}[t]{0.3\linewidth}
        \centering
        \includegraphics[width=\linewidth*4/4]{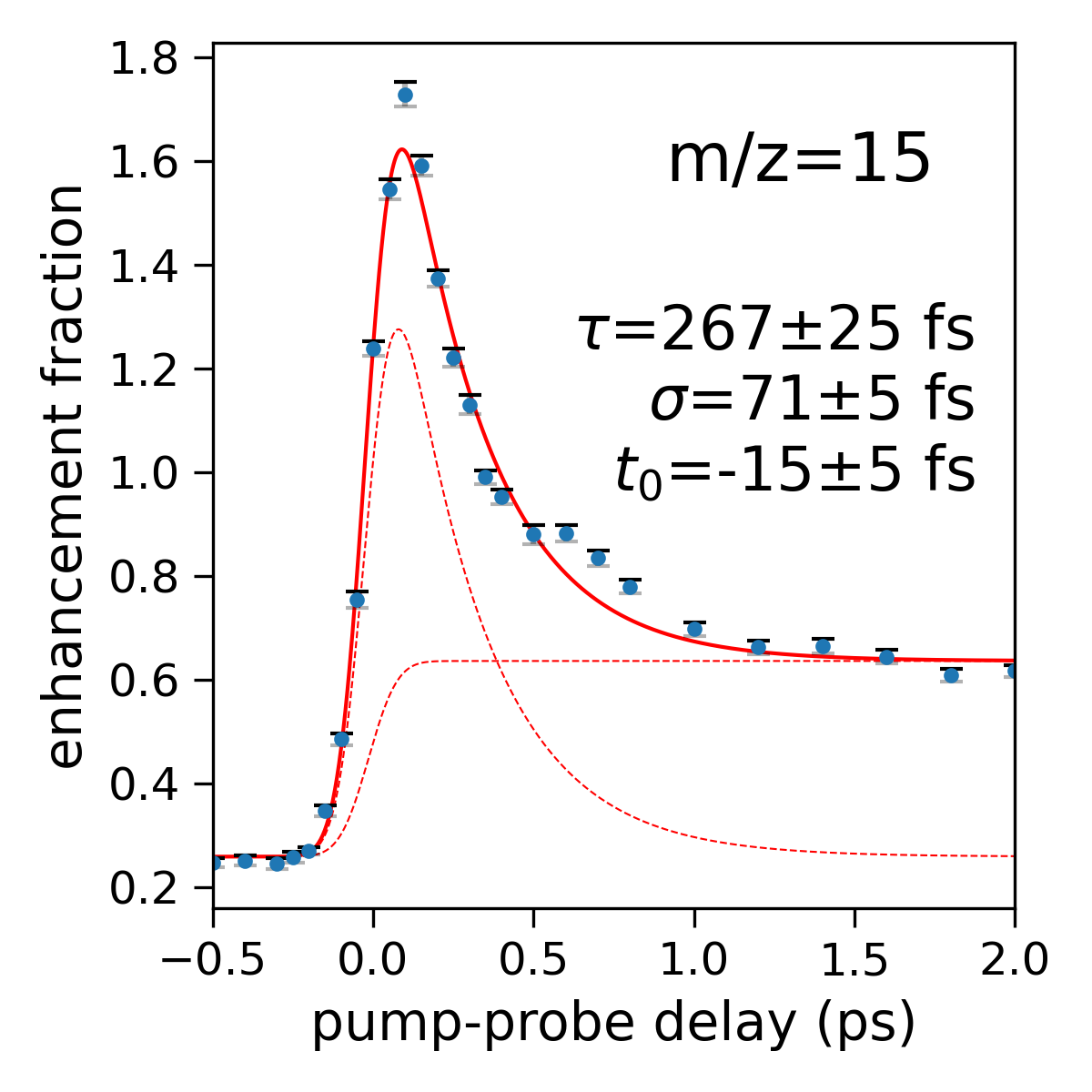}
        \caption[D0 H07 mq15]{\label{D0_H07_mq15} }
    \end{subfigure}%

    \begin{subfigure}[t]{0.3\linewidth}
        \centering
        \includegraphics[width=\linewidth*4/4]{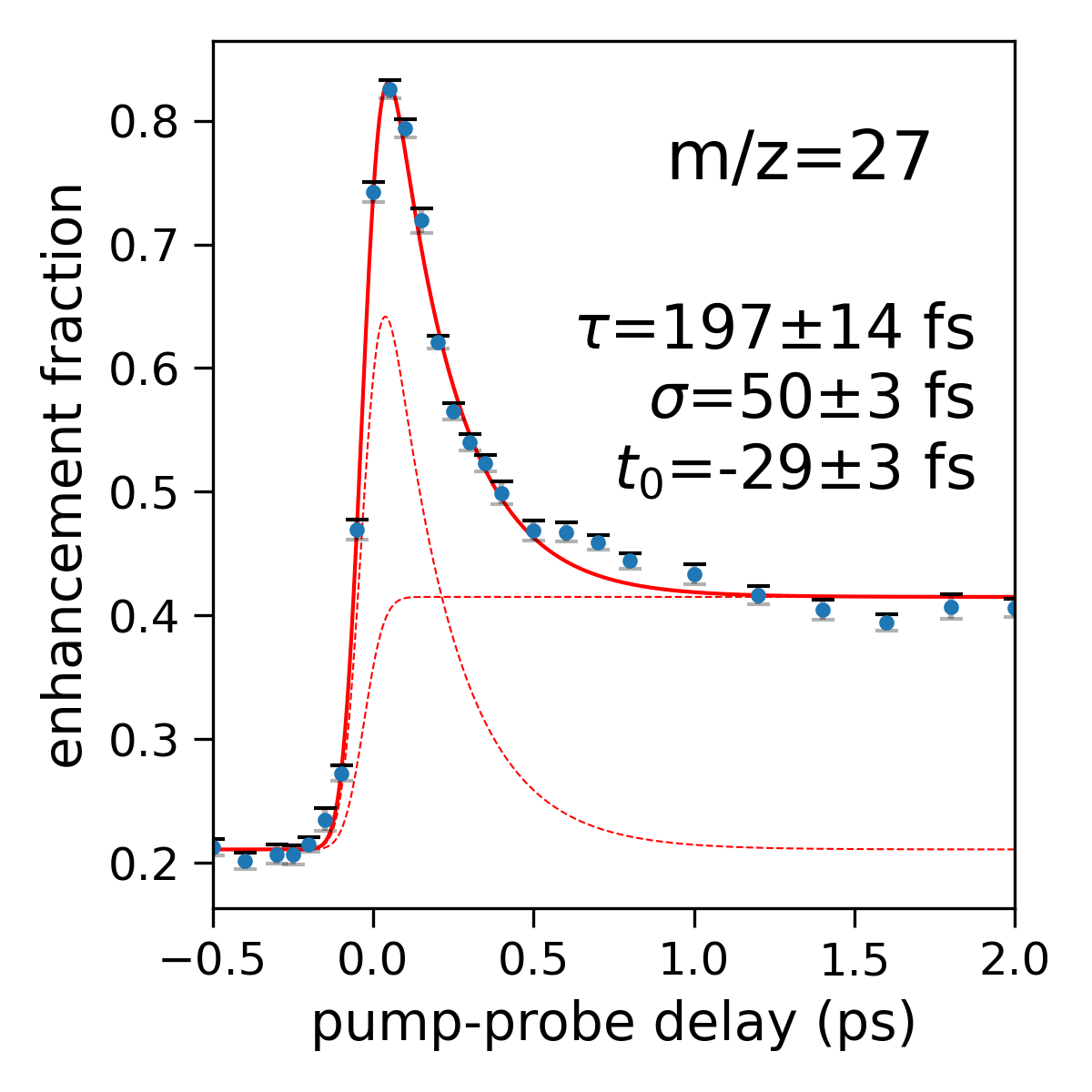}
        \caption[D0 H07 mq27]{\label{D0_H07_mq27} }
    \end{subfigure}%
    ~
    \begin{subfigure}[t]{0.3\linewidth}
        \centering
        \includegraphics[width=\linewidth*4/4]{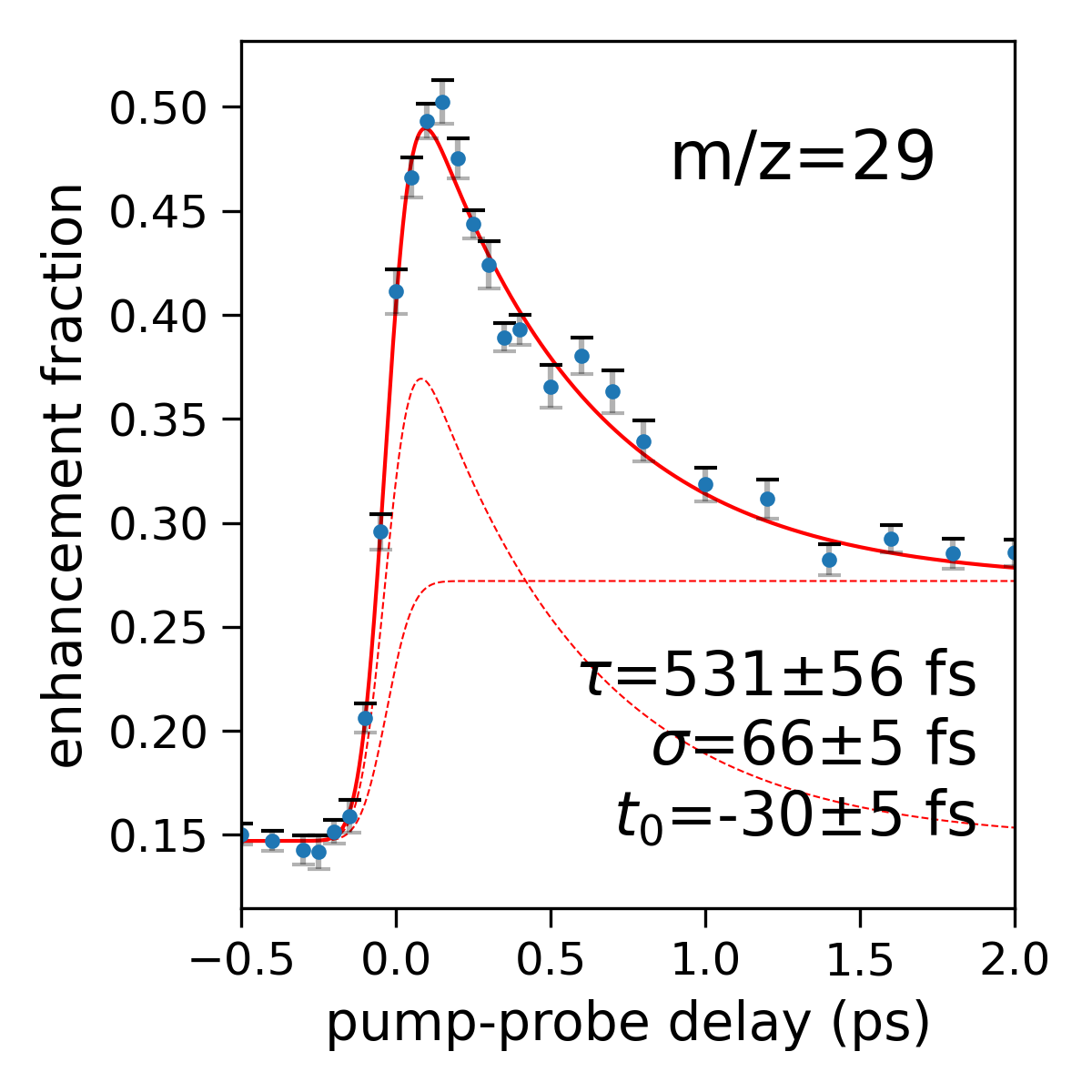}
        \caption[D0 H07 mq29]{\label{D0_H07_mq29} }
    \end{subfigure}%
    ~
    \begin{subfigure}[t]{0.3\linewidth}
        \centering
        \includegraphics[width=\linewidth*4/4]{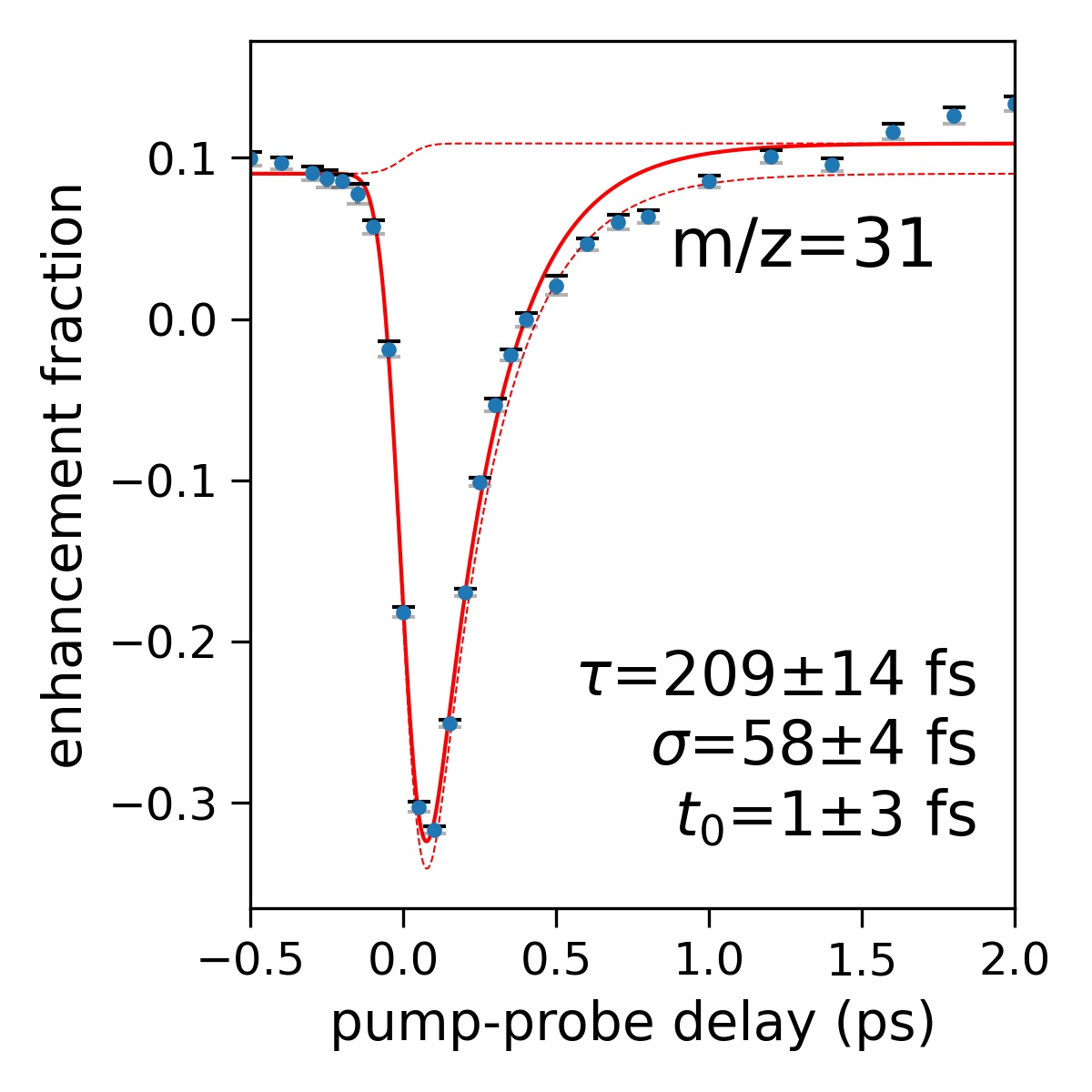}
        \caption[D0 H07 mq31]{\label{D0_H07_mq31} }
    \end{subfigure}%
    
    \begin{subfigure}[t]{0.3\linewidth}
        \centering
        \includegraphics[width=\linewidth*4/4]{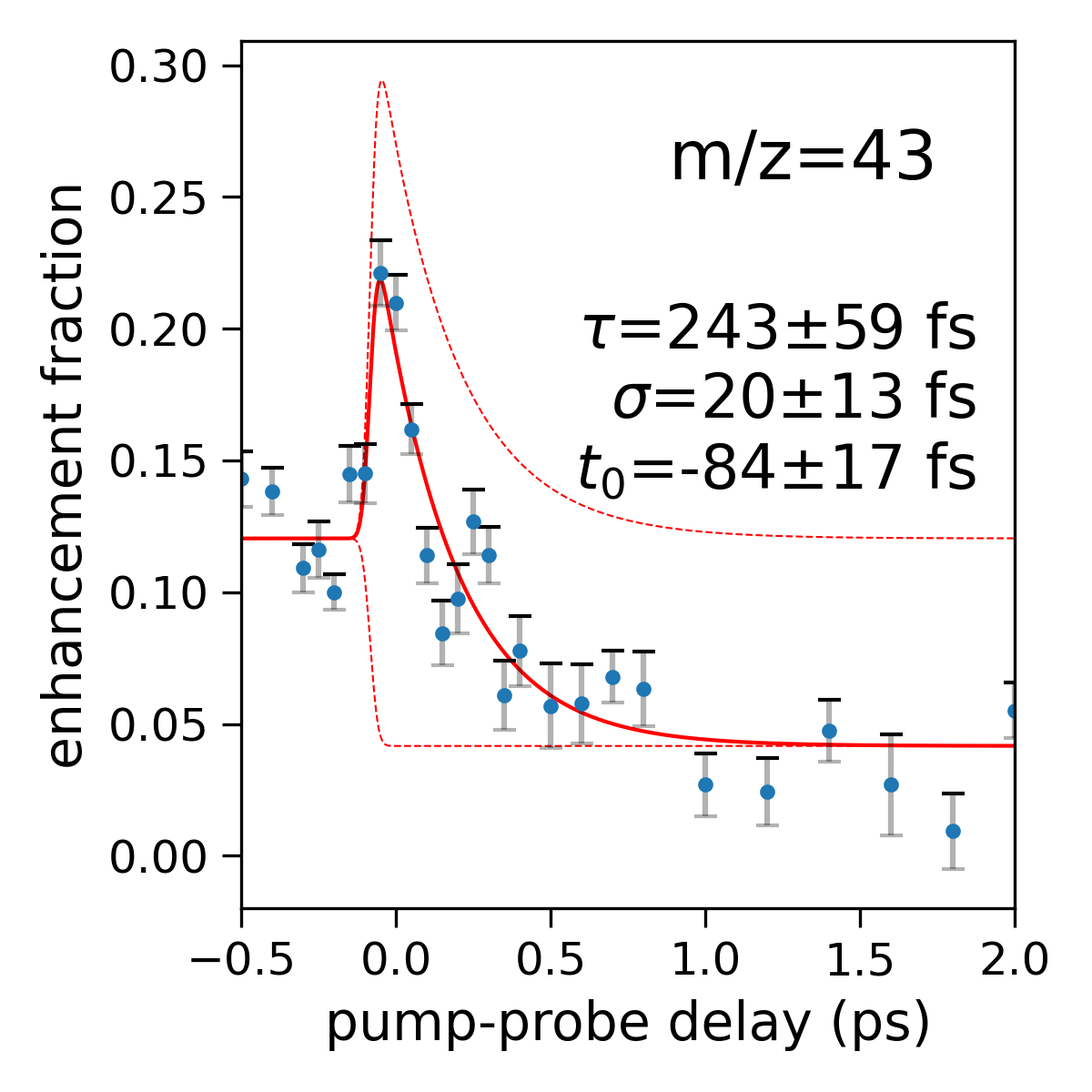}
        \caption[D0 H07 mq43]{\label{D0_H07_mq43} }
    \end{subfigure}%
    ~
    \begin{subfigure}[t]{0.3\linewidth}
        \centering
        \includegraphics[width=\linewidth*4/4]{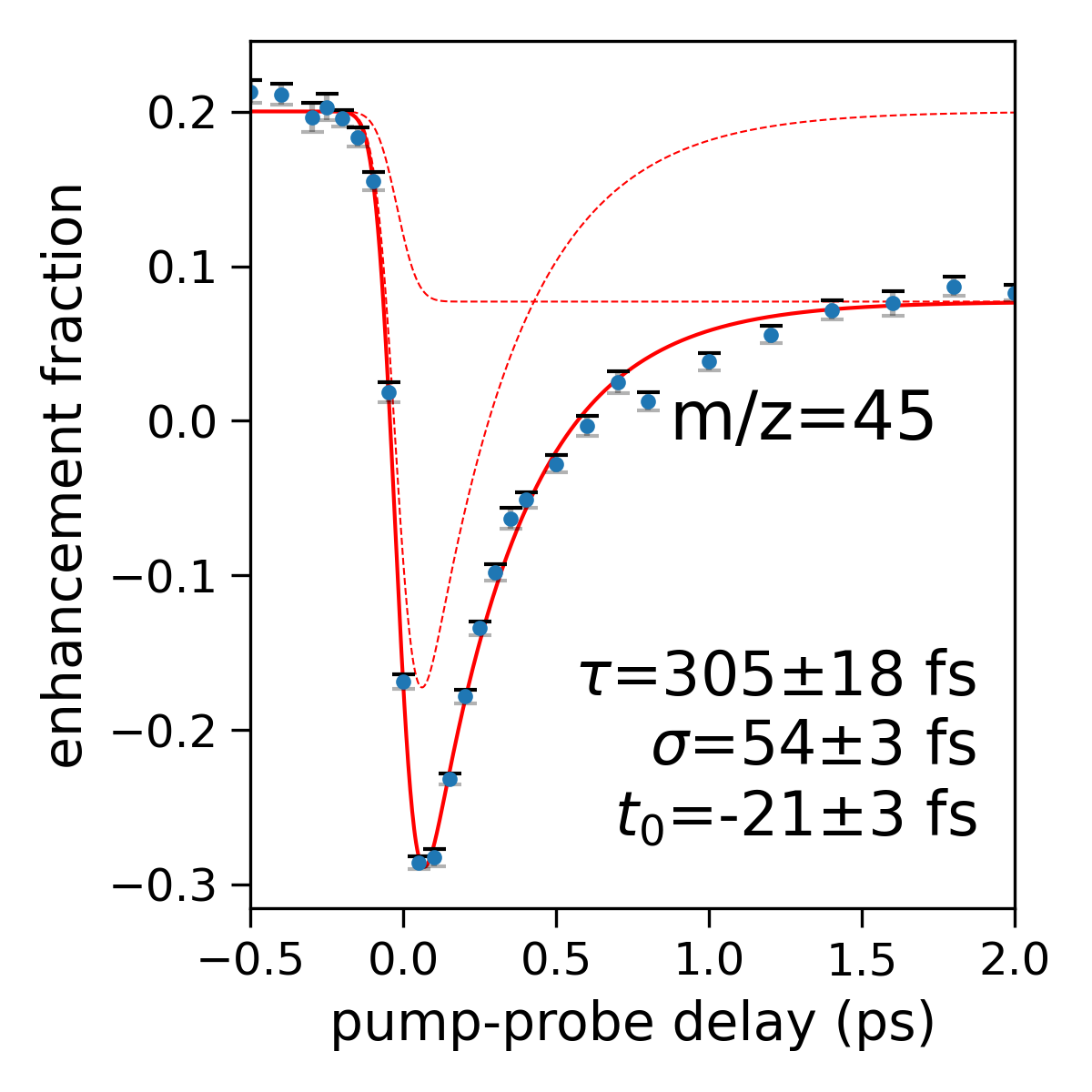}
        \caption[D0 H07 mq45]{\label{D0_H07_mq45} }
    \end{subfigure}%
    ~
    \begin{subfigure}[t]{0.3\linewidth}
        \centering
        \includegraphics[width=\linewidth*4/4]{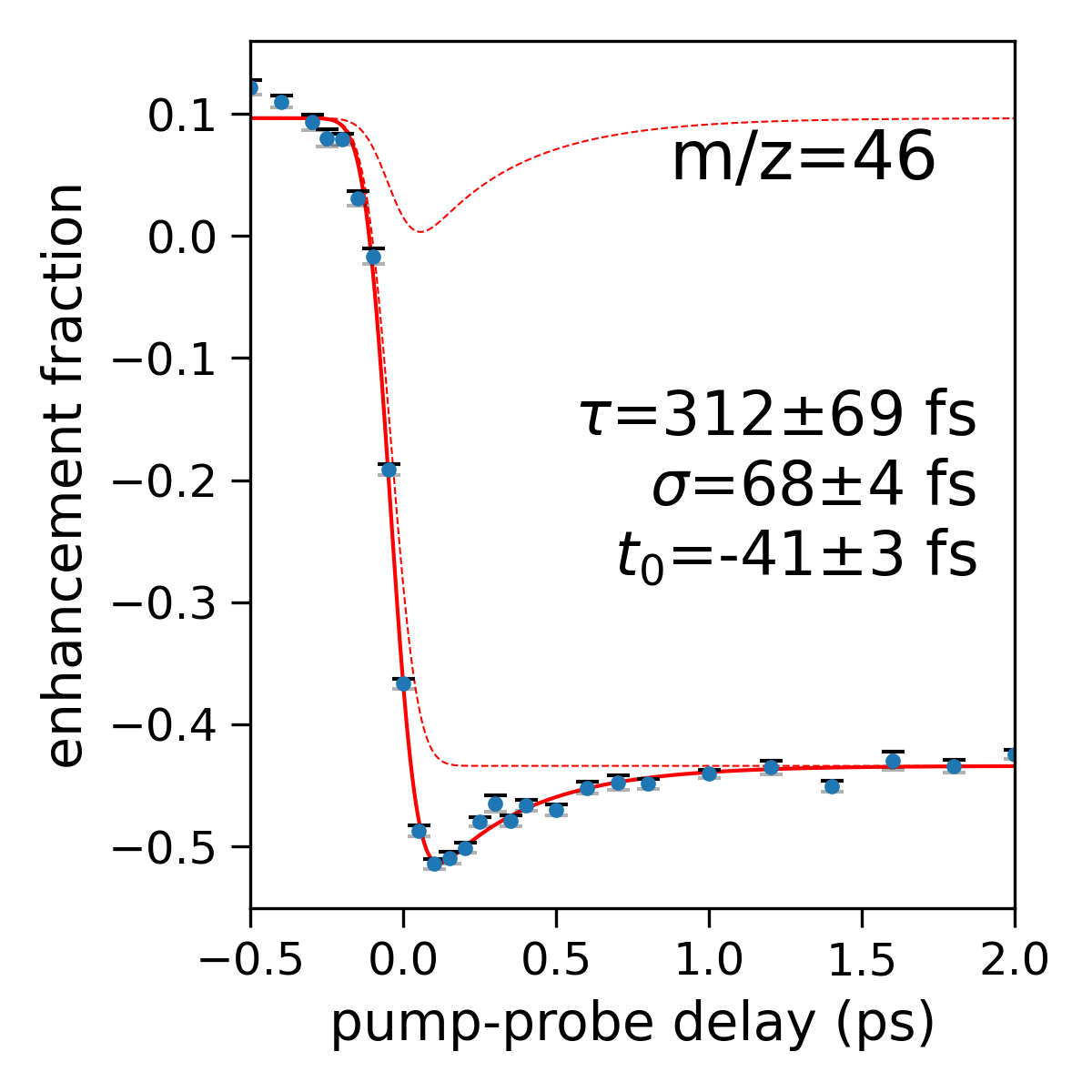}
        \caption[D0 H07 mq46]{\label{D0_H07_mq46} }
    \end{subfigure}%
		
    \caption[]{\label{fig:D0_H7} Pump-probe delay-dependent ion yields from the photoionization of non-deuterated ethanol, at an XUV photon energy $hv$=24.7\ eV. The data was fit by the model in Eq.~\eqref{eq:depletion_fit}.}
\end{figure}

\begin{figure}[t]
    \centering
    \begin{subfigure}[t]{0.3\linewidth}
        \centering
        \includegraphics[width=\linewidth*4/4]{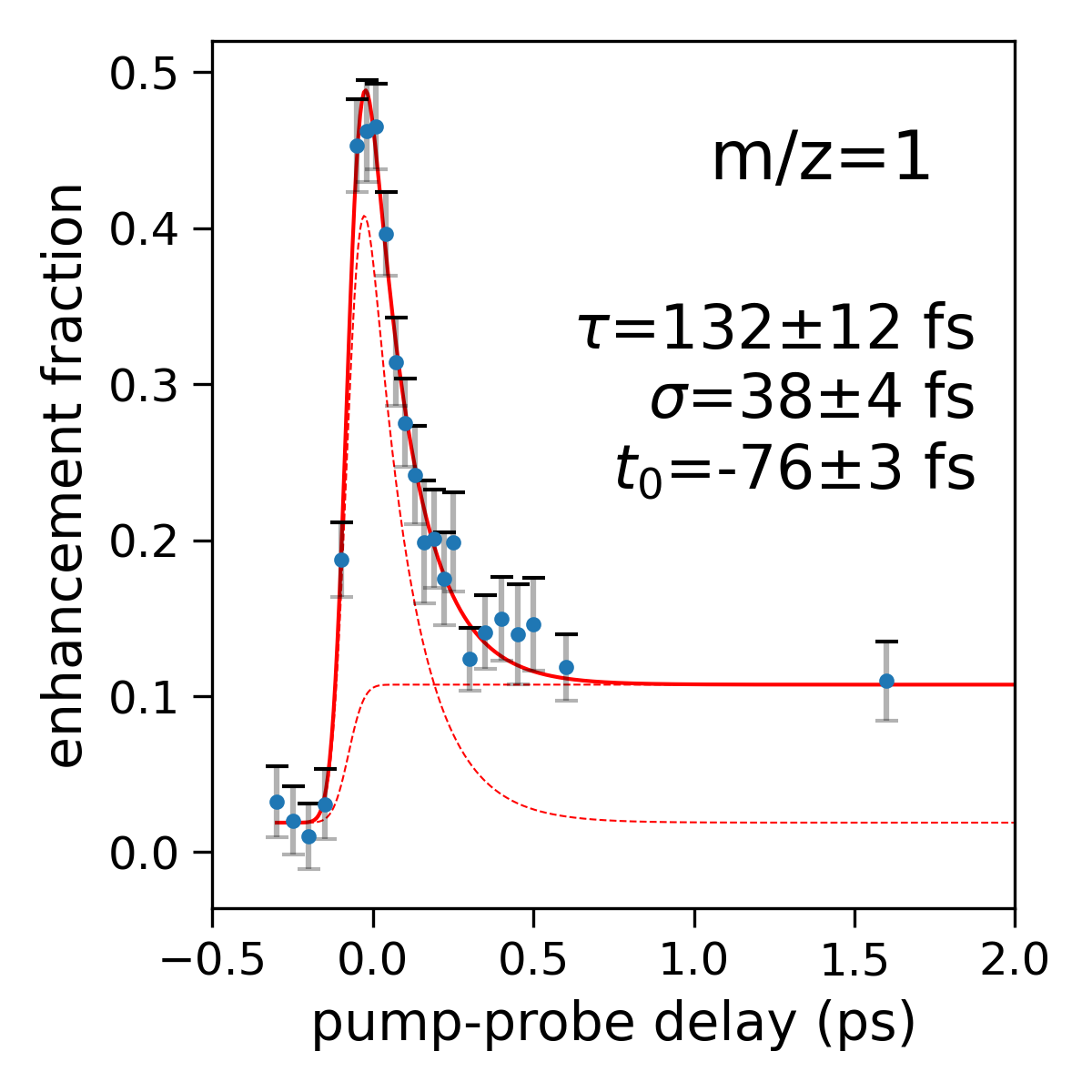}
        \caption[ami D0 H09 mq1]{\label{ami_D0_H09_mq1} }
    \end{subfigure}%
    ~
    \begin{subfigure}[t]{0.3\linewidth}
        \centering
        \includegraphics[width=\linewidth*4/4]{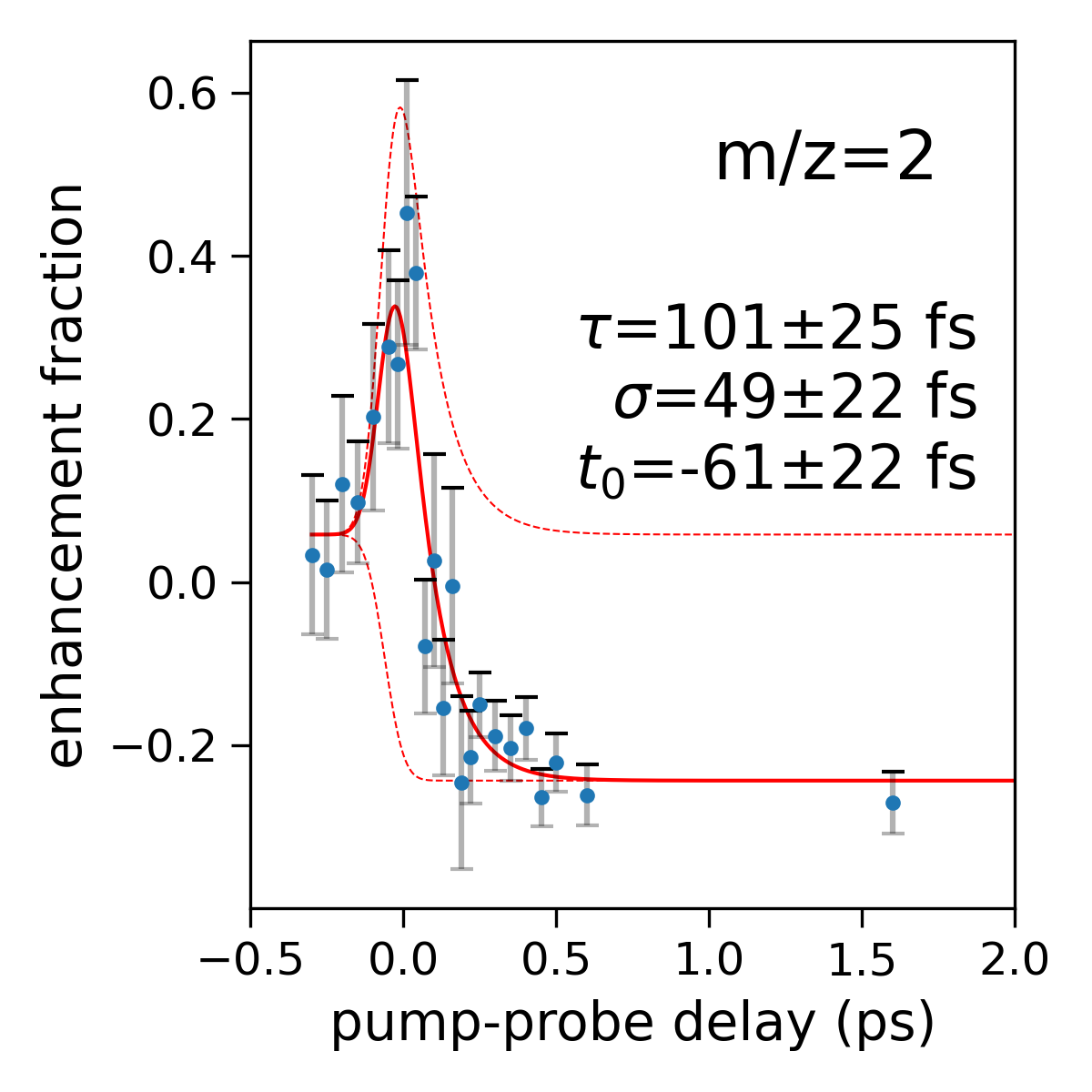}
        \caption[amin D0 H09 mq2]{\label{ami_D0_H09_mq2} }
    \end{subfigure}%
    ~
    \begin{subfigure}[t]{0.3\linewidth}
        \centering
        \includegraphics[width=\linewidth*4/4]{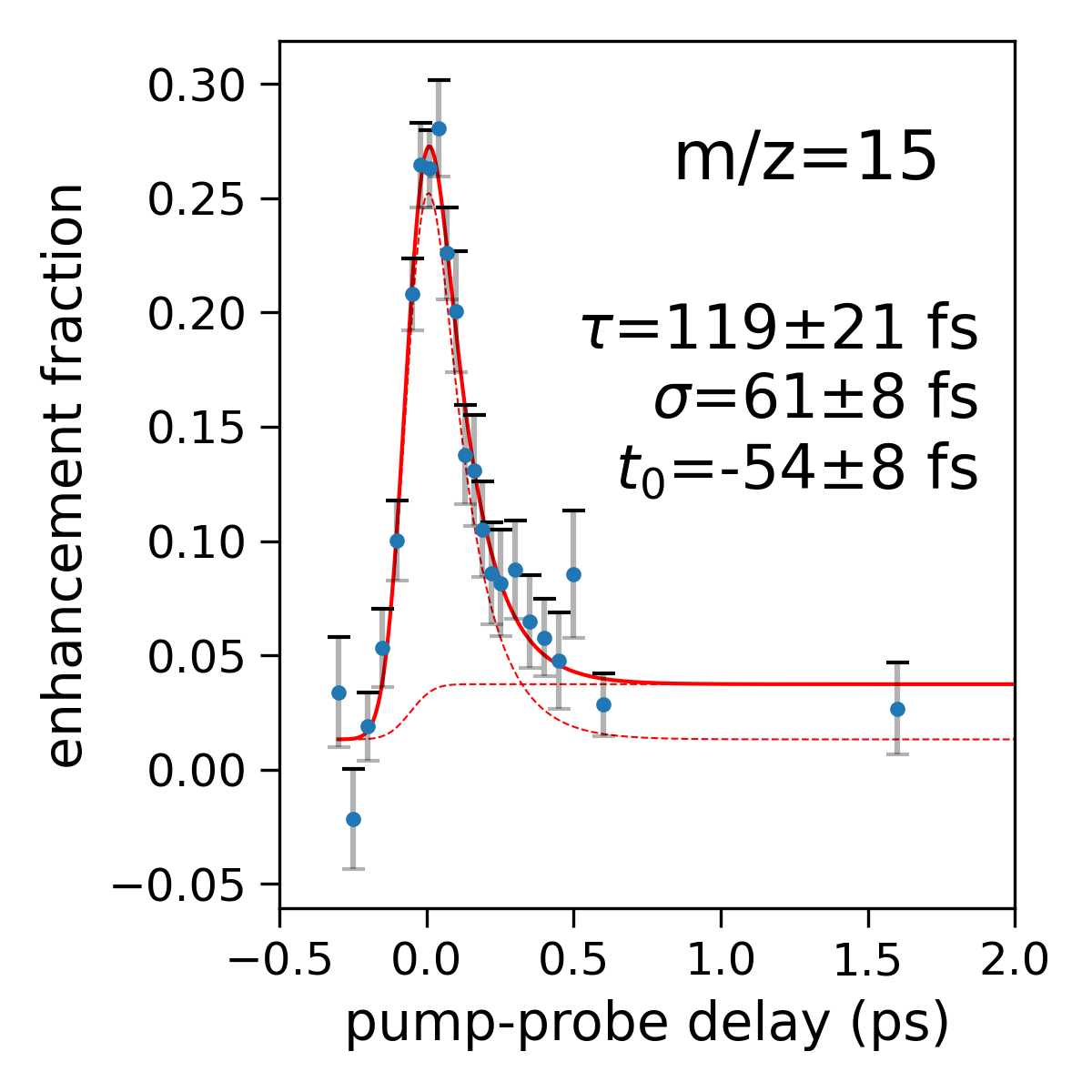}
        \caption[amin D0 H09 mq15]{\label{ami_D0_H09_mq15} }
    \end{subfigure}%
    
    \begin{subfigure}[t]{0.3\linewidth}
        \centering
        \includegraphics[width=\linewidth*4/4]{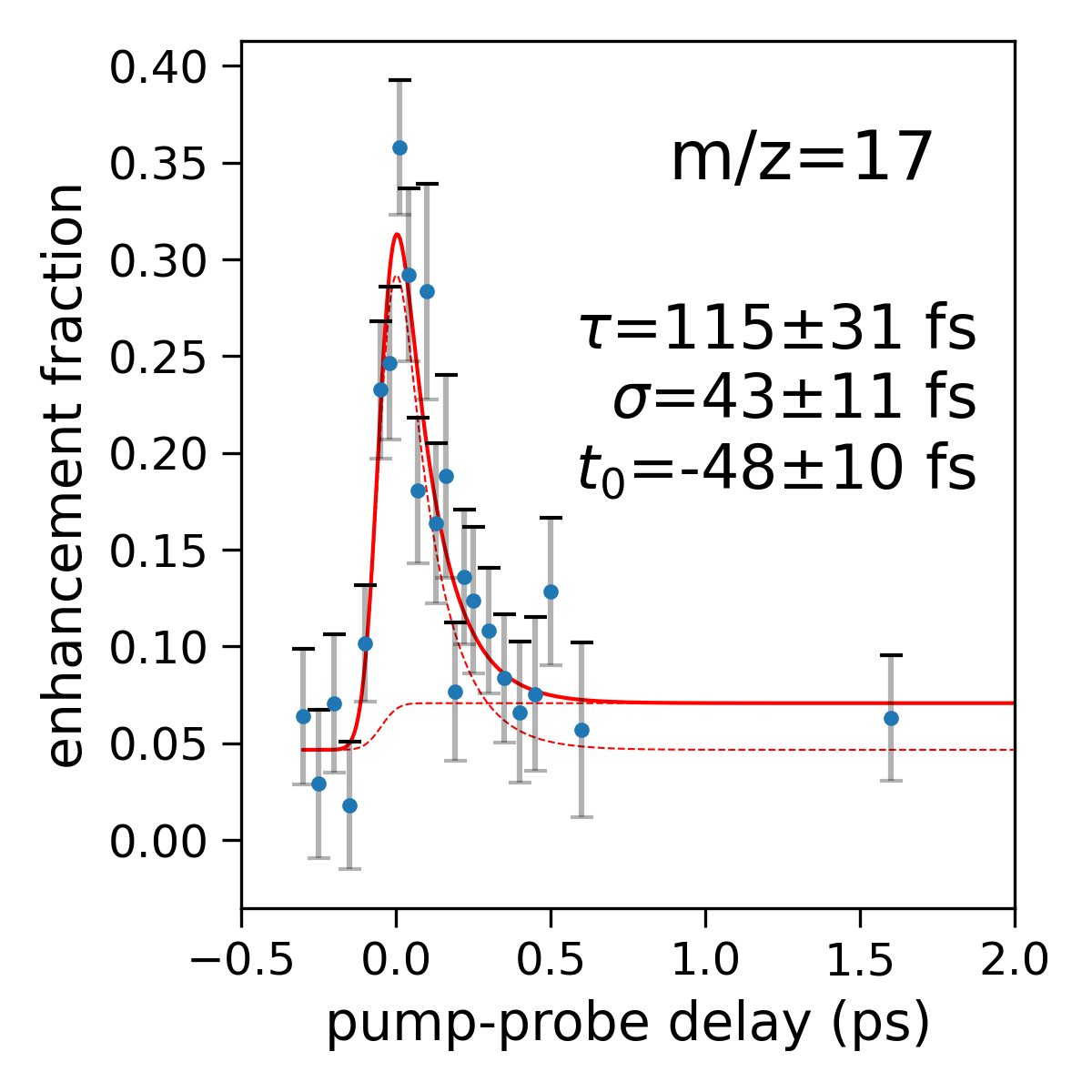}
        \caption[amin D0 H09 mq27]{\label{ami_D0_H09_mq17} }
    \end{subfigure}%
    ~
    \begin{subfigure}[t]{0.3\linewidth}
        \centering
        \includegraphics[width=\linewidth*4/4]{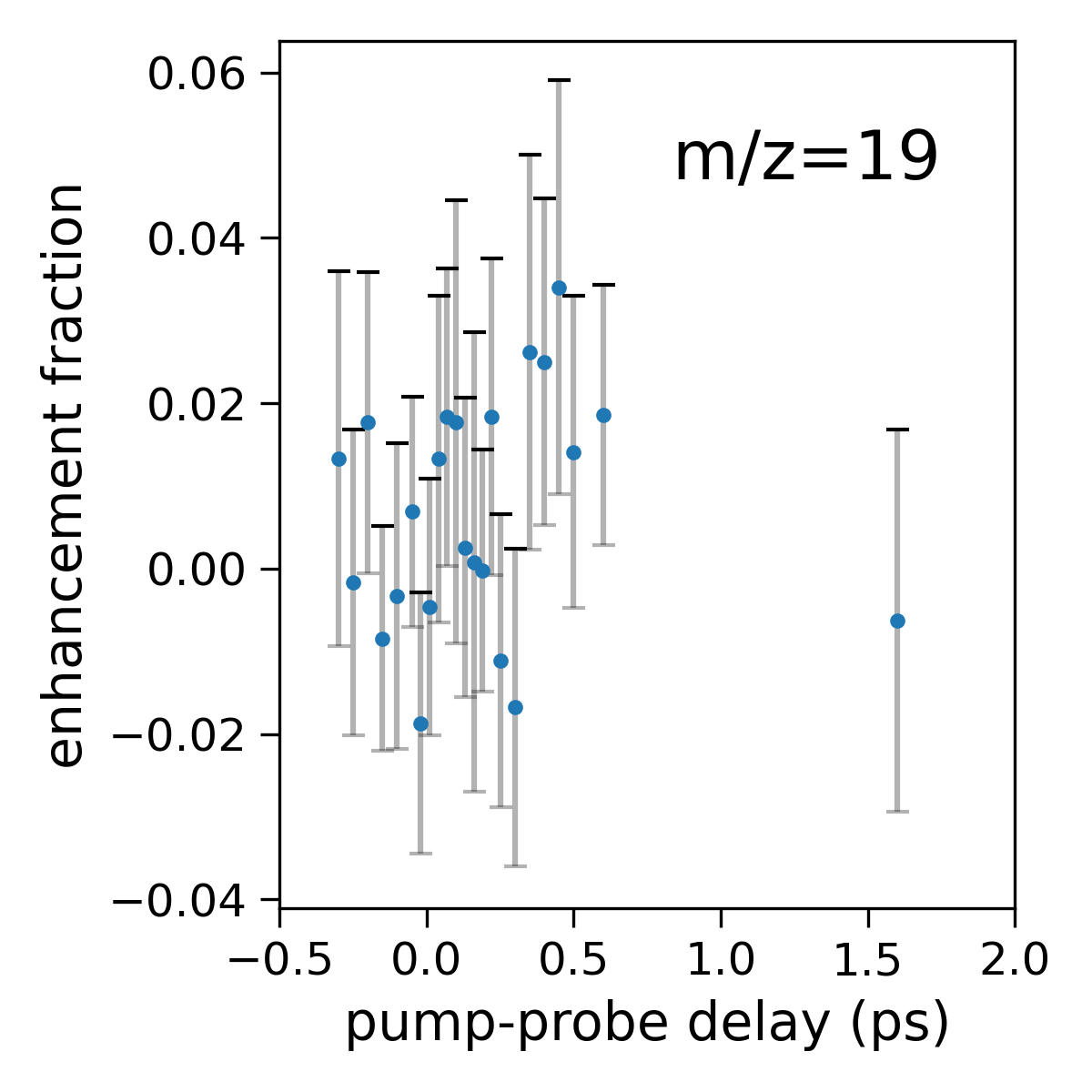}
        \caption[amin D0 H09 mq29]{\label{ami_D0_H09_mq19} }
    \end{subfigure}%
    ~
    \begin{subfigure}[t]{0.3\linewidth}
        \centering
        \includegraphics[width=\linewidth*4/4]{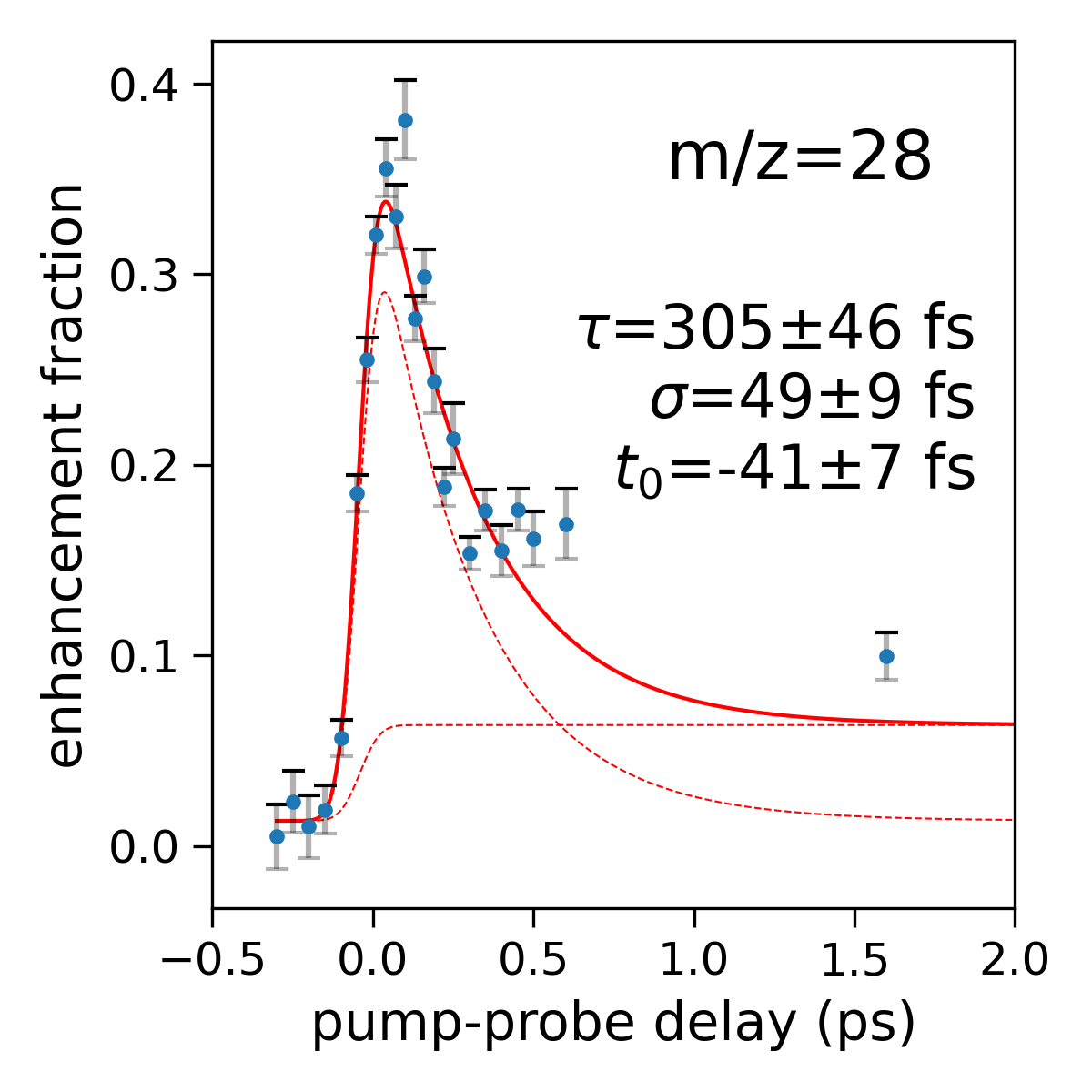}
        \caption[amin D0 H09 mq31]{\label{ami_D0_H09_mq28} }
    \end{subfigure}%
    
    \begin{subfigure}[t]{0.3\linewidth}
        \centering
        \includegraphics[width=\linewidth*4/4]{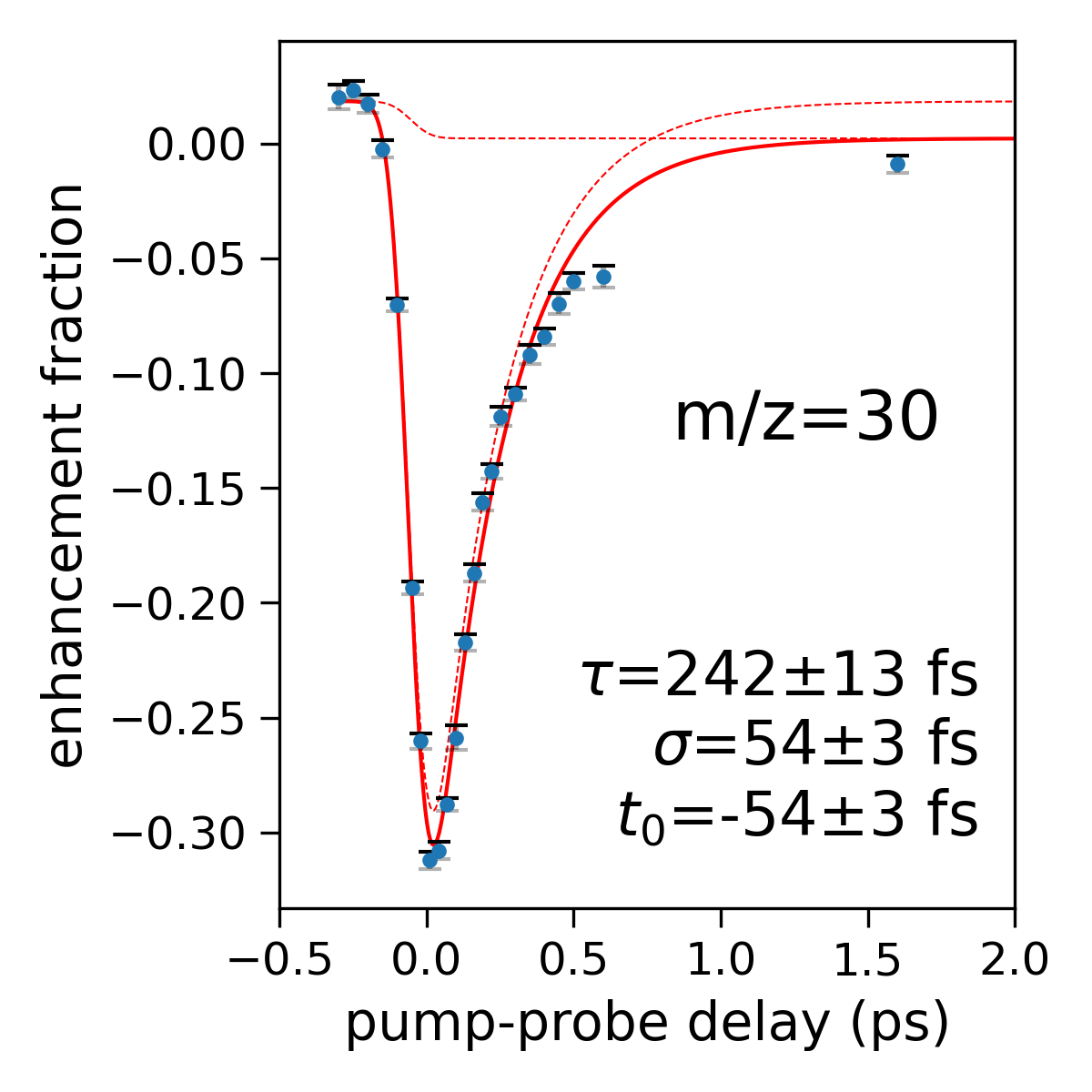}
        \caption[amin D0 H09 mq43]{\label{ami_D0_H09_mq30} }
    \end{subfigure}%
    ~
    \begin{subfigure}[t]{0.3\linewidth}
        \centering
        \includegraphics[width=\linewidth*4/4]{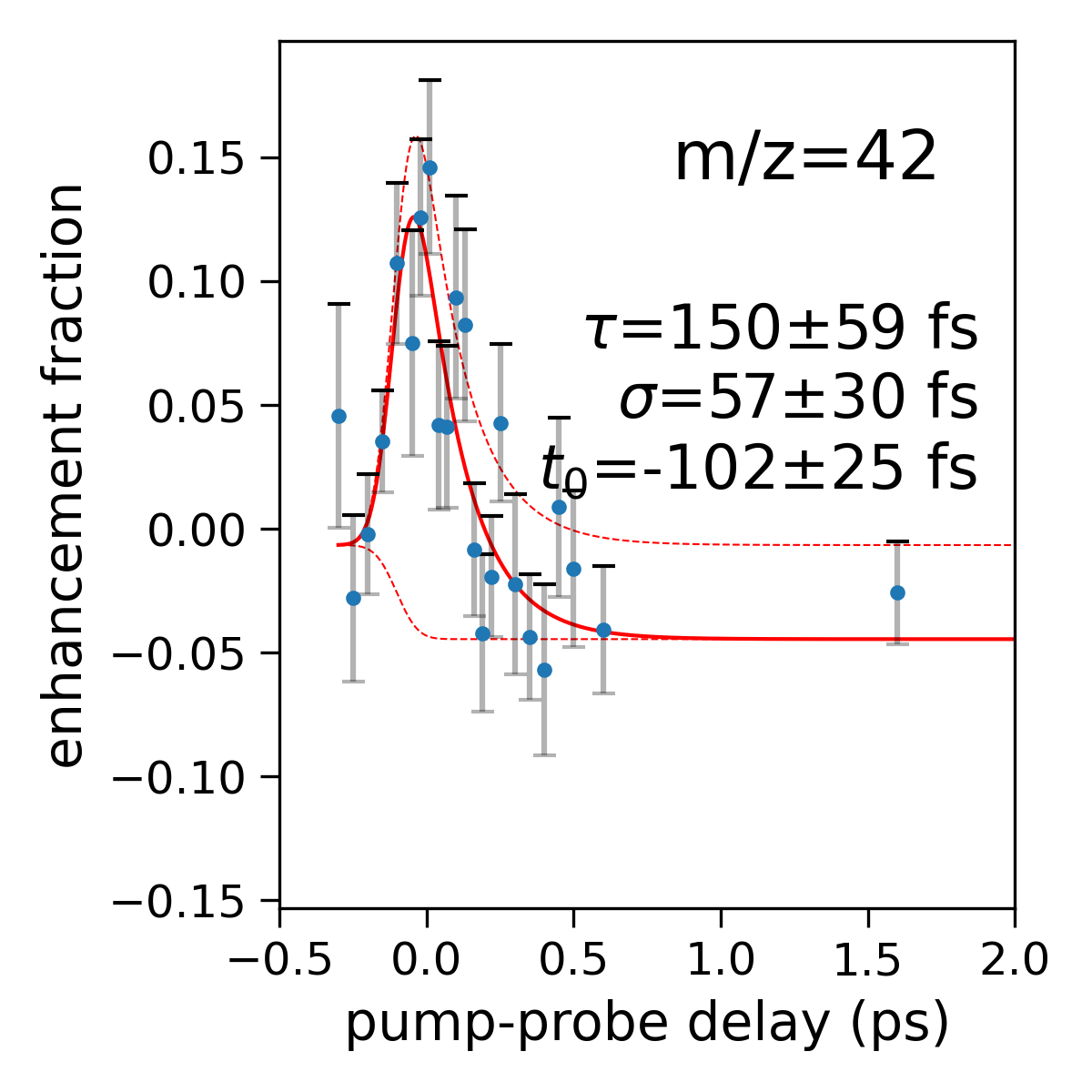}
        \caption[amin D0 H09 mq45]{\label{ami_D0_H09_mq42} }
    \end{subfigure}%
    ~
    \begin{subfigure}[t]{0.3\linewidth}
        \centering
        \includegraphics[width=\linewidth*4/4]{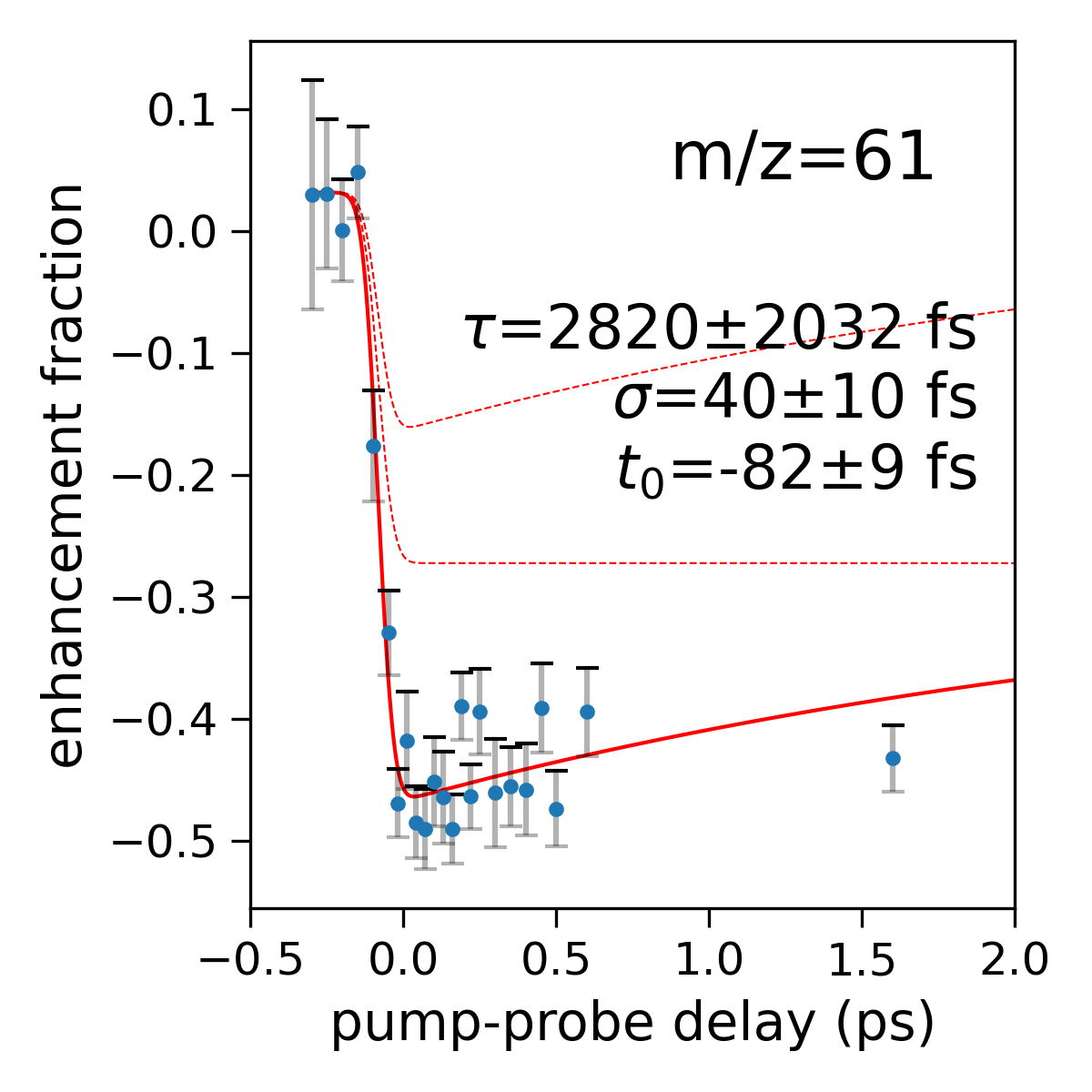}
        \caption[amin D0 H09 mq46]{\label{ami_D0_H09_mq61} }
    \end{subfigure}%
    
    \caption[]{\label{fig:ami_D0_H9} Pump-probe delay-dependent ion yields from the photoionization of non-deuterated 2-aminoethanol, at an XUV photon energy $hv$=31.7\ eV. The data was fit by the model in Eq.~\eqref{eq:depletion_fit}.}
\end{figure}

\begin{figure}[t]
    \centering
    \begin{subfigure}[t]{0.3\linewidth}
        \centering
        \includegraphics[width=\linewidth*4/4]{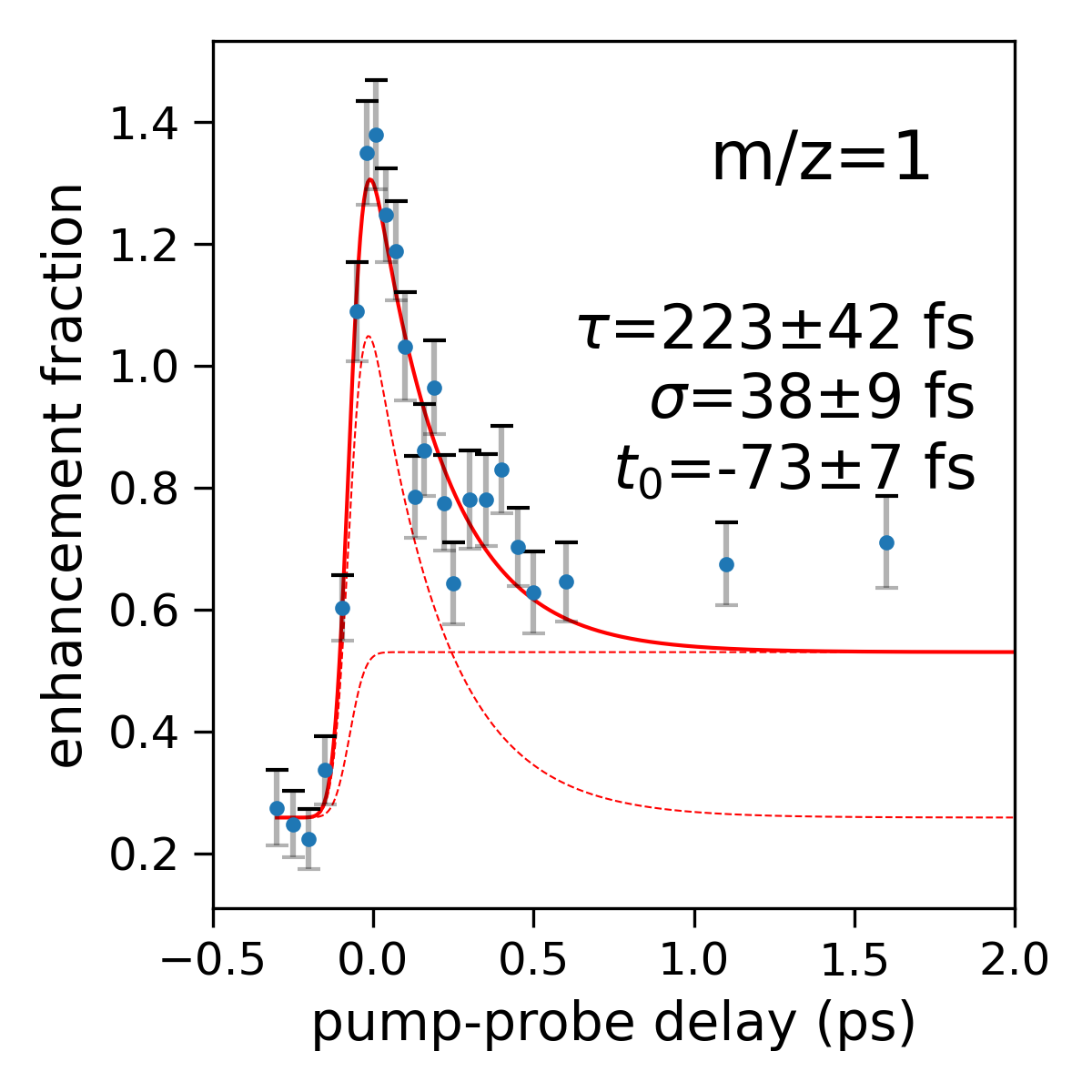}
        \caption[ami D0 H06 mq1]{\label{ami_D0_H06_mq1} }
    \end{subfigure}%
    ~
    \begin{subfigure}[t]{0.3\linewidth}
        \centering
        \includegraphics[width=\linewidth*4/4]{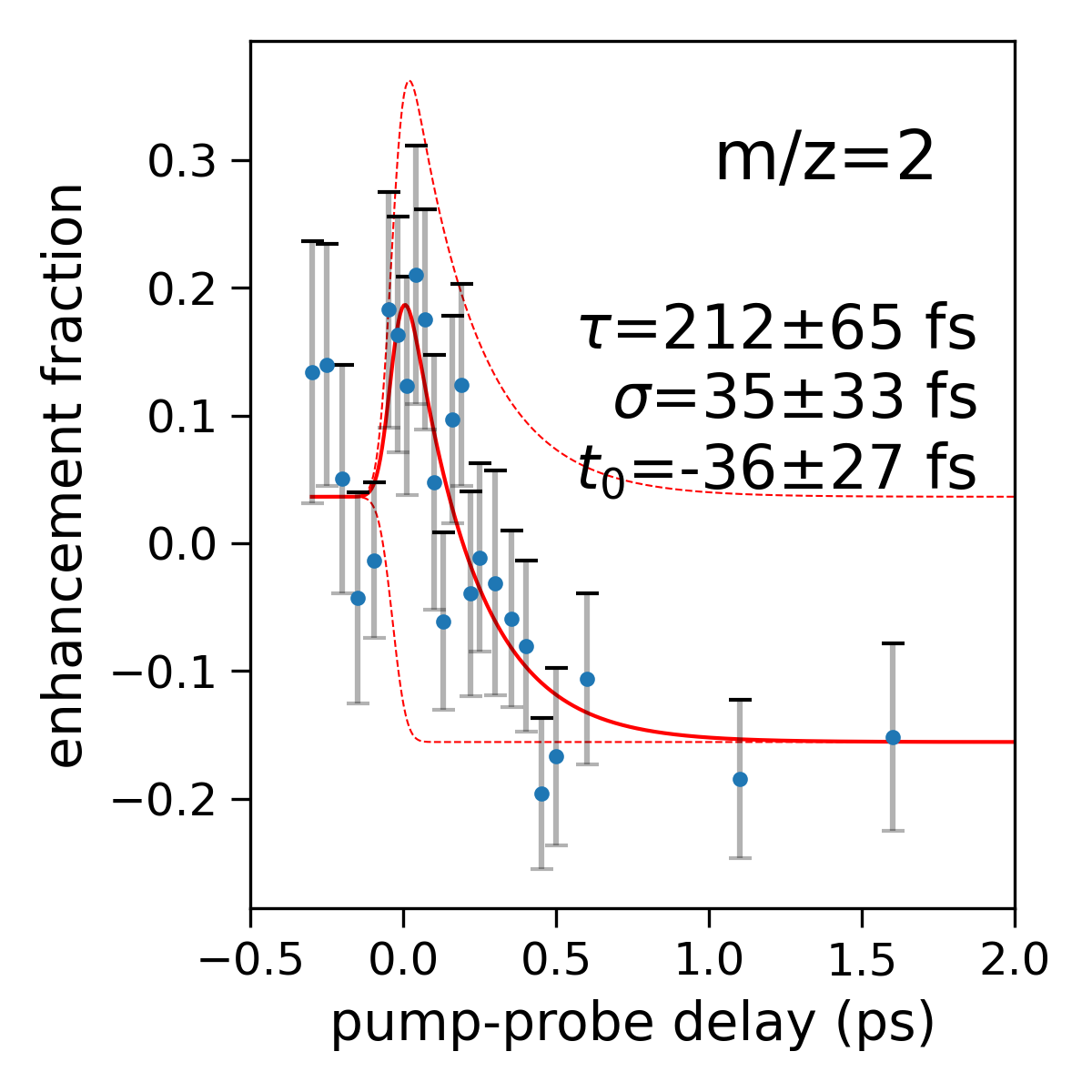}
        \caption[amin D0 H06 mq2]{\label{ami_D0_H06_mq2} }
    \end{subfigure}%
    ~
    \begin{subfigure}[t]{0.3\linewidth}
        \centering
        \includegraphics[width=\linewidth*4/4]{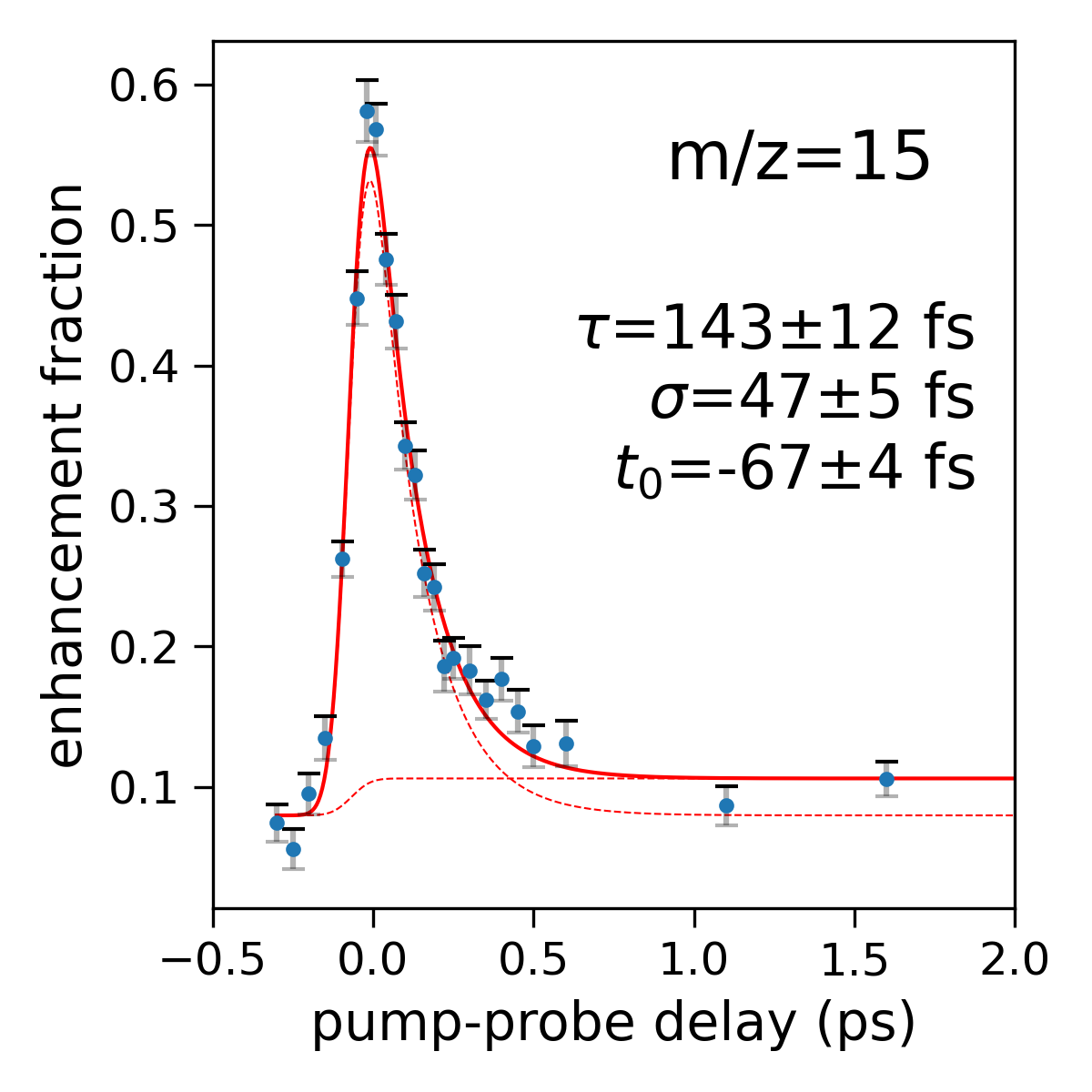}
        \caption[amin D0 H06 mq15]{\label{ami_D0_H06_mq15} }
    \end{subfigure}%
    
    \begin{subfigure}[t]{0.3\linewidth}
        \centering
        \includegraphics[width=\linewidth*4/4]{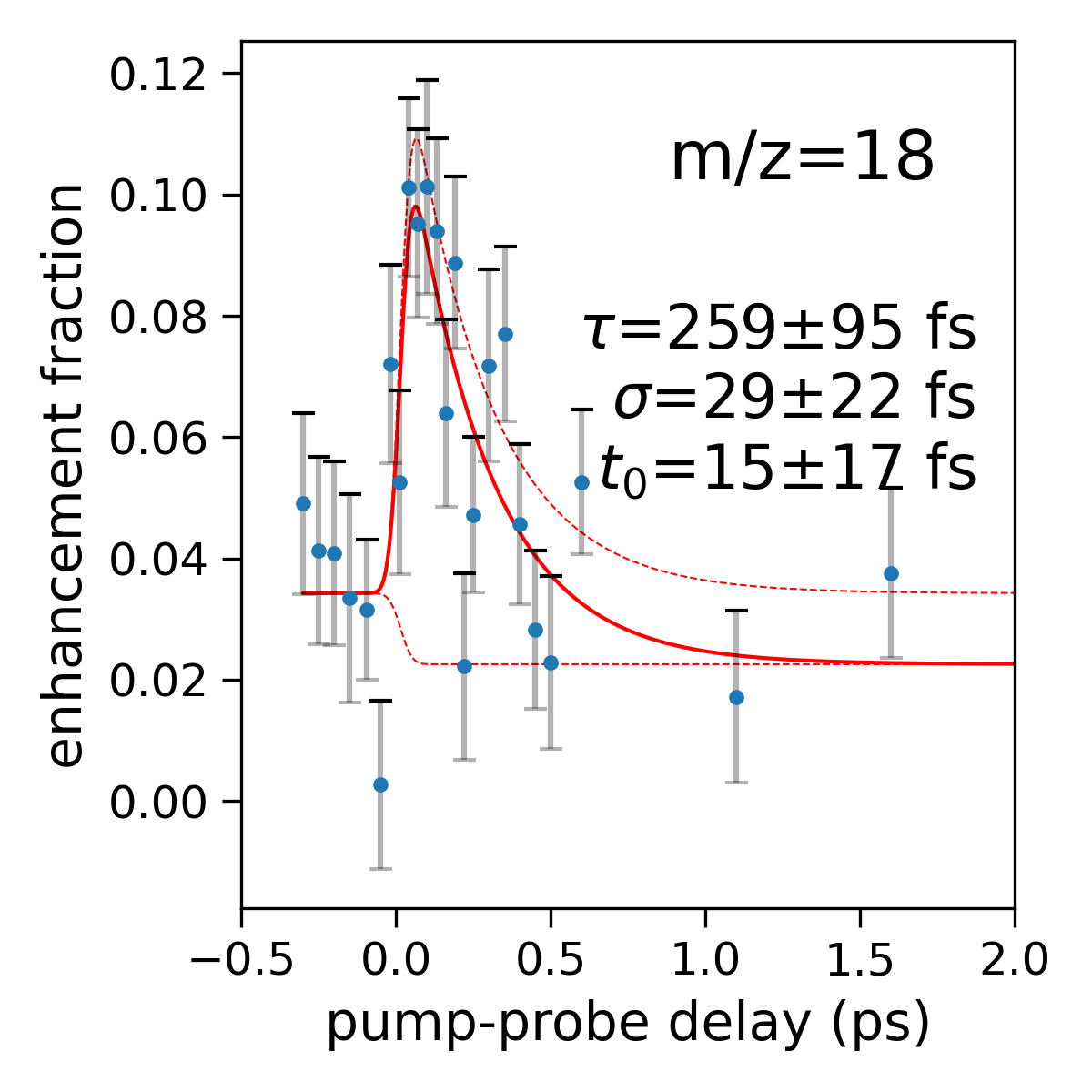}
        \caption[amin D0 H06 mq27]{\label{ami_D0_H06_mq18} }
    \end{subfigure}%
    ~
    \begin{subfigure}[t]{0.3\linewidth}
        \centering
        \includegraphics[width=\linewidth*4/4]{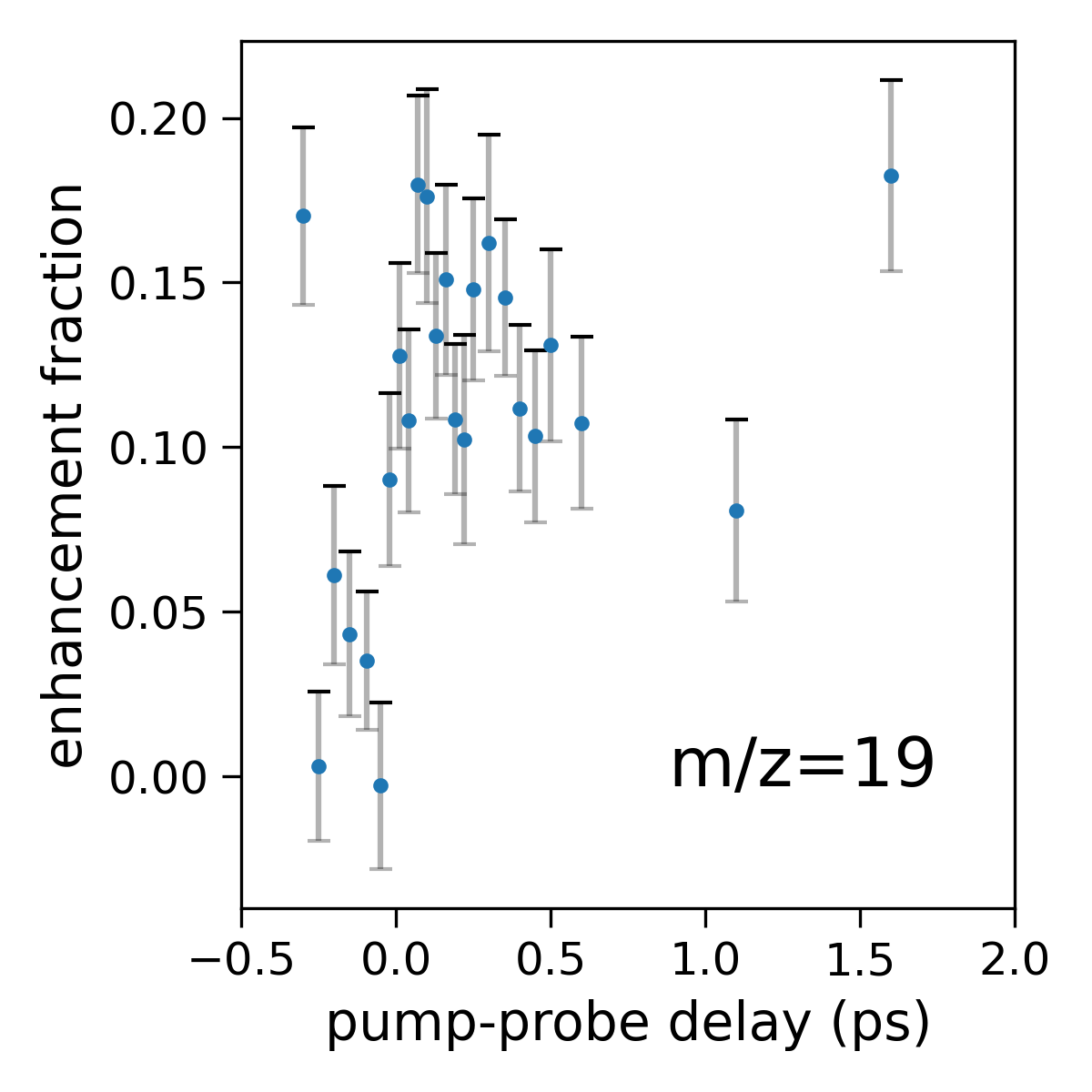}
        \caption[amin D0 H06 mq29]{\label{ami_D0_H06_mq19} }
    \end{subfigure}%
    ~
    \begin{subfigure}[t]{0.3\linewidth}
        \centering
        \includegraphics[width=\linewidth*4/4]{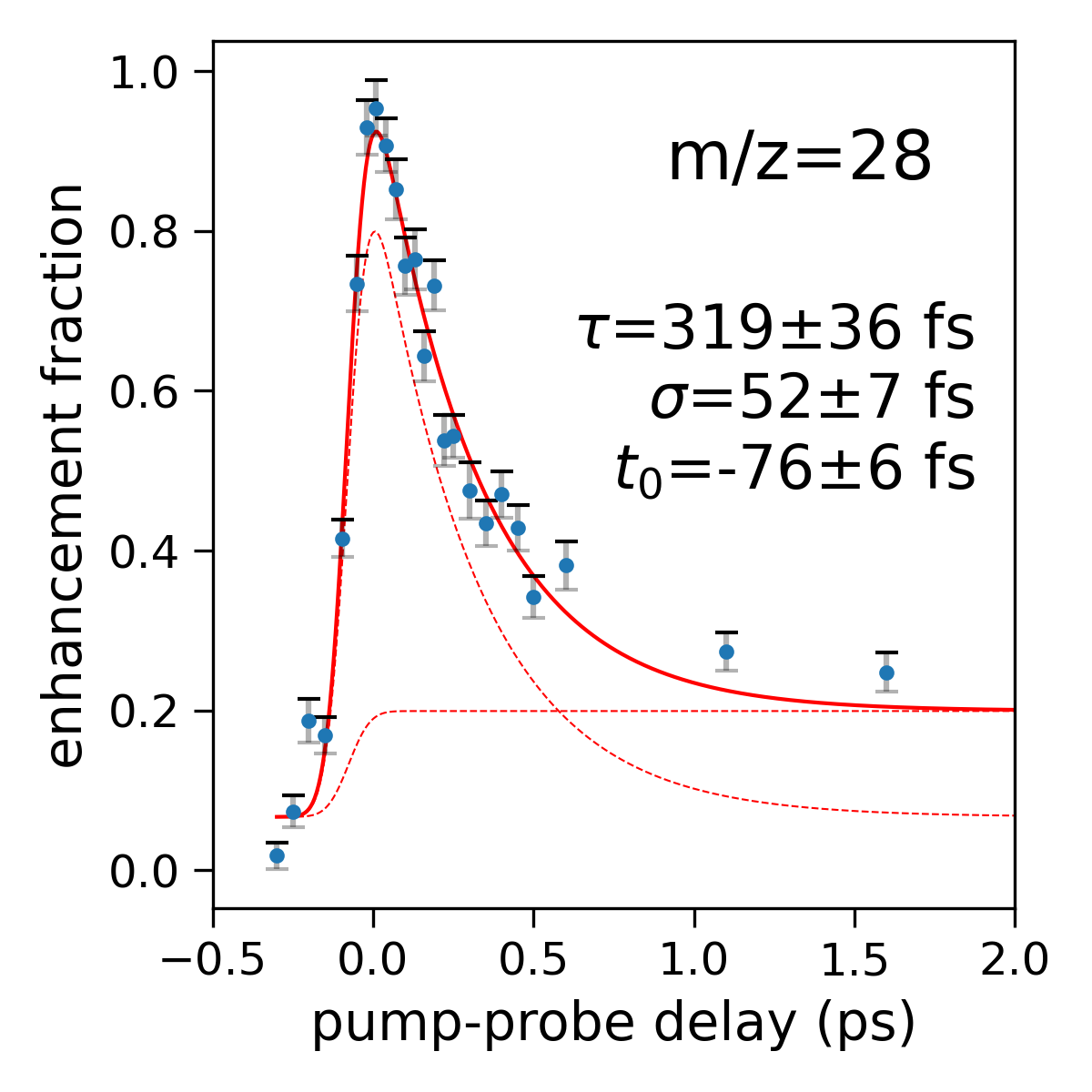}
        \caption[amin D0 H06 mq31]{\label{ami_D0_H06_mq28} }
    \end{subfigure}%
    
    \begin{subfigure}[t]{0.3\linewidth}
        \centering
        \includegraphics[width=\linewidth*4/4]{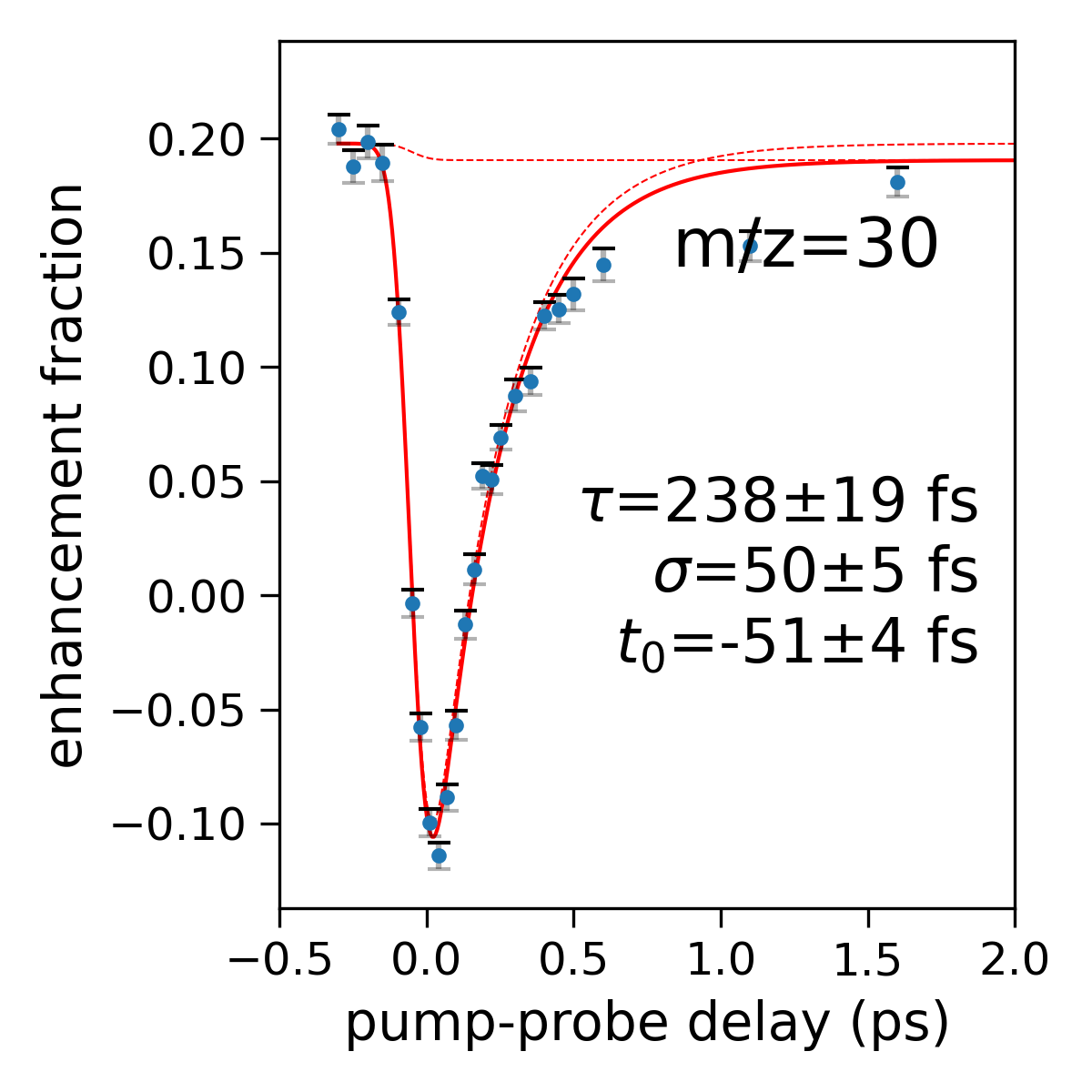}
        \caption[amin D0 H06 mq43]{\label{ami_D0_H06_mq30} }
    \end{subfigure}%
    ~
    \begin{subfigure}[t]{0.3\linewidth}
        \centering
        \includegraphics[width=\linewidth*4/4]{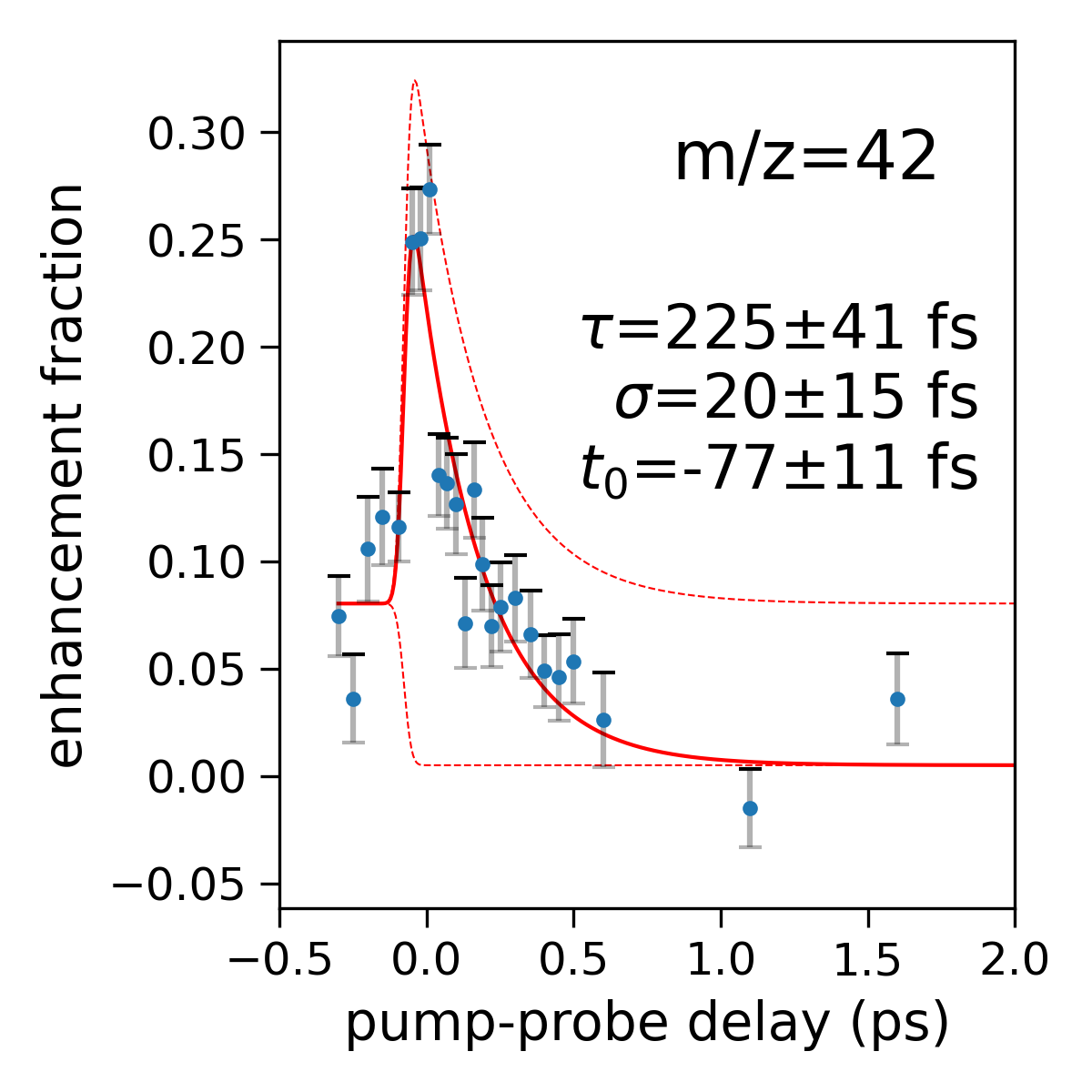}
        \caption[amin D0 H06 mq45]{\label{ami_D0_H06_mq42} }
    \end{subfigure}%
    ~
    \begin{subfigure}[t]{0.3\linewidth}
        \centering
        \includegraphics[width=\linewidth*4/4]{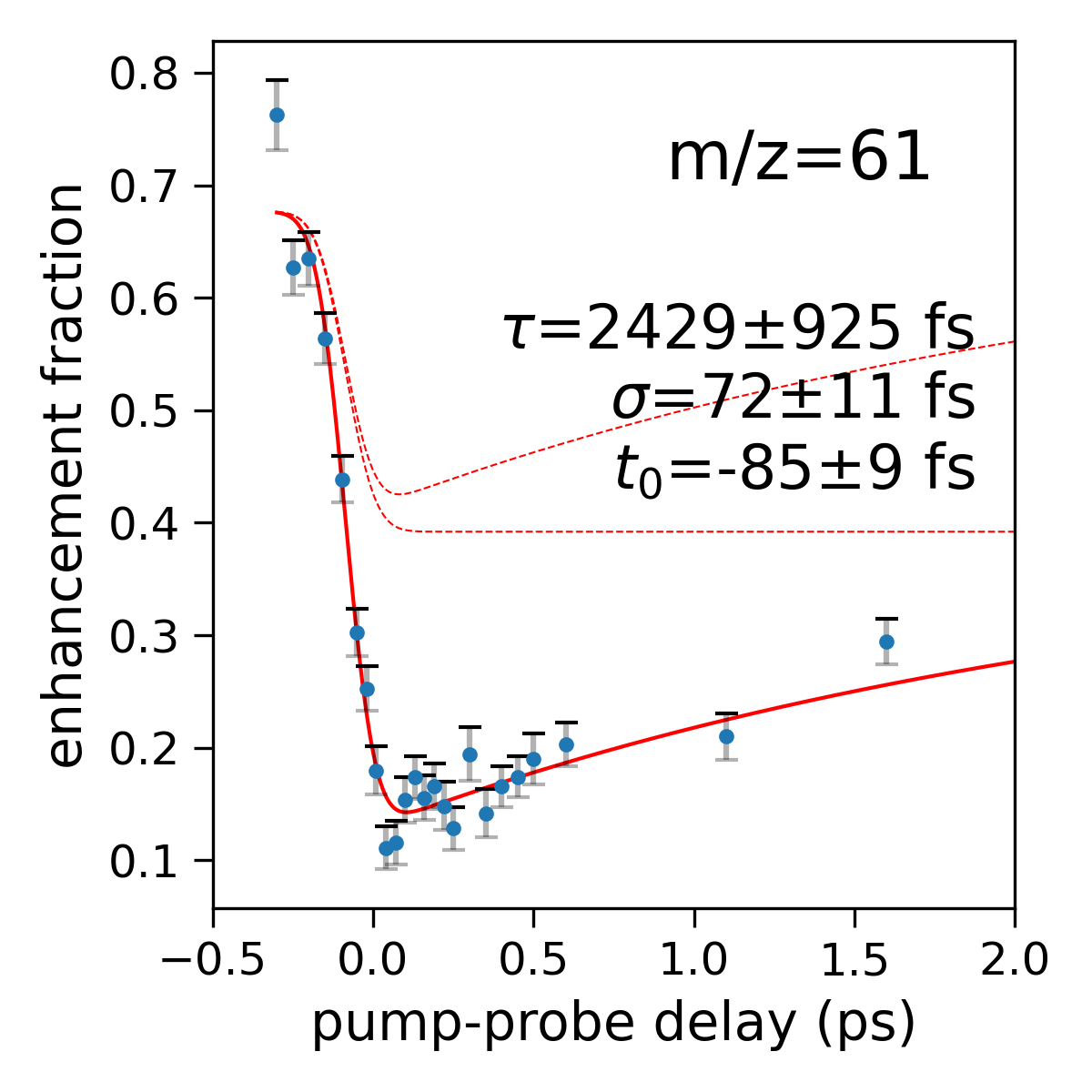}
        \caption[amin D0 H06 mq46]{\label{ami_D0_H06_mq61} }
    \end{subfigure}%
		
    \caption[]{\label{fig:ami_D0_H6} Pump-probe delay-dependent ion yields from the photoionization of non-deuterated 2-aminoethanol, at an XUV photon energy $hv$=21.1\ eV. The data was fit by the model in Eq.~\eqref{eq:depletion_fit}.}
\end{figure}

For non-deuterated ethanol, we show a variety of mass fragments at XUV photon energies of 31.7\, eV (Supplementary Figure~\ref{fig:D0_H9}) and 24.7\, eV (Supplementary Figure~\ref{fig:D0_H7}).
For fully-deuterated ethanol, mass fragments obtained at an XUV photon energy of 31.7\, eV are shown in Supplementary Figure~\ref{fig:D6_H9}.
For 2-aminoethanol, we have obtained fragment mass at XUV photon energies of 31.7\,eV (Supplementary Figure~\ref{fig:ami_D0_H9}) and 21.1\,eV (Supplementary Figure~\ref{fig:ami_D0_H6}).

\begin{figure}[t]
    \centering
    \begin{subfigure}[t]{0.3\linewidth}
        \centering
        \includegraphics[width=\linewidth*4/4]{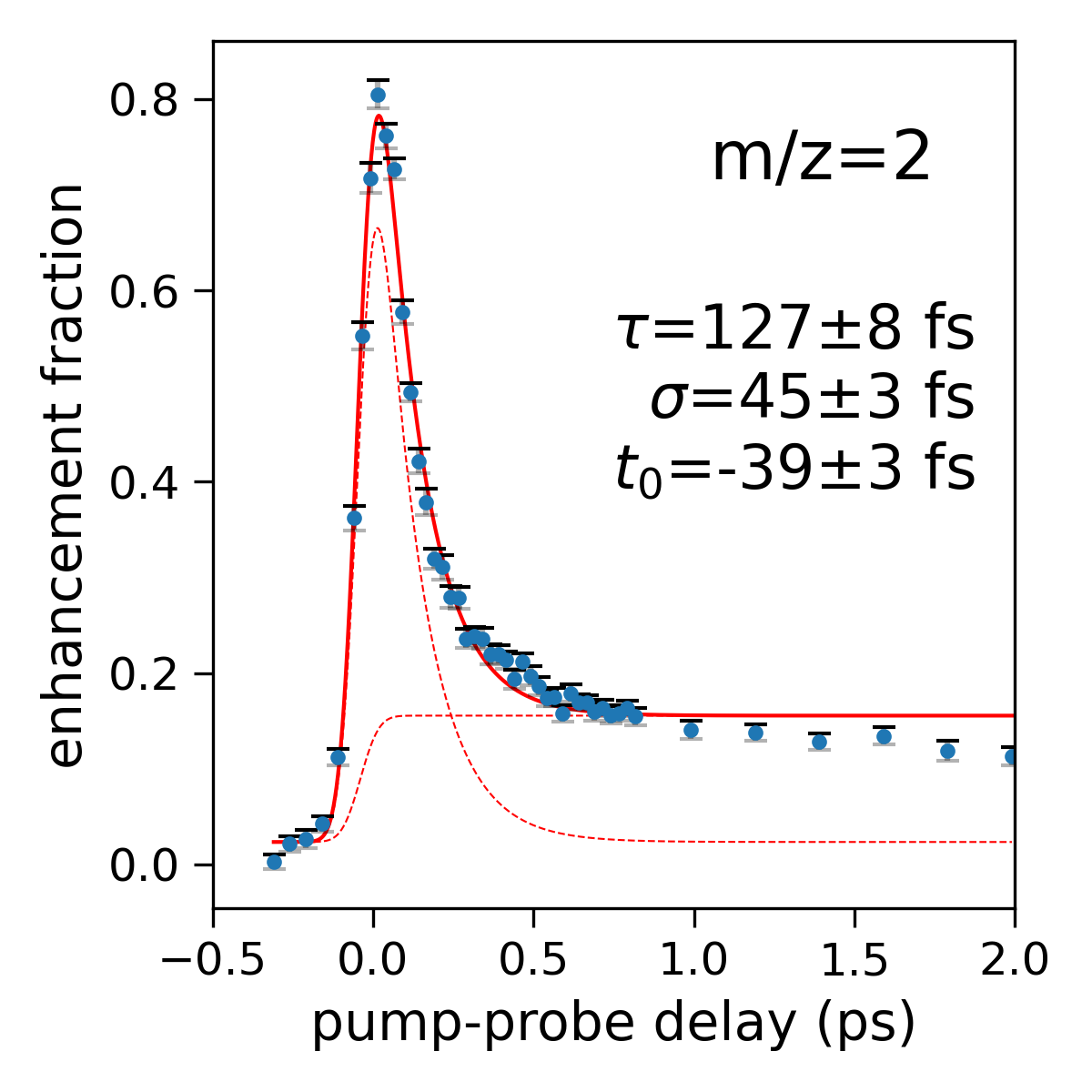}
        \caption[D6 H09 mq2n]{\label{D6_H09_mq2n} }
    \end{subfigure}%
    ~
    \begin{subfigure}[t]{0.3\linewidth}
        \centering
        \includegraphics[width=\linewidth*4/4]{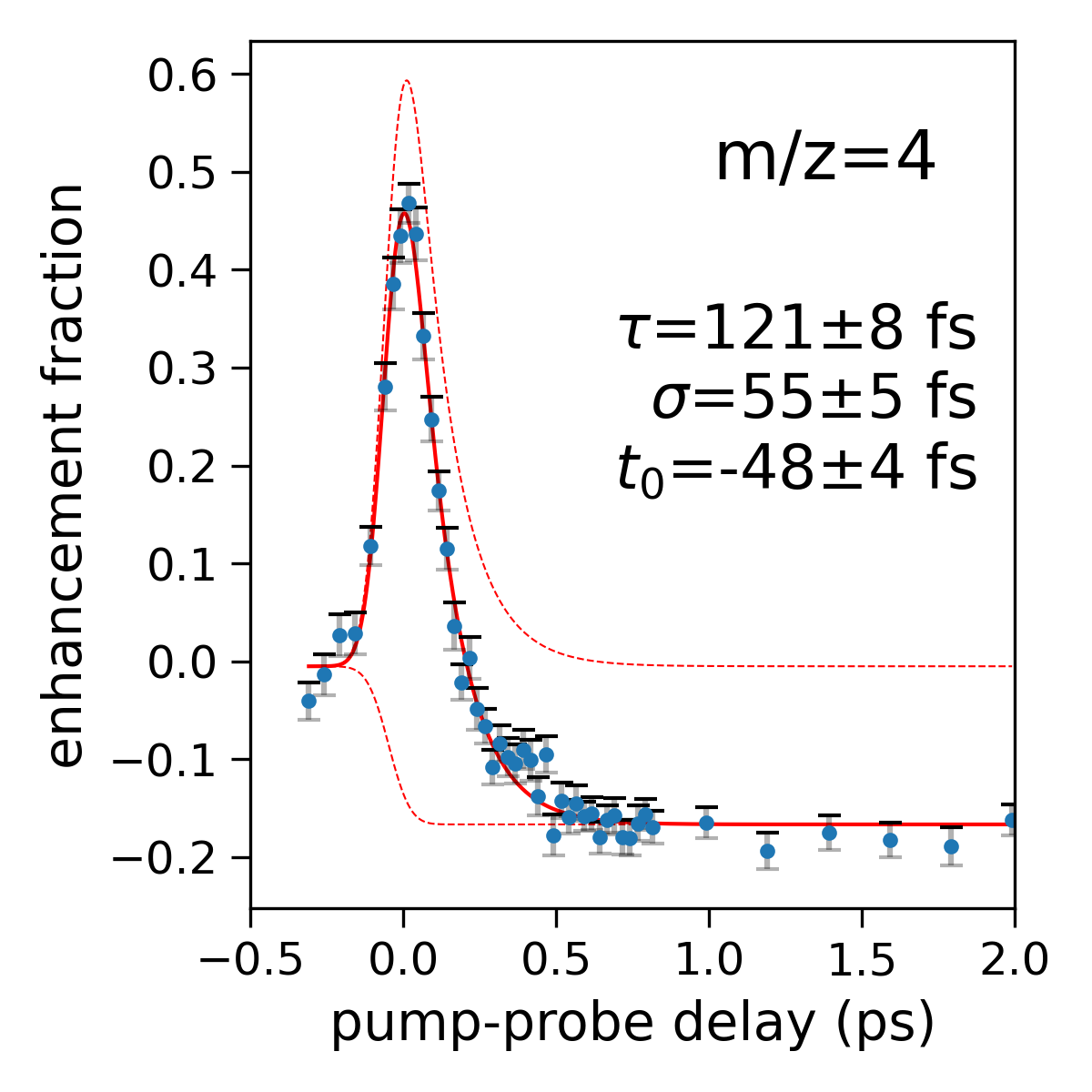}
        \caption[D6 H09 mq2]{\label{D6_H09_mq2} }
    \end{subfigure}%
    ~
    \begin{subfigure}[t]{0.3\linewidth}
        \centering
        \includegraphics[width=\linewidth*4/4]{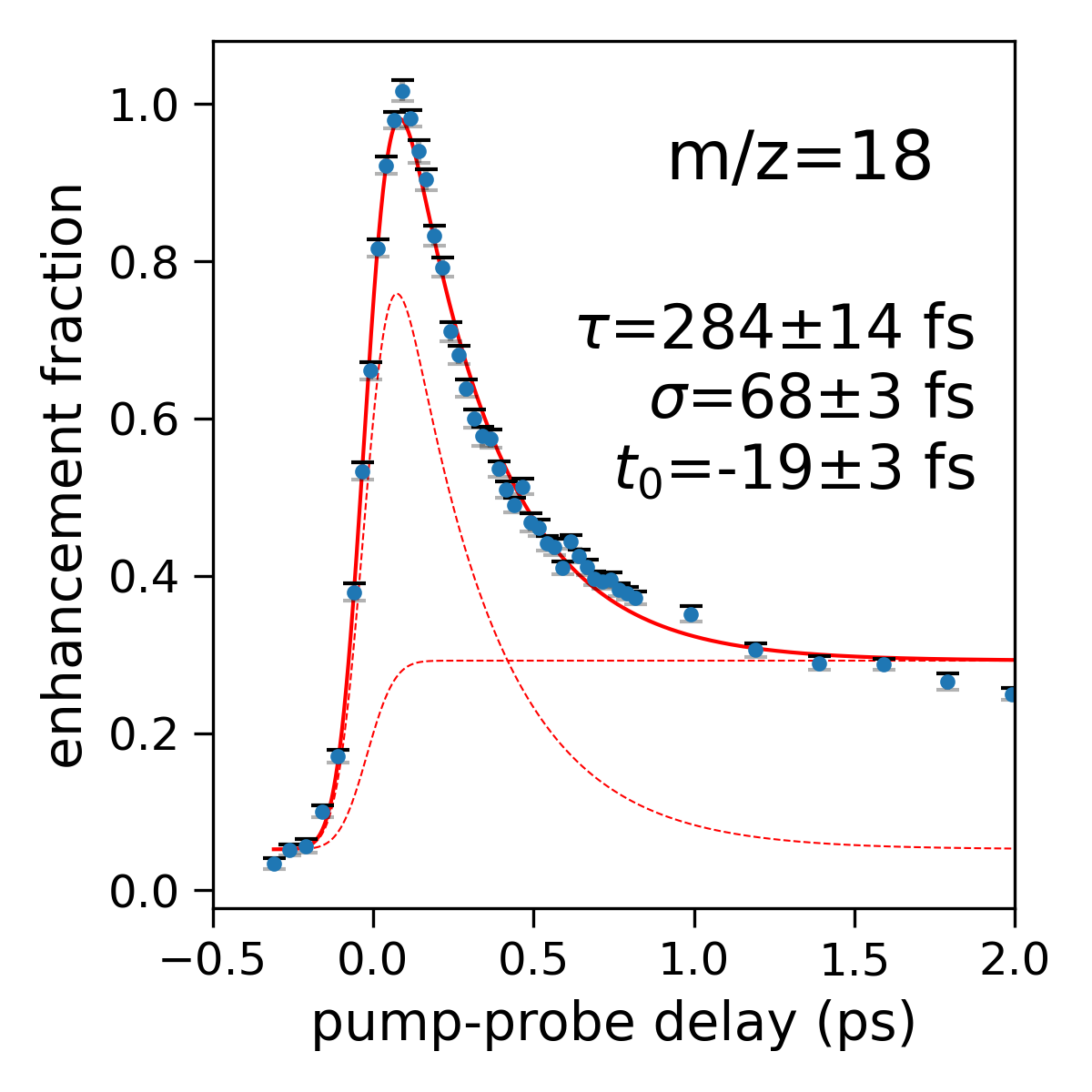}
        \caption[D6 H09 mq4]{\label{D6_H09_mq4} }
    \end{subfigure}%
    
    \begin{subfigure}[t]{0.3\linewidth}
        \centering
        \includegraphics[width=\linewidth*4/4]{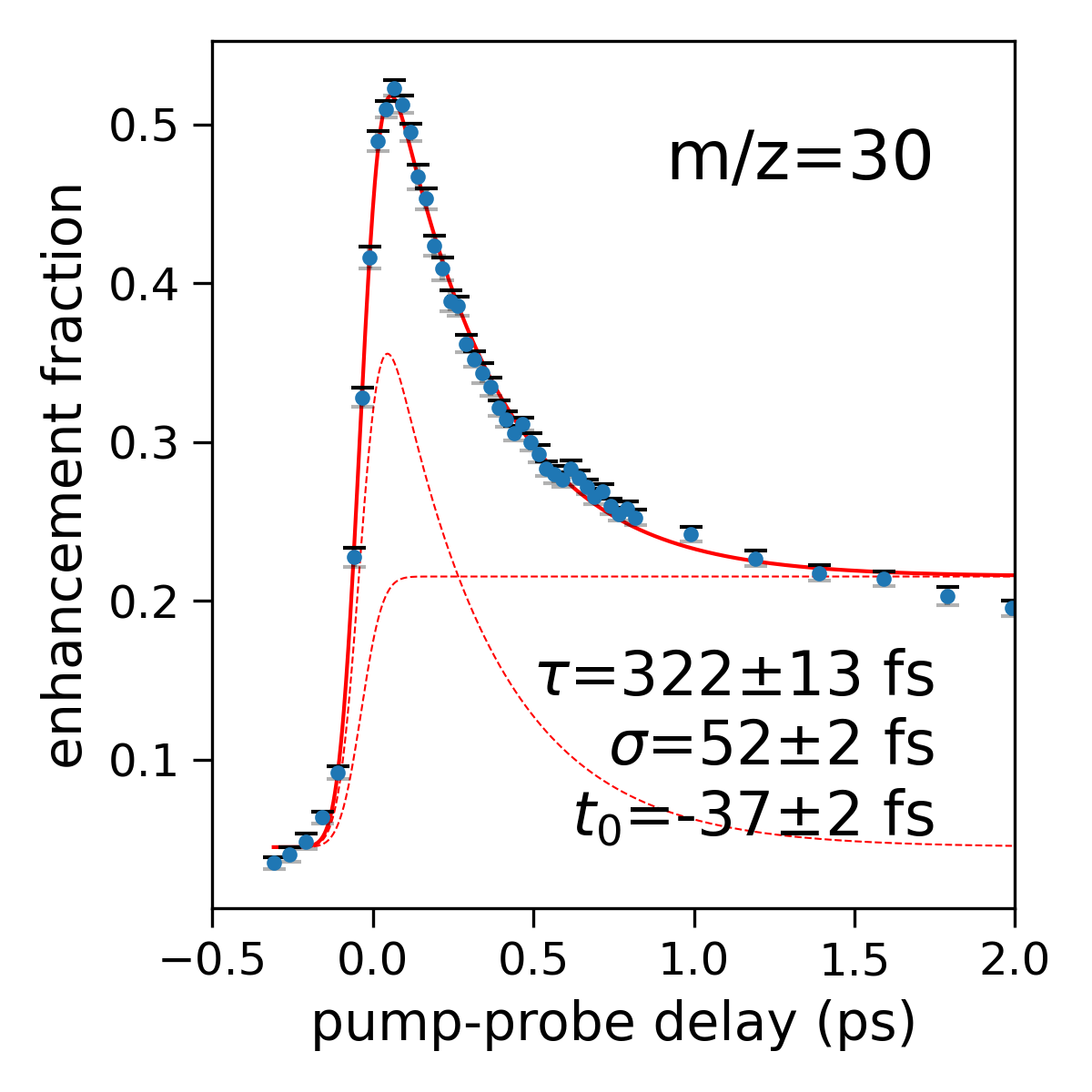}
        \caption[D6 H09 mq30]{\label{D6_H09_mq30} }
    \end{subfigure}%
    ~
    \begin{subfigure}[t]{0.3\linewidth}
    \end{subfigure}%
    ~
    \begin{subfigure}[t]{0.3\linewidth}
        \centering
        \includegraphics[width=\linewidth*4/4]{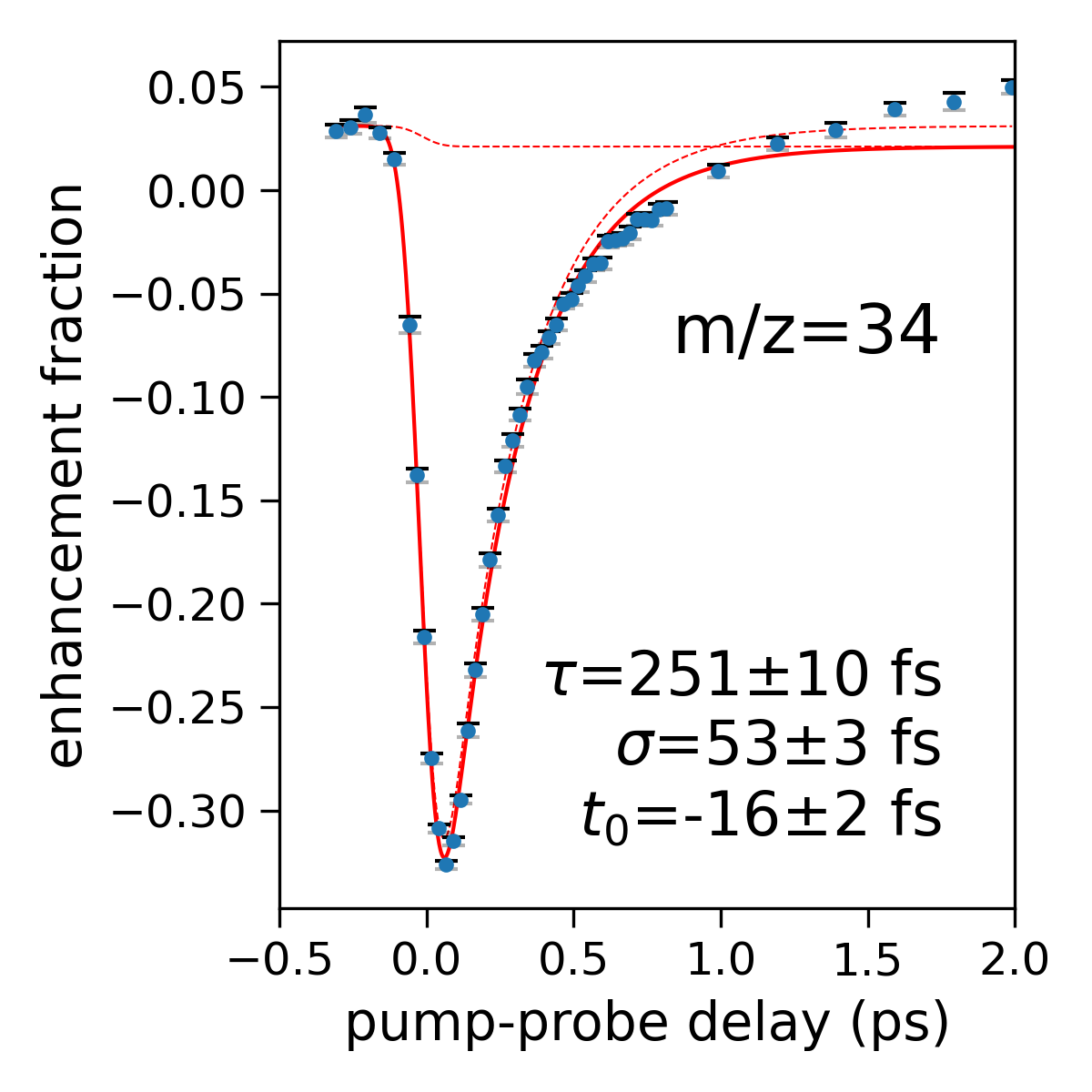}
        \caption[D6 H09 mq34]{\label{D6_H09_mq34} }
    \end{subfigure}%
    
    \begin{subfigure}[t]{0.3\linewidth}
        \centering
        \includegraphics[width=\linewidth*4/4]{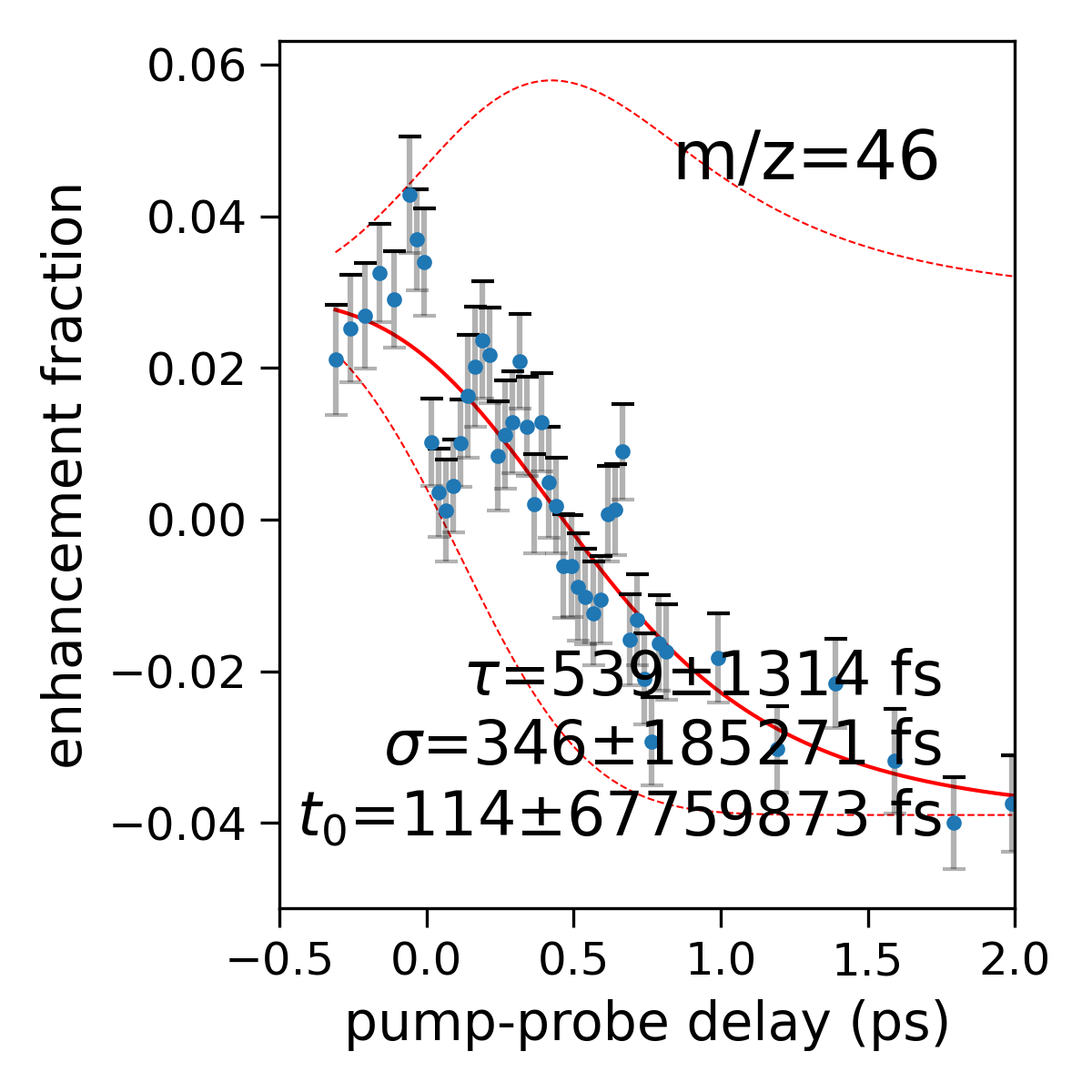}
        \caption[D6 H09 mq46]{\label{D6_H09_mq46} }
    \end{subfigure}%
    ~
    \begin{subfigure}[t]{0.3\linewidth}
        \centering
        \includegraphics[width=\linewidth*4/4]{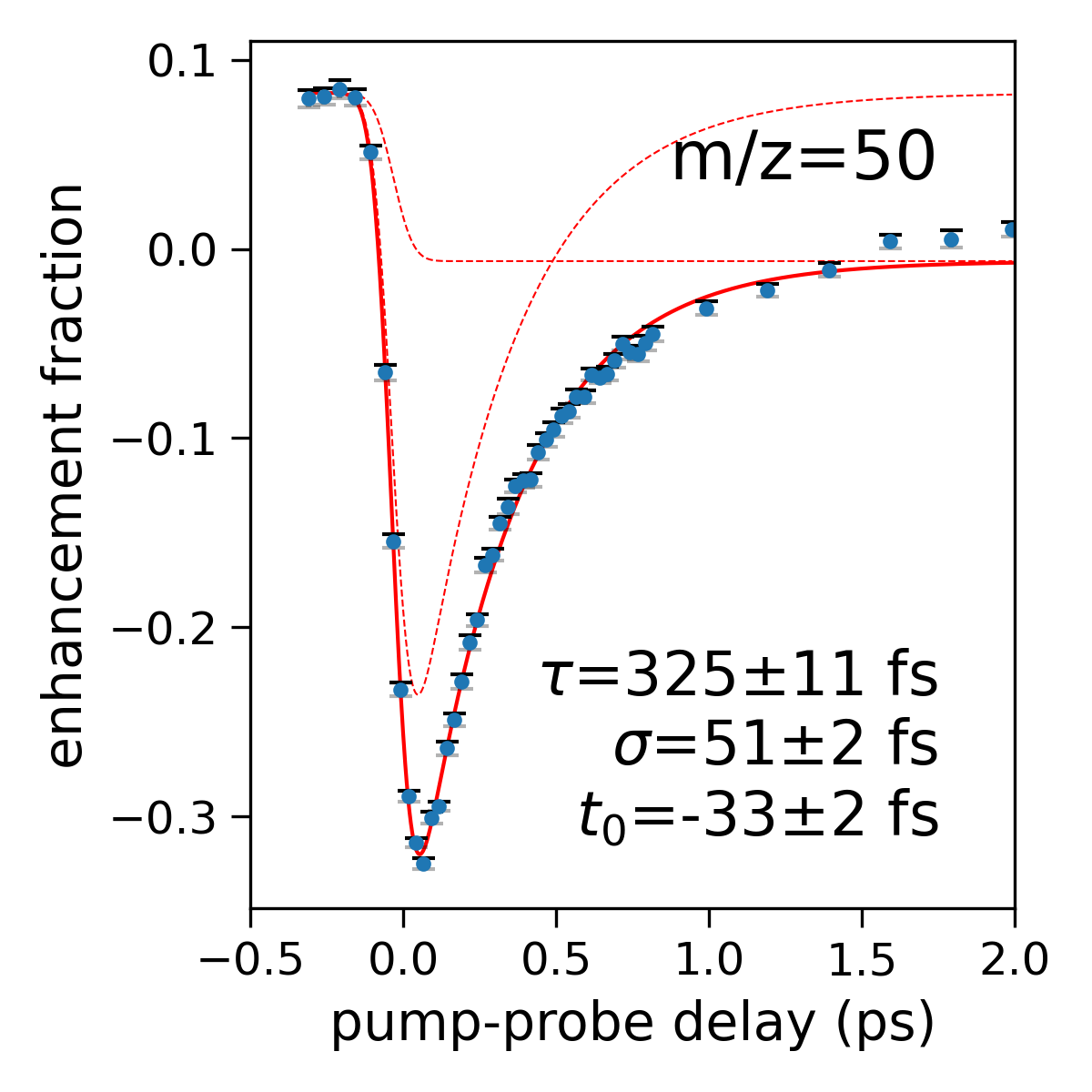}
        \caption[D6 H09 mq50]{\label{D6_H09_mq50} }
    \end{subfigure}%
    ~
    \begin{subfigure}[t]{0.3\linewidth}
        \centering
        \includegraphics[width=\linewidth*4/4]{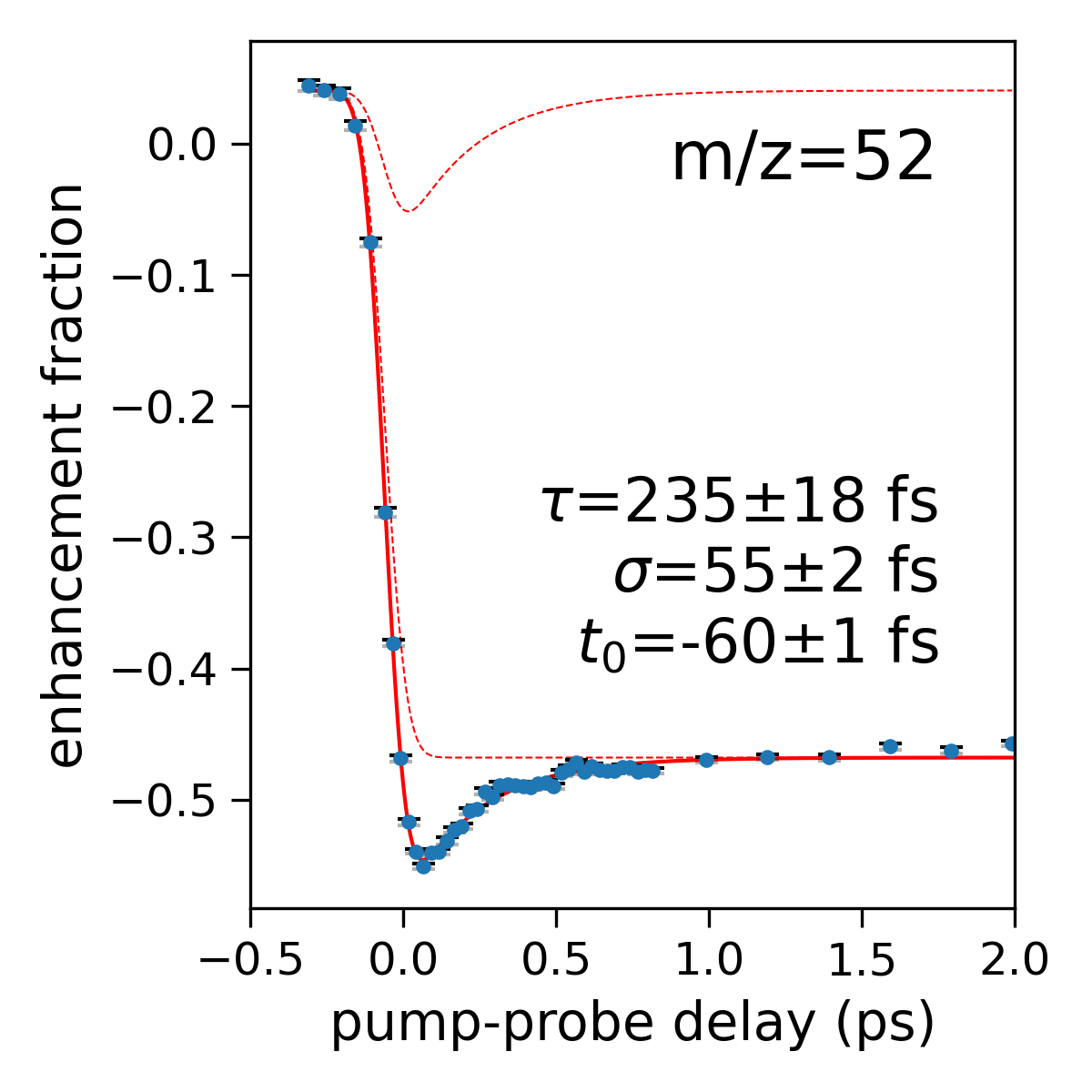}
        \caption[D6 H09 mq52]{\label{D6_H09_mq52} }
    \end{subfigure}%
		
    \caption[]{\label{fig:D6_H9} Pump-probe delay-dependent ion yields from the photoionization of fully-deuterated ethanol, at an XUV photon energy $hv$=31.7\ eV. The data was fit by the model in Eq.~\eqref{eq:depletion_fit}. }
\end{figure}

\section{Depletion data for Ne$^{+}$ ($m/z$=20, 22) and Ne$^{2+}$ ($m/z$=10)}

\begin{figure}[t]
    \centering
    \includegraphics[width=\linewidth*4/4]{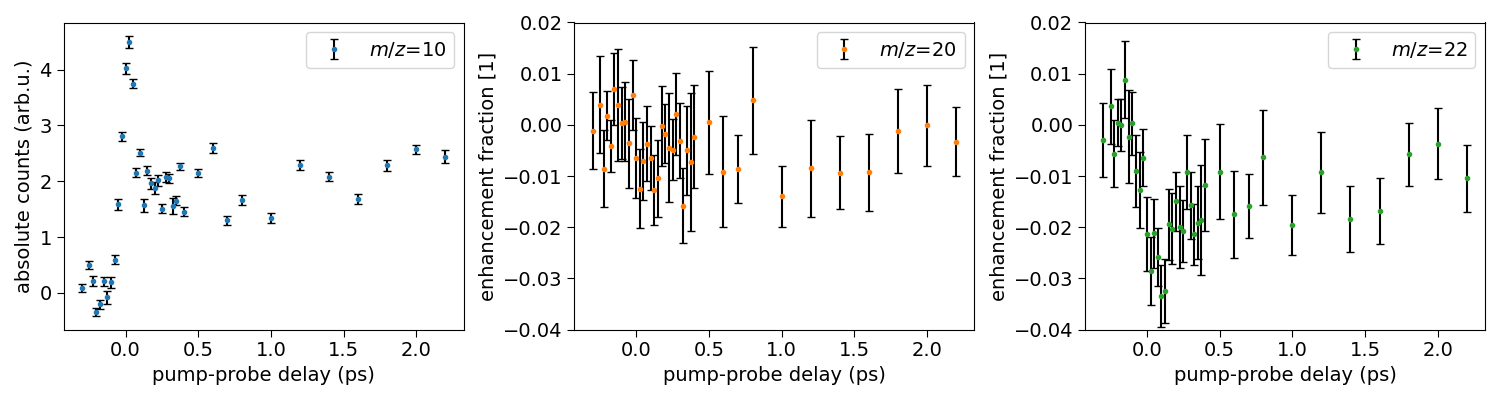}
    \caption[]{\label{fig:neon_D0_H9} Pump-probe delay-dependent ion yields of Neon obtained concurrently with non-deuterated ethanol (c.f. Supplementary Figure.~\ref{fig:D0_H9}, at an XUV photon energy $hv$=31.7\ eV. Absolute ion-yield shown for $m/z$=10, while the enhancement ratio for $m/z$=20, 22 are given.}
\end{figure}

\begin{figure}[t]
    \centering
    \includegraphics[width=\linewidth*4/4]{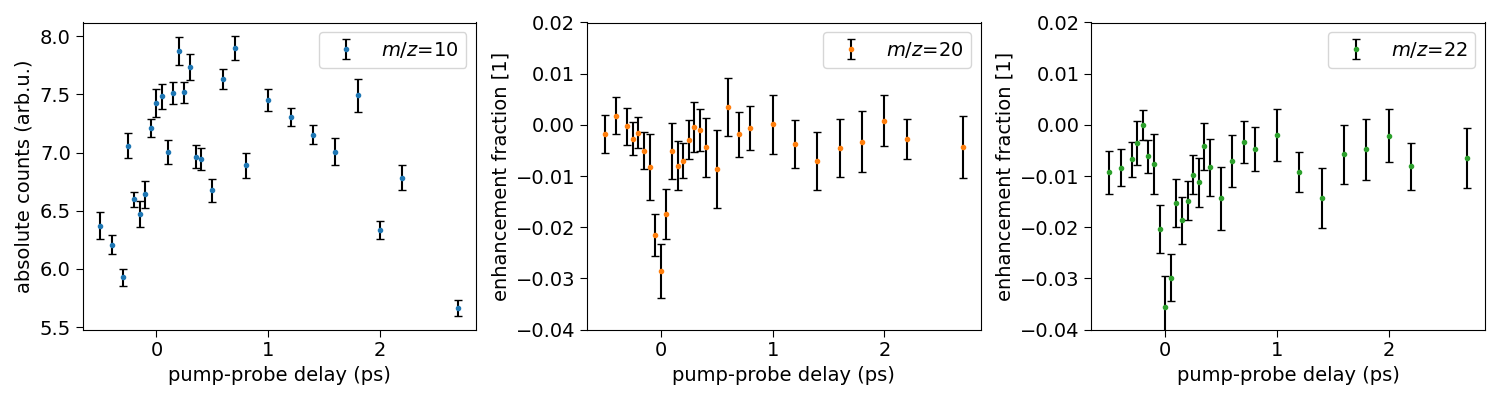}
    \caption[]{\label{fig:neon_D0_H7} Pump-probe delay-dependent ion yields of Neon obtained concurrently with non-deuterated ethanol (c.f. Supplementary Figure.~\ref{fig:D0_H7}, at an XUV photon energy $hv$=24.7\ eV. Absolute ion-yield shown for $m/z$=10, while the enhancement ratio for $m/z$=20, 22 are given.}
\end{figure}

\begin{figure}[t]
    \centering
    \includegraphics[width=\linewidth*4/4]{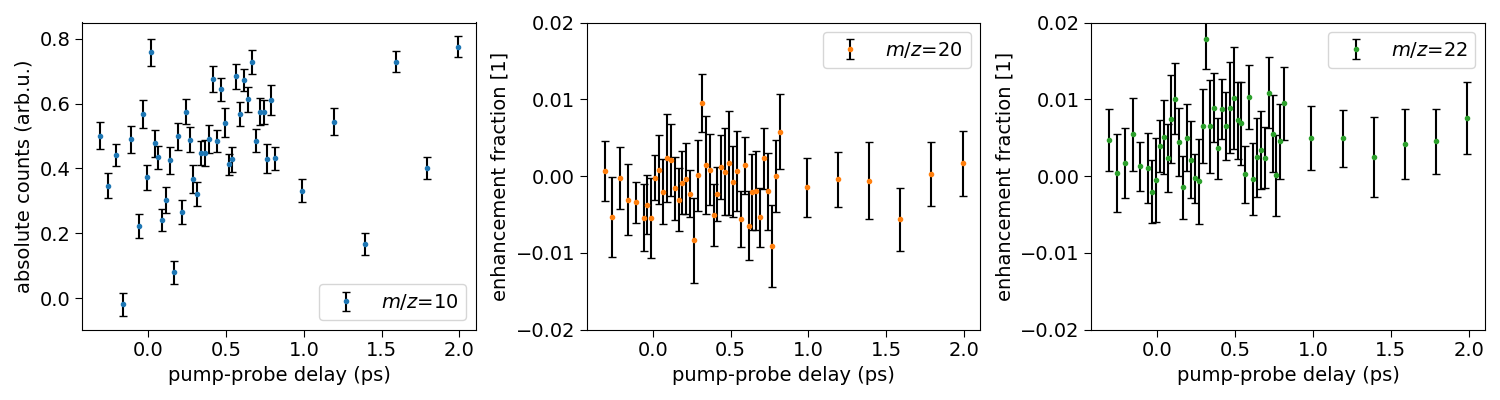}
    \caption[]{\label{fig:neon_D6_H9} Pump-probe delay-dependent ion yields of Neon obtained concurrently with fully-deuterated ethanol (c.f. Supplementary Figure.~\ref{fig:D0_H9}, at an XUV photon energy $hv$=31.7\ eV. Absolute ion-yield shown for $m/z$=10, while the enhancement ratio for $m/z$=20, 22 are given.}
\end{figure}

For the measurements performed on ethanol, we additionally include data for the Neon ion yields $^{20}$Ne$^+$, $^{22}$Ne$^+$, and $^{20}$Ne$^{2+}$, (Supplementary Figures~\ref{fig:neon_D0_H9}, \ref{fig:neon_D0_H7} and \ref{fig:neon_D6_H9}). 
In a previous study \cite{Livshits_2020_CommunChem} the Ne$^{2+}$ yield has been used to determine the cross correlation between a broad band XUV pump pulse and an NIR probe pulse.
Our data show that the variation of the experimental conditions between measurements (see Supplementary Table 1) influence the observed dynamics considerably. 
For the measurements on H$_6$-ethanol with 31.7 eV, the Ne$^{2+}$ signal is enhanced by a delayed UV pulse, that induces multi-photon ionization of excited states of Ne$^+$. The corresponding depletion of the Ne$^+$ signal is only observed for the $^{22}$Ne isotope, since the more abundant $^{20}$Ne isotope slightly saturated the detector. 
In the data set on D$_6$-ethanol at the same photon energies (Supplementary Figure \ref{fig:neon_D6_H9}) the lower UV intensity (see Supplementary Table 1) makes these multi-photon ionization processes unfeasible, and thus explains the lack of dynamics in the neon ion yields.
For the measurement at 24.7 eV (Supplementary Figure \ref{fig:neon_D0_H7}) the different XUV photon energy changes the populated excited states of Ne$^+$ and relative yields of Ne$^+$ and Ne$^{2+}$. Here only the Ne$^+$ signals show dynamics in their yields.

\bibliography{references}